\documentclass[11pt,a4paper]{article}

\usepackage{ascmac}
\usepackage{amsmath}
\usepackage{amssymb}
\usepackage{bm}
\usepackage[footnotesize,bf]{caption}
\usepackage{fullpage}
\usepackage{color}
\usepackage{float}
\usepackage{graphicx}
\usepackage[hidelinks]{hyperref} 
\usepackage[]{natbib}
\usepackage{subfigure}
\usepackage{txfonts}
\usepackage{threeparttable}
\usepackage{wrapfig}
\usepackage[]{multicol}
\usepackage{wrapfig}
\usepackage{authblk}

\usepackage{floatflt}

\title{\large Complete list of the ASTRO-H Science Working Group}
\date{\vspace{-0.5cm}}
\newcommand{\MakeWhitePaperTitle}{
	\begin{center}
		\begin{figure}
			\vspace{1cm}
			\begin{center}
				\includegraphics[width=0.2\hsize]{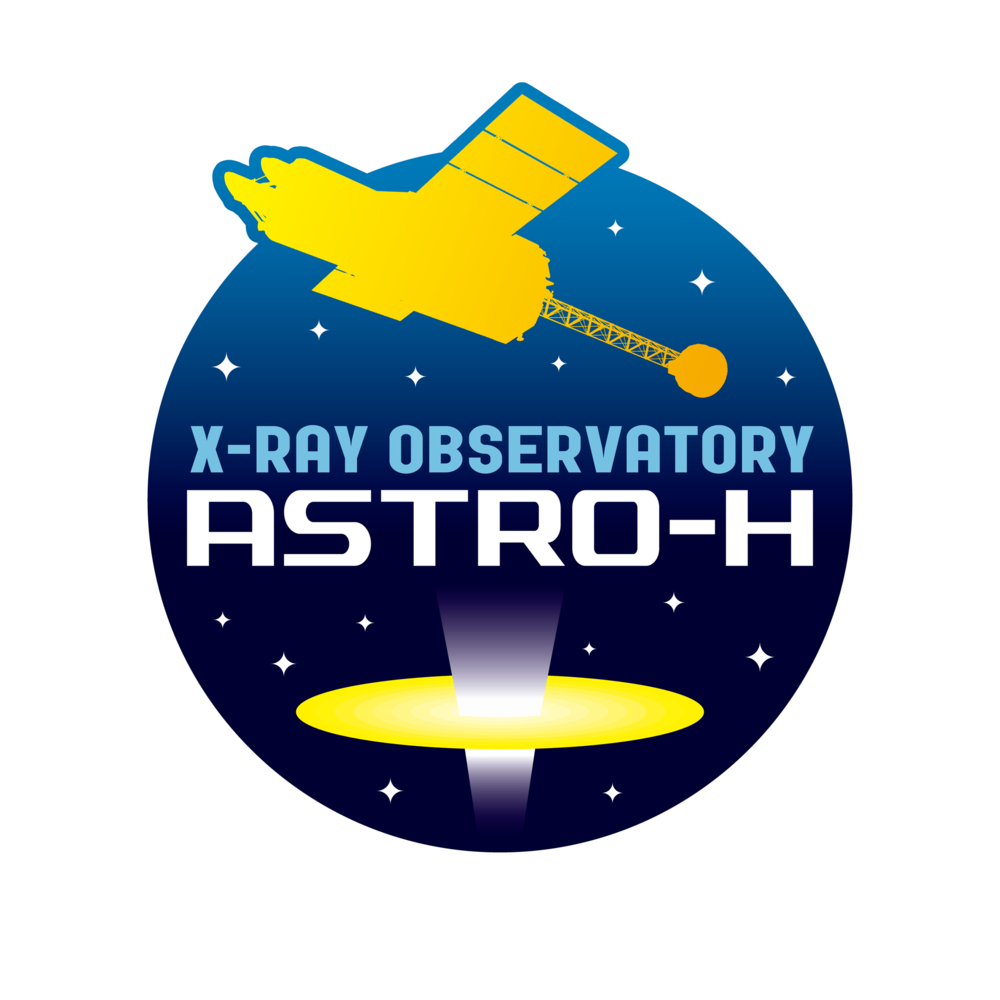}
			\end{center}
		\end{figure}
		\vspace{1cm}
		{\LARGE
		ASTRO-H Space X-ray Observatory\\
		White Paper\\
		}
		\vspace{5mm}
		{\large
		\WhitePaperTitle\\
		}
		\vspace{1cm}
		{
		\WhitePaperAuthors\\
		on behalf of the ASTRO-H Science Working Group
		}
	\end{center}
}

\usepackage{authblk}
\author[a]{Tadayuki~Takahashi}
\author[a]{Kazuhisa~Mitsuda}
\author[b]{Richard~Kelley}
\author[c]{Felix~Aharonian}
\author[d]{Hiroki~Akamatsu}
\author[e]{Fumie~Akimoto}
\author[f]{Steve~Allen}
\author[g]{Naohisa~Anabuki}
\author[b]{Lorella~Angelini}
\author[h]{Keith~Arnaud}
% ...
\affil[a]{Institute of Space and Astronautical Science (ISAS), Japan Aerospace Exploration Agency (JAXA), Kanagawa 252-5210, Japan}
\affil[b]{NASA/Goddard Space Flight Center, MD 20771, USA}
\affil[c]{Astronomy and Astrophysics Section, Dublin Institute for Advanced Studies, Dublin 2, Ireland}
\affil[d]{SRON Netherlands Institute for Space Research, Utrecht, The Netherlands}
\affil[e]{Department of Physics, Nagoya University, Aichi 338-8570, Japan}
\affil[f]{Kavli Institute for Particle Astrophysics and Cosmology, Stanford University, CA 94305, USA}
\affil[g]{Department of Earth and Space Science, Osaka University, Osaka 560-0043, Japan}
\affil[h]{Department of Astronomy, University of Maryland, MD 20742, USA}
% ...

\begin{document}

\newcommand{\WhitePaperTitle}{Broad-band Spectroscopy and Polarimetry}
\newcommand{\WhitePaperAuthors}{
	P.~Coppi~(Yale~University),
	{\L}.~Stawarz~(JAXA),
	C.~Done~(Durham~University),
	Y.~Fukazawa~(Hiroshima~University),
	P.~Gandhi~(Durham~University\footnote{Also at University of Southampton}),
	K.~Hagino~(JAXA),
	S.~LaMassa~(Yale~University),
	P.~Laurent~(CEA~Saclay),
	G.~Madejski~(KIPAC/SLAC)
	T.~Mizuno~(Hiroshima~University),
	K.~Mukai~(NASA/GSFC/CRESST \& UMBC),
	H.~Odaka~(JAXA),
	H.~Tajima~(Nagoya~University),
	Y.~Tanaka~(Hiroshima~University),
	F.~Tombesi~(University~of~Maryland),~and
	M.~Urry~(Yale~University)
}
\MakeWhitePaperTitle

\begin{abstract}
The broad energy range spanned by {\it ASTRO-H} instruments, from 
$\sim 0.3$ to $600$\,keV, with its high spectral resolution calorimeter and
sensitive hard X-ray imaging, offers unique opportunities to study
black holes and their environments. The ability to measure polarization
is particularly novel, with potential sources including blazars, Galactic 
pulsars and X-ray binaries. In this White Paper, we present an
overview of the synergistic instrumental capabilities and the
improvements over prior missions. We also show how {\it ASTRO-H} fits 
into the multi-wavelength landscape. We present in more detail examples 
and simulations of key science {\it ASTRO-H} can achieve in a typical 100\,ksec 
observation when data from all four instruments are combined. Specifically, 
we consider observations of black-hole source (Cyg\,X-1 and GRS\,1915+105), 
blazars (Mrk\,421 and Mrk\,501), a quasar (3C\,273), radio galaxies (Centaurus\, A
and 3C\,120), and active galaxies with a strong starburst (Circinus and
NGC\,4945). We will also address possible new discoveries expected from
{\it ASTRO-H}.
\end{abstract}

\maketitle
\clearpage

\tableofcontents
\clearpage

\section{Introduction}
\label{S:intro}

Any comprehensive insight into the nature of astrophysical sources of
high-energy radiation and particles requires a \emph{multi-wavelength}
approach. That is because the high-temperature and non-thermal particles
energized in extreme astrophysical environments emit over an extremely
broad range of the electromagnetic spectrum, from the low-frequency radio
domain up to high and very high energy $\gamma$-rays. Hence, high
quality, multiwavelength data is required to 
access the source energetics and content, and to identify robustly
the dominant particle acceleration and radiative processes
involved. Many of these sources are highly variable, so these data
need to be simultaneous, and repeated so as to sample different source
activity levels. Only
recently such a task has become feasible, due to the development of
modern ground-based and space-borne instruments, ranging from
high-resolution radio interferometers (EVLA, ALMA, soon SKA) up to
sensitive $\gamma$-ray satellites (AGILE, {\it Fermi}-LAT) and
Cherenkov Telescopes (H.E.S.S., MAGIC, VERITAS, soon also CTA). The
general scope of this white paper is to discuss the
X-ray domain as seen by {\it ASTRO-H}
in this multiwavelength context. We
argue in particular that the operation of the Hard X-ray Imager (HXI)
and the Soft Gamma-ray Detector (SGD), in \emph{synergy} and not in
conflict with the operation of the Soft X-ray Spectrometer System
(SXS), assures the sensitivity, energy resolution, and spectral
coverage necessary for achieving a quantitative progress in the field
\citep[see also][]{Takahashi13}.

One of the most pressing open problems in high-energy astrophysics
which can be resolved, at least partly, by future {\it ASTRO-H}
observations, is the exact coupling between the accretion processes
fueling supermassive black holes (SMBHs) in active galactic nuclei
(AGN), and their jet activity. While it is now widely accepted that
extremely luminous, relativistic jets associated with \emph{some} AGN
are formed via an efficient extraction of the energy and angular
momentum from the SMBH/accretion disk system \citep{Begelman84},
several crucial aspects of jet formation process are still under
debate, including the exact relation between disk state
transitions and jet production efficiency, and the role of any
external circum-nuclear medium in jet collimation. As discussed in
\S\,\ref{S:jets} below, the data gathered with {\it ASTRO-H} can
provide valuable and novel input on these subjects, 
robustly disentangling the AGN jet, accretion disk, and disk wind
radiative signatures in several targets via detailed spectroscopy
and timing analysis, augmented by multi-wavelength correlation studies
involving simultaneous radio and $\gamma$-ray observations. 
Analogous studies have been successfully conducted in the past for 
Galactic jet sources associated with X-ray binaries, but only rarely
for AGN, mostly due to the fact that jetted (`radio-loud') AGN are
much dimmer in X-rays than nearby XRBs. Thus, only with the
new-generation X-ray telescopes such as {\it ASTRO-H} can the
jet-accretion disk coupling in active galaxies be analyzed extensively
enough to reach robust conclusions.

The other relevant topic calling for in-depth observational studies
with {\it ASTRO-H} is the interplay between the SMBH activity and the
nuclear starburst activity in evolving galaxies. There is
general consensus that there is a strong connection between these
processes, and that this link is crucial for understanding
galaxy formation processes in general \citep[see, e.g.,][]{Kormendy13}. 
However, previous investigations of the problem were
hampered by the fact that available X-ray data with limited energy
resolution and spectral coverage did not allow for a robust
partition between the AGN and the starburst emission components in the
observed spectra. As discussed in \S\,\ref{S:agn}, broad-band X-ray
spectroscopy of even particularly complex and therefore particularly
interesting systems simultaneously with SXS, HXI, and SGD, will enable
address this issue properly.

The AGN-starburst connection and the jet-disk coupling in active
galaxies are not the only topics of interest in the context of
`broad-band' studies with {\it ASTRO-H}, and in \S\,\ref{S:other} we
briefly mention some other selected problems, focusing there on the
potential of hard X-ray instruments HXI and SGD operating jointly in
the $5-600$\,keV range. The hard X-ray/soft $\gamma$-ray astronomy
\emph{is} the field for making new discoveries indeed, simply because
this electromagnetic window is the least studied one so far, due to
severe instrumental limitations and challenges. The joint HXI and SGD
observations are therefore very relevant for constraining poorly
known properties of various types of astrophysical systems which
release bulk of their radiative output around MeV photon energies,
including $\gamma$-ray detected novae, high-$z$ blazars, or tidal
disruption events. However, the particularly exciting and unique hallmark
of {\it ASTRO-H} is the potential for the hard X-ray polarimetry with
the SGD. This issue is discussed first in \S\,\ref{S:polar} below.

\section{X-ray Polarimetry of Astrophysical Sources}
\label{S:polar}

\begin{figure}[!t]
\begin{center}
\includegraphics[width=1.0\textwidth]{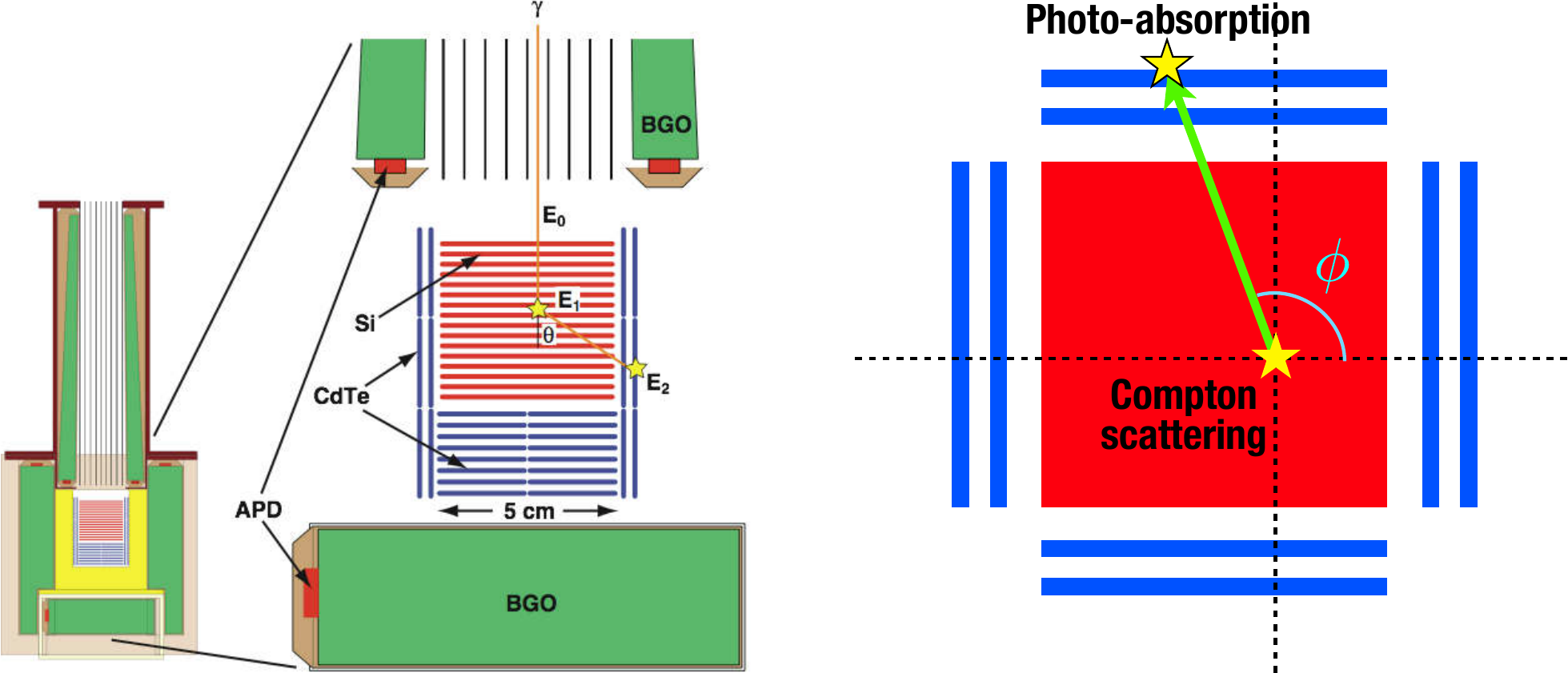}
\caption{Schematic drawing of a single Compton camera unit of the SGD. The left panel shows a side view and the right panel presents the top view. The system measures two interactions (normally, Compton scattering and photo-absorption) and determines geometrical information on the scattering, namely the scattering angle $\theta$ and the azimuth angle $\phi$.}
\label{fig:broadband/polarization/SGD}
\end{center}
\end{figure}

In addition to imaging, spectroscopy, and timing measurements,
polarimetry in the X-ray and $\gamma$-ray bands is a new promising
probe of high-energy phenomena in the Universe. In general,
polarimetry provides us with crucial information about the radiation
mechanisms involved, as well as  the structure and content of the
emitting regions. That is especially true in the X-ray domain, since
polarization signatures in this range may arise from vastly different
processes, including bremsstrahlung emission from anisotropic electron
distribution, synchrotron emission in ordered magnetic fields,
anisotropic Compton scattering, inverse-Comptonization of a polarized
photon field, or finally, photon propagation through a highly magnetized
plasma ($B \geq 10^{12}$\,G). Hence, substantially polarized X-ray
emission may in principle be expected in a variety of astrophysical
systems such as stellar flares, pulsars, pulsar nebulae, magnetars,
accreting white dwarfs, supernova remnants, black hole accretion disks
and coronae, jetted AGN, microquasars, or gamma-ray bursts
\citep[see][and also the ``Pulsars \& Magnetars'' {\it ASTRO-H} White
Paper]{Lei97,Blandford02,Bellazzini10,Krawczynski11}.

The problem is, however, that X-ray polarization measurements are much
more challenging compared with those at other wavelengths such as
optical, infrared, or radio bands. The only satellite instruments with
significant polarization capability were {\it OSO-8}, launched in
1975. which detected the Crab nebula \citep{Weisskopf76}, and provided
upper limits for several other X-ray bright Galactic sources at 2.6
and 5.2\,keV, and {\it INTEGRAL} SPI and IBIS ((20\,keV--1\,MeV) which
detected the Crab pulsar and nebulae \citep{Dean08,Forot08}, the
gamma-ray burst GRB\,041219 \citep{McGlynn07,Goetz09}, and
(marginally) the microquasar Cygnus\,X-1
\citep{Laurent11,Jourdain12}. However, these measurements suffer from
large uncertainties due to low photon statistics and high in-orbit
background. Similar issues apply to current balloon-borne X-ray
polarimeters including PoGoLite \citep[2--100\,keV;][]{Pearce12},
X-Calibur \citep[20--80\,keV;][]{Beilicke12}, POLAR
\citep[50--500\,keV;][]{Orsi11}, and GRAPE
\citep[50--500\,keV;][]{Bloser09}. Nevertheless, the importance of
polarimetry is recognised by the multiple proposals for 
new small satellite missions, such as POET \citep[2--500\,keV;][]{Hill08}, GEMS
\citep[2--10\,keV;][]{Black10}, POLARIX
\citep[2--10\,keV;][]{Costa10}, or XIPE
\citep[2--10\,keV;][]{Soffitta13}, with 
SPHiNX (50--500\,keV) and PolariS (2--80\,keV) being selected for
further study. 

\subsection{X-ray Polarimetry with SGD}
\label{S:sgd}

\begin{figure}[!t]
\begin{center}
\includegraphics[width=0.4\textwidth]{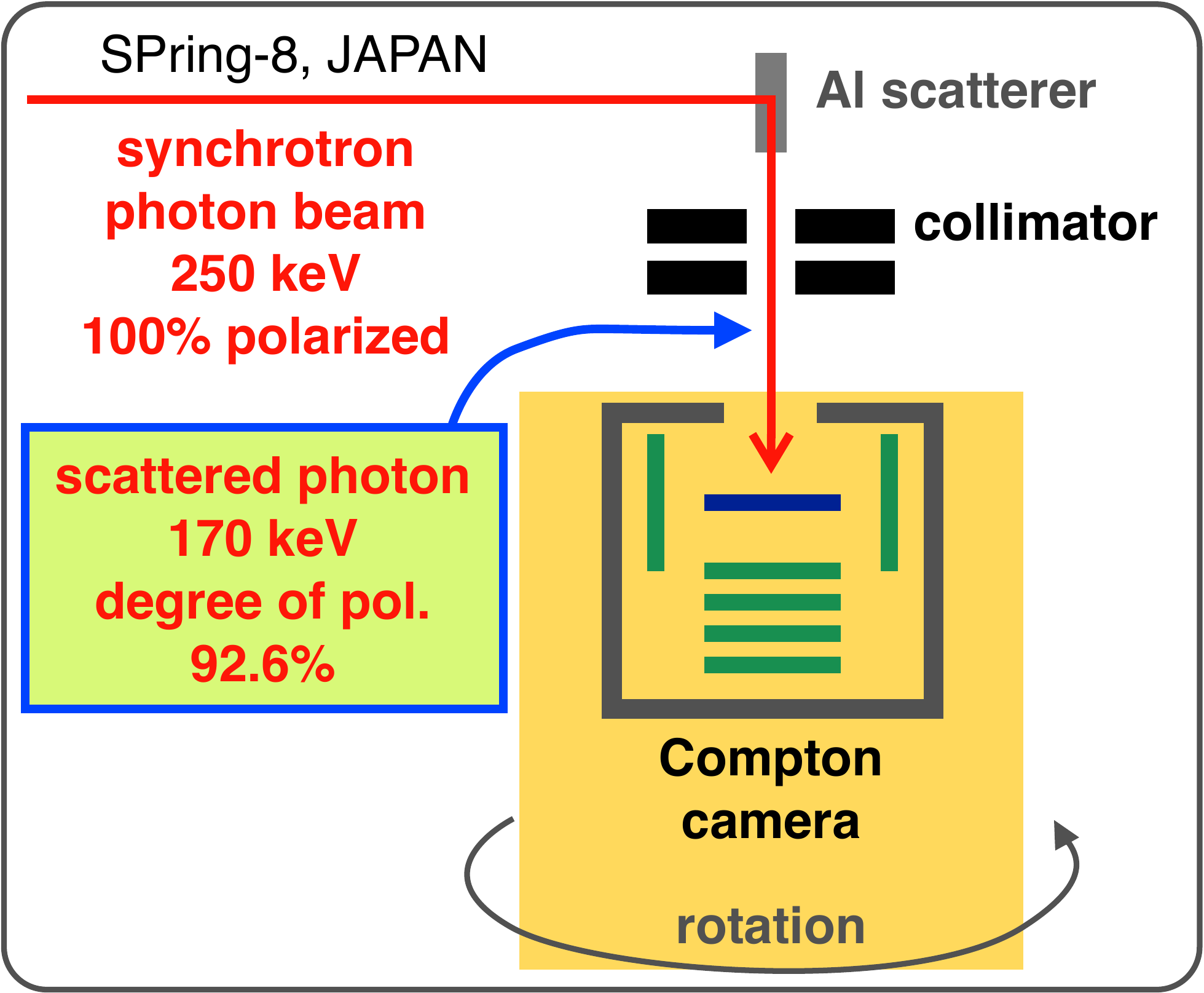}
\quad
\includegraphics[width=0.55\textwidth]{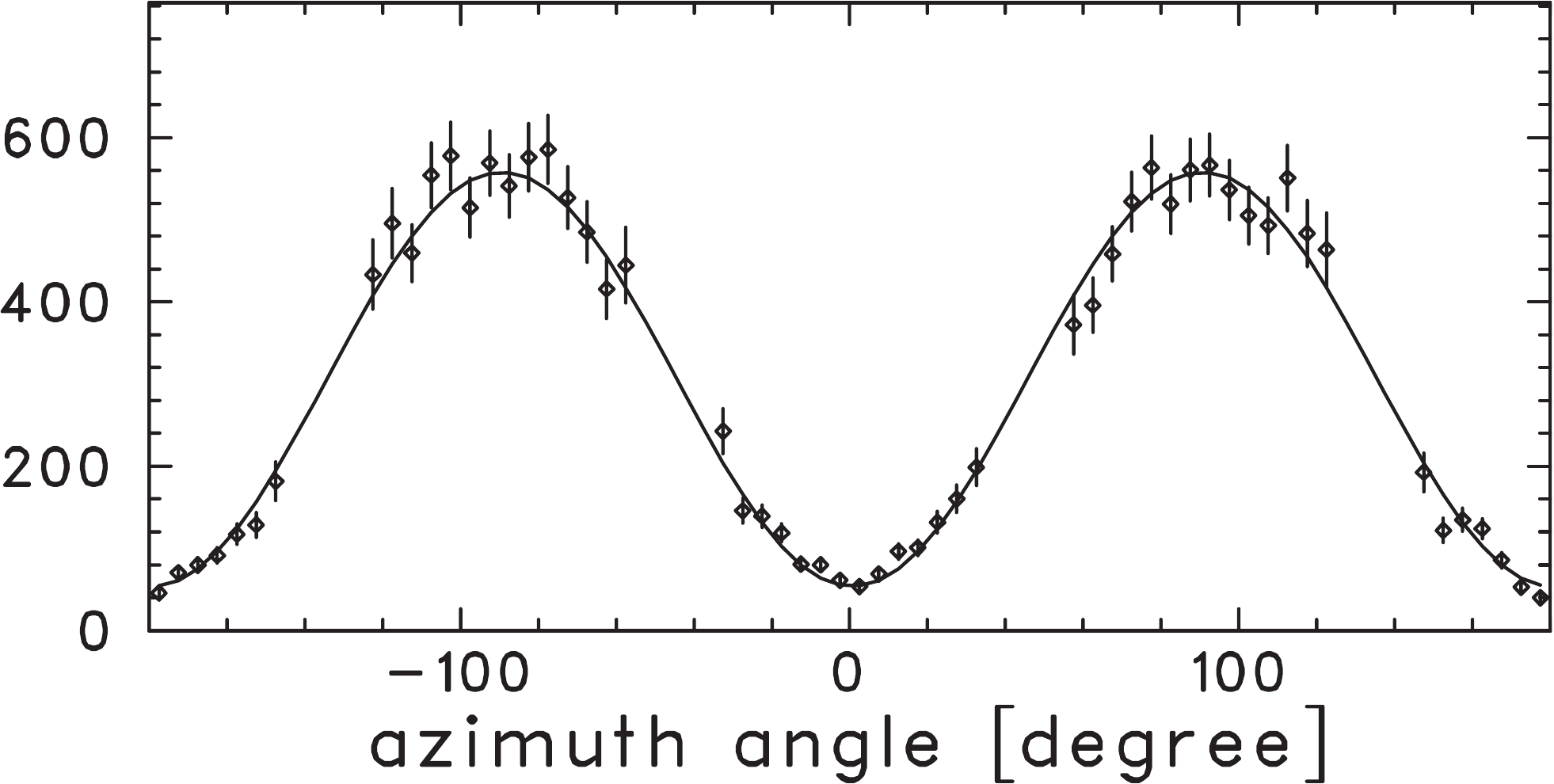}
\caption{Left: the experimental setup of the polarization measurement at SPring-8. Synchrotron beam photons (250\,keV) are scattered by an aluminum block and then collimated before incidence to the Compton camera. The camera system can be rotated upon the incident direction so that systematic uncertainties due to roll angle can be evaluated. Right: the obtained distribution of the azimuth angle after the response correction.}
\label{fig:broadband/polarization/spring8}
\end{center}
\end{figure}

The SGD is designed not only for spectroscopy but also for
high-precision polarimetry by measuring Compton scattering in the
detector system
\citep{Tajima:2010a,Tajima:2010b,Watanabe:2012,Watanabe:2014}. Moreover,
this instrument will be the first hard X-ray/soft $\gamma$-ray
polarimeter in orbit that has moderate collection area
and low instrumental background. For this reason, the SGD will improve
the sensitivity and precision of polarization measurements even in
a relatively short exposure time.\footnote{Conventional instruments that
use the principle of Compton polarimetry require longer exposure and
nontrivial treatment of evaluating systematic uncertainties.} The SGD
covers the energy range from about 50\,keV up to 600\,keV. The
sensitivity peak is at 80--100\,keV, depending on the signal
brightness.

The polarization measurement by the SGD utilizes the anisotropy of
the scattering direction of Compton scattering. As shown in
Figure\,\ref{fig:broadband/polarization/SGD}, the instrument
determines the scattering process of an incident photon in the
sensitive detectors by measuring the deposited energies and positions of
the interactions. The differential cross section of Compton scattering
is given by the Klein-Nishina formula,
\begin{equation}
\frac{d\sigma}{d\Omega} = \frac{r_e{}^2}{2} \, \left(\dfrac{E'}{E}\right) \, \left(\dfrac{E}{E'}+\dfrac{E'}{E}-2 \, \sin^2\!\theta \, \cos^2\!\chi \right) \, ,
\end{equation}
where $r_e$ denotes the classical electron radius, $\theta$ and $\chi$ denote the scattering angle and the azimuth angle of scattering measured from the polarization angle, respectively, while $E$ and $E'$ are energies before and after the scattering:
\begin{equation}
E' = \frac{E}{1+\dfrac{E}{m_e c^2} \, \left(1-\cos\theta\right)} \, .
\end{equation}
These formulae show that the azimuthal angle of Compton scattering has sinusoidal distribution, peaking at $\pm$90$^\circ$ to the polarization plane.

\citet{Takeda:2010} demonstrated the capability of the SGD for
high-precision polarimetry with a prototype model in the synchrotron
photon beam at SPring-8, Japan. The left panel of
Figure\,\ref{fig:broadband/polarization/spring8} shows the
experimental setup, where the Compton camera measures the collimated
170\,keV photons which are scattered by an aluminum block and have
polarization fraction of $92.6\%$. To extract polarization
information, the azimuthal angle distribution has to be corrected for
the detector response, including both the detection efficiency and
shielding by passive material. Monte Carlo simulations are used to
quantify the effect of the complicated structure of the Compton camera
system. The obtained distribution
of azimuthal angle after the response correction, shown in the right
panel of Figure\,\ref{fig:broadband/polarization/spring8}. This gives a
polarization fraction of $91\pm1$\% and polarization angle of
$0.2^\circ\pm 0.4^\circ$, consistent with the input 
with small uncertainties.\footnote{All the errors in
this section are presented in $1\sigma$ values.}

\begin{figure}[!t]
\begin{center}
\includegraphics[width=0.5\textwidth]{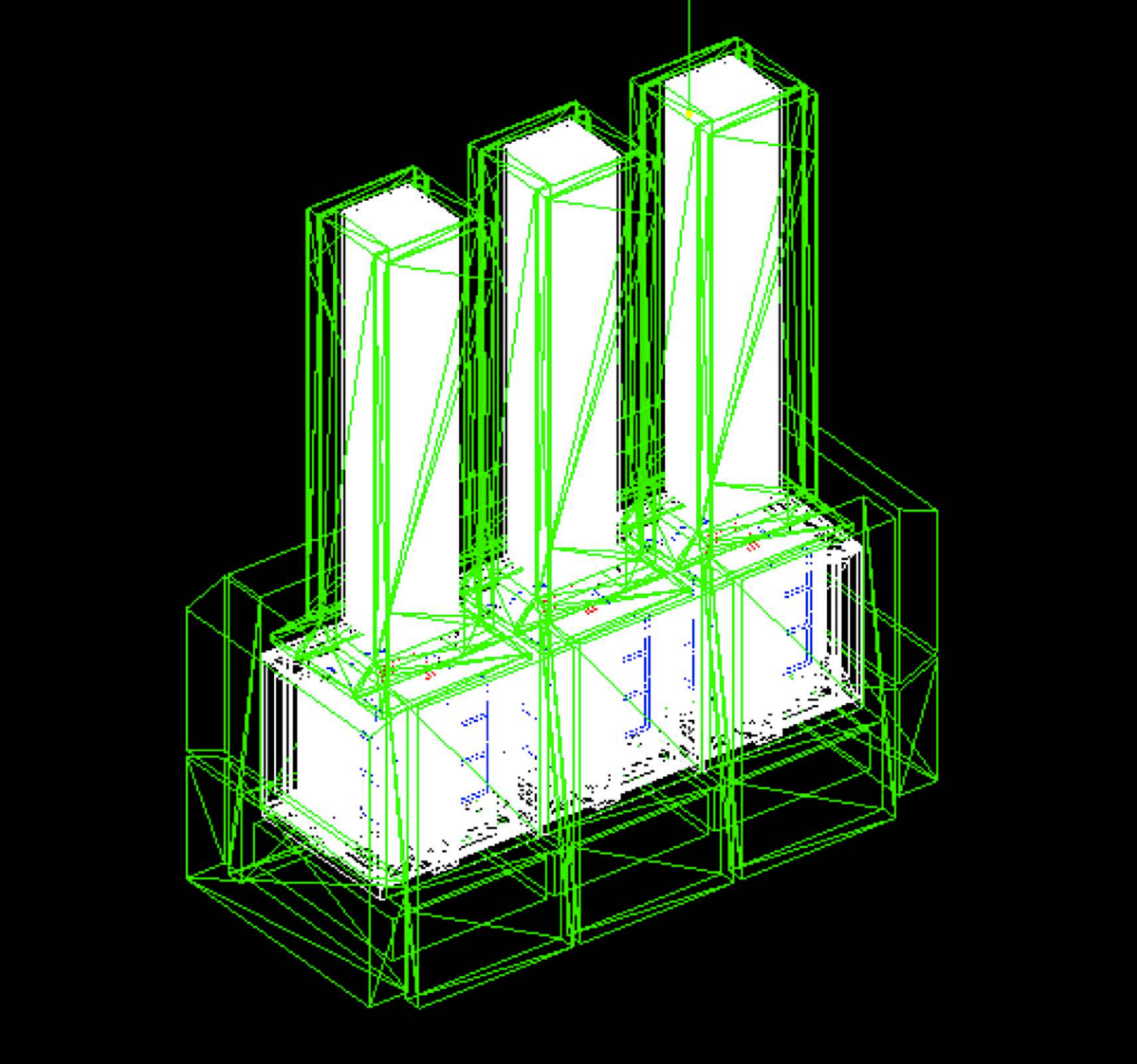}
\quad
\includegraphics[width=0.46\textwidth]{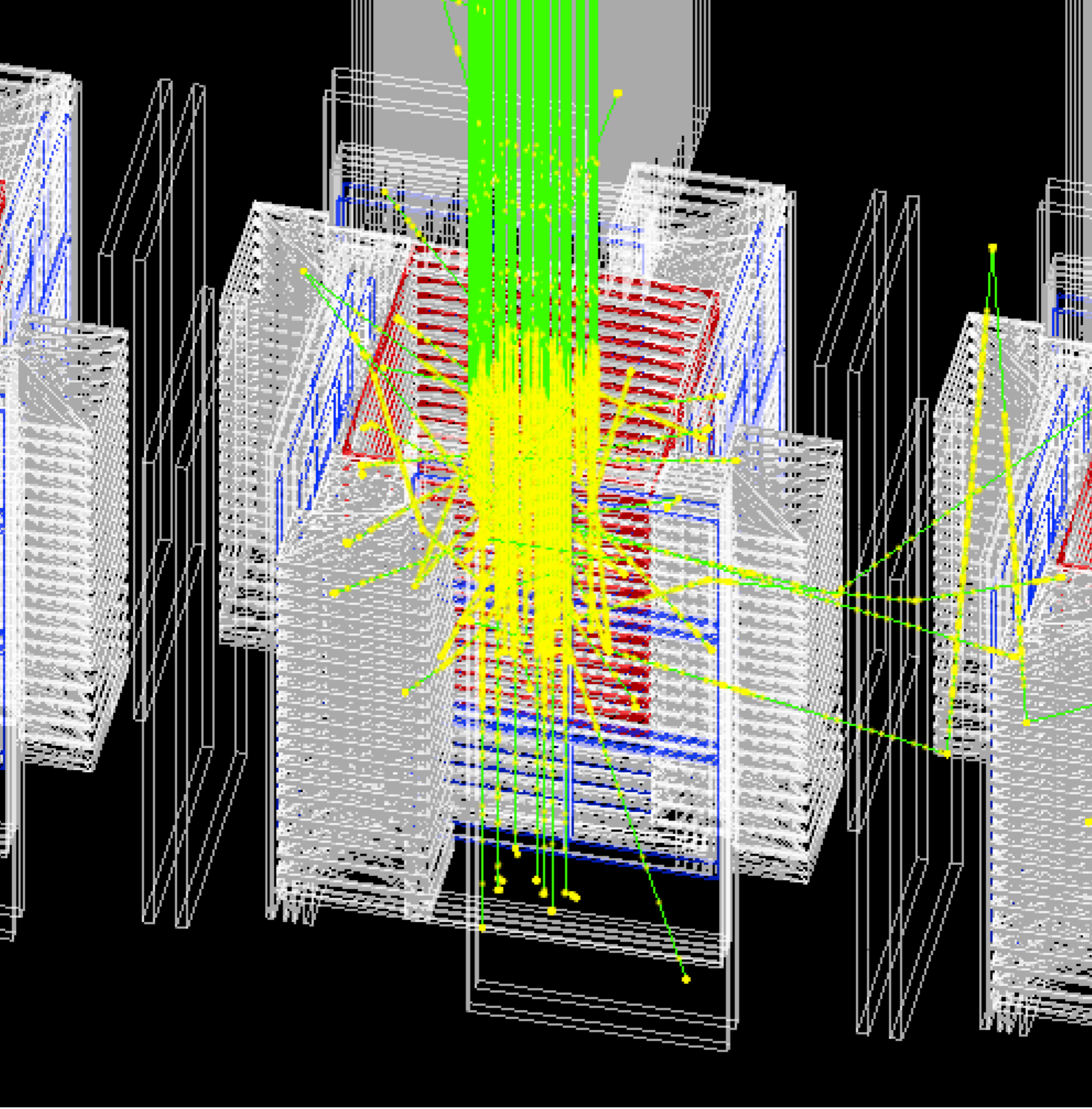}
\caption{Mass model of the SGD for Monte Carlo simulations. Left: the whole SGD unit. Right: zoom up about one of the Compton cameras. In the right panel, photon trajectories are drawn as green lines and interactions are drawn in yellow.}
\label{fig:broadband/polarization/mass_model}
\end{center}
\end{figure}

We have developed an SGD simulator using the Geant4 toolkit. This
evaluates the detector response and estimates in-orbit background, so
it can be used for optimizing observatory operations as well as data
analysis
\citep{Odaka:2010,Ozaki:2012}. Figure\,\ref{fig:broadband/polarization/mass_model}
shows the mass model of the SGD for the Monte Carlo simulations. This
describes the geometry and material of the sensitive detectors and the
surrounding passive structures.  Physical processes taking place in
the instrument are accurately tracked by the simulator, which also
takes account of charge transport in semiconductor sensors, signal
readout electronics and data acquisition system in order to reproduce
the detector response. We use this simulator to assess the
polarisation signal which can be detected against the background,
including the additional polarised light produced from scattering within the
detector and its housing. 
However, there are still systematic uncertainties due to uncertainties
in the mass model and background.  It is important to investigate
these using both the simulations and on-ground calibration
measurements. Further systematic uncertainties can be reduced by
performing multiple observations with different roll angle of the spacecraft.

\subsection{Crab Pulsar and Nebula}
\label{S:crab}

The Crab pulsar and nebula, as the X-ray and $\gamma$-ray brightest
source in sky, is the most feasible target for the polarimetry with
the SGD. Importantly, the pulsar and nebular emission components can
be decomposed despite the pulsar being only $\sim 10\%$ of the nebula
flux as they have different spectral, timing and polarisation behaviour.
These two components have different
origins: the pulsar emits via curvature radiation in the pulsar
magnetosphere while the pulsar wind nebula is dominated by the
synchrotron radiation. The SGD has sufficient timing resolution
($<1$\,ms) to resolve the behaviour on the pulsar spin period of 33\,ms.

\begin{figure}[!t]
\begin{center}
\includegraphics[width=0.495\textwidth]{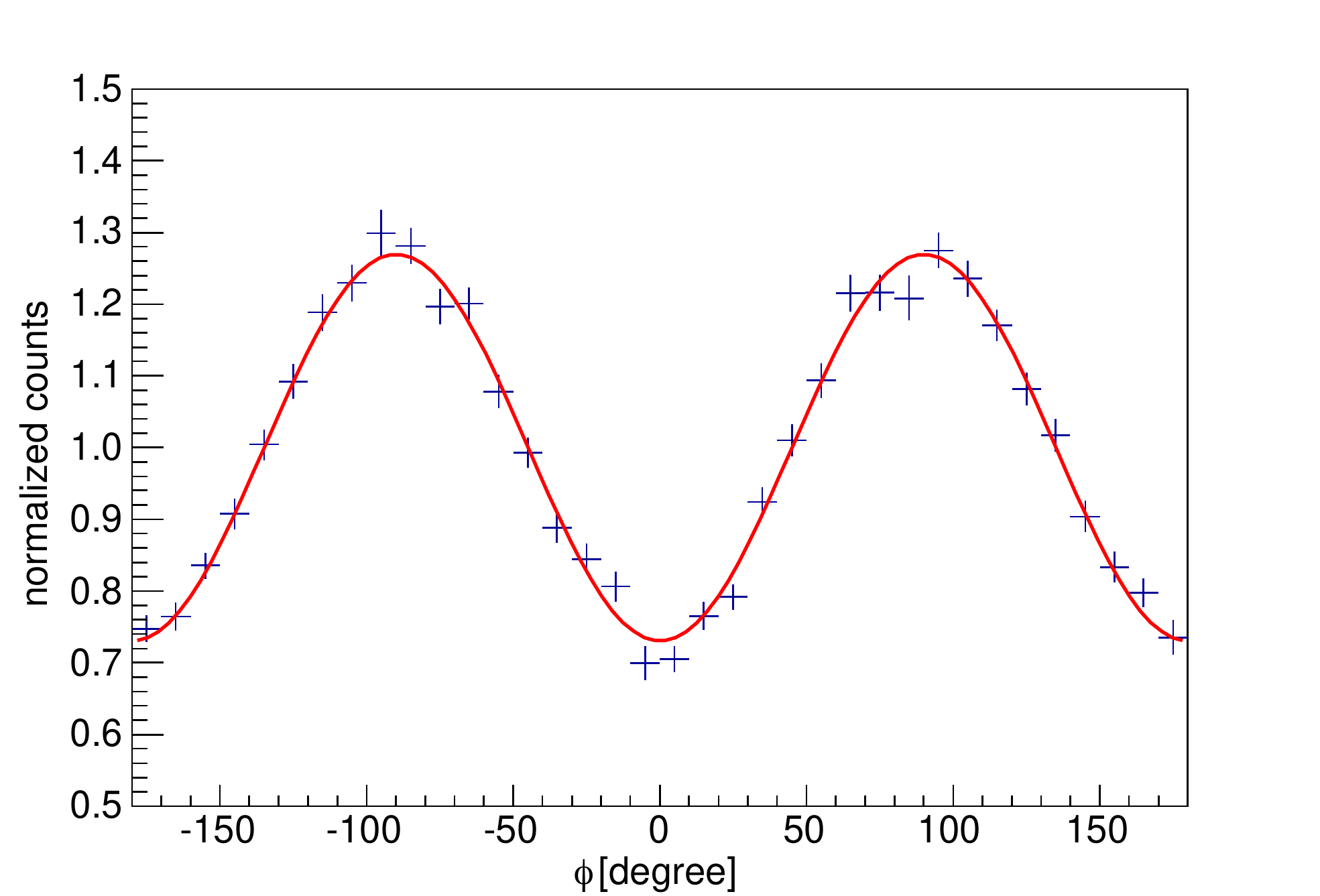}
\includegraphics[width=0.495\textwidth]{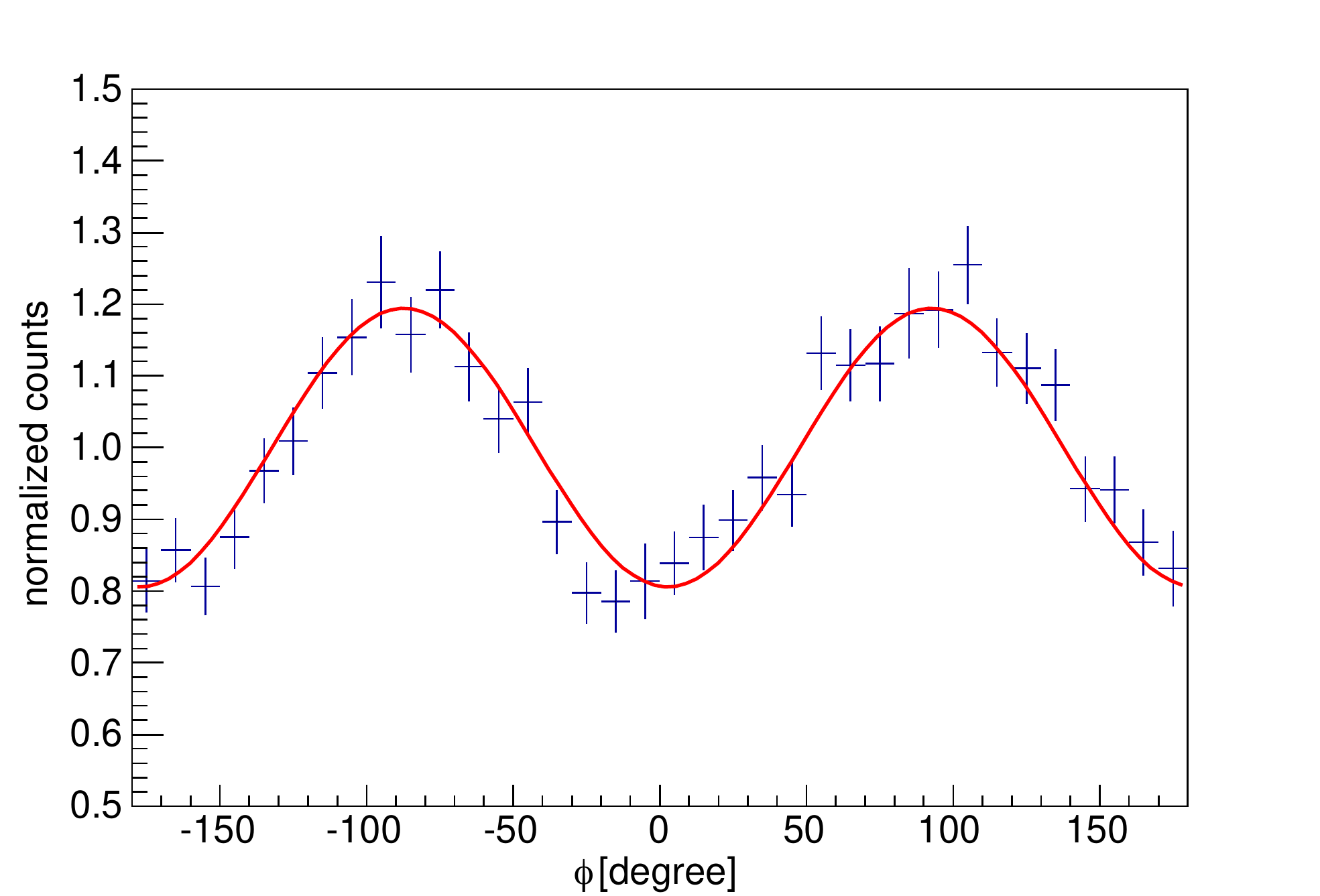}
\caption{Simulated polarimetry results of 100\,ks observation of Crab in energy bands of 60--100\,keV (left) and 180--330\,keV (right). Both assume flux of 1 Crab, photon index of 2.1, and polarization fraction of 50\%. These azimuth angle distribution are corrected by the detector response for unpolarized radiation.}
\label{fig:broadband/polarization/observation_crab}
\end{center}
\end{figure}

We conducted simulations to evaluate the technical
feasibility of the SGD polarimetry. First, we show simulation results
for the  two SGD units for a 100\,ks observation of the Crab nebula in
Figure\,\ref{fig:broadband/polarization/observation_crab}. The input
spectrum is assumed to be a power law, $F(E)=K \times E^{-\Gamma}$,
with the normalization $K=11.6\ \mathrm{photons\ s^{-1}\ cm^{-2}}$ at
1 keV, and the photon index $\Gamma = 2.1$. Assuming a polarization
fraction of 50\%, we expect clear detection of the polarization from a
1 Crab source. The
fitted values are $48\pm 1\%$ for the polarization fraction and
$0.1^\circ\pm 0.5^\circ$ for the polarization angle in the energy
range of 60--100\,keV. At the higher band between 180\,keV and
330\,keV, photon statistics becomes limited due to decreasing
detection efficiency, but the result still reveals a significant
detection; the fitted values are $48\pm 3\%$ for the polarization
fraction and $2.7^\circ\pm 1.6^\circ$ for the polarization
angle. These agree well with the assumed values for the simulation
(50\% polarization and $0^\circ$ angle). Moreover, the statistical
uncertainty will be much smaller than observational uncertainties of
currently reported measurements by {\it INTEGRAL} \citep{Dean08,
Forot08, Moran:2013}, though we need to be careful about systematic
errors. On the other hand, in the case of Crab the polarization
information in the non-peak phase can be used as a background in order
to reduce the background systematics; as a result, the sensitivity to
the Crab pulsar polarization can be better than for other
sources with comparable brightness.

\citet{Moran:2013} studied the linear polarization of the
Crab Nebula and pulsar in both the optical (HST/ACS) and in the
$\gamma$-rays ({\it INTEGRAL}/IBIS). For both wavelengths,
they found evidence of an alignment between the polarization position
angle of the pulsar and its rotation axis. They confirmed that the
synchrotron knot \citep{Hester95} located 0.65'' SE of the pulsar is
responsible for the highly polarized off-pulse emission seen in both
the optical \citep{Smith88,Slowikowska09} and $\gamma$-rays
\citep{Forot08}. Hence, further multi-wavelength studies of this
astrophysical system with {\it ASTRO-H} would enable a more precise
check for any correlations between the two energy ranges.
This could show whether the 
polarization of the system is linked to the highly variable flux from
the nebula seen in the 100--300\,MeV range by the {\it Fermi} and
AGILE telescopes \citep[][and references
therein]{Tavani11,Striani11,Mayer13} --- the so-called $\gamma$-ray
flares. Such monitoring could be achieved through simultaneous/joint
observations using optical polarimeters such as the Galway
Astronomical Stokes Polarimeter \citep[GASP;][]{Collins13}. These
multi-wavelength studies will benefit from the good {\it ASTRO-H}
timing capabilities, and will be nicely complemented by high time
resolution optical polarimeters. This will enable detailed
phase-resolved spectro-polarimetry analysis.

\subsection{Microquasars}
\label{S:XRBs}

Jetted black hole XRBs, or `microquasars', are the other promising
targets for the X-ray polarimetry with the SGD, since the reflection
at the accretion disk, thermal Comptonization in a hot disk corona,
and also synchrotron emission of the jet electrons, can all lead to
distinct polarization signatures in the hard X-ray/soft
$\gamma$-ray bands. Hence, energy dependent polarization measurements
(`spectropolarimetry') of such systems is in principle capable of
constraining the geometry of their accretion flows, and the magnetic
structure of their jets.

\begin{figure}[!t]
\begin{center}
\includegraphics[width=0.49\textwidth]{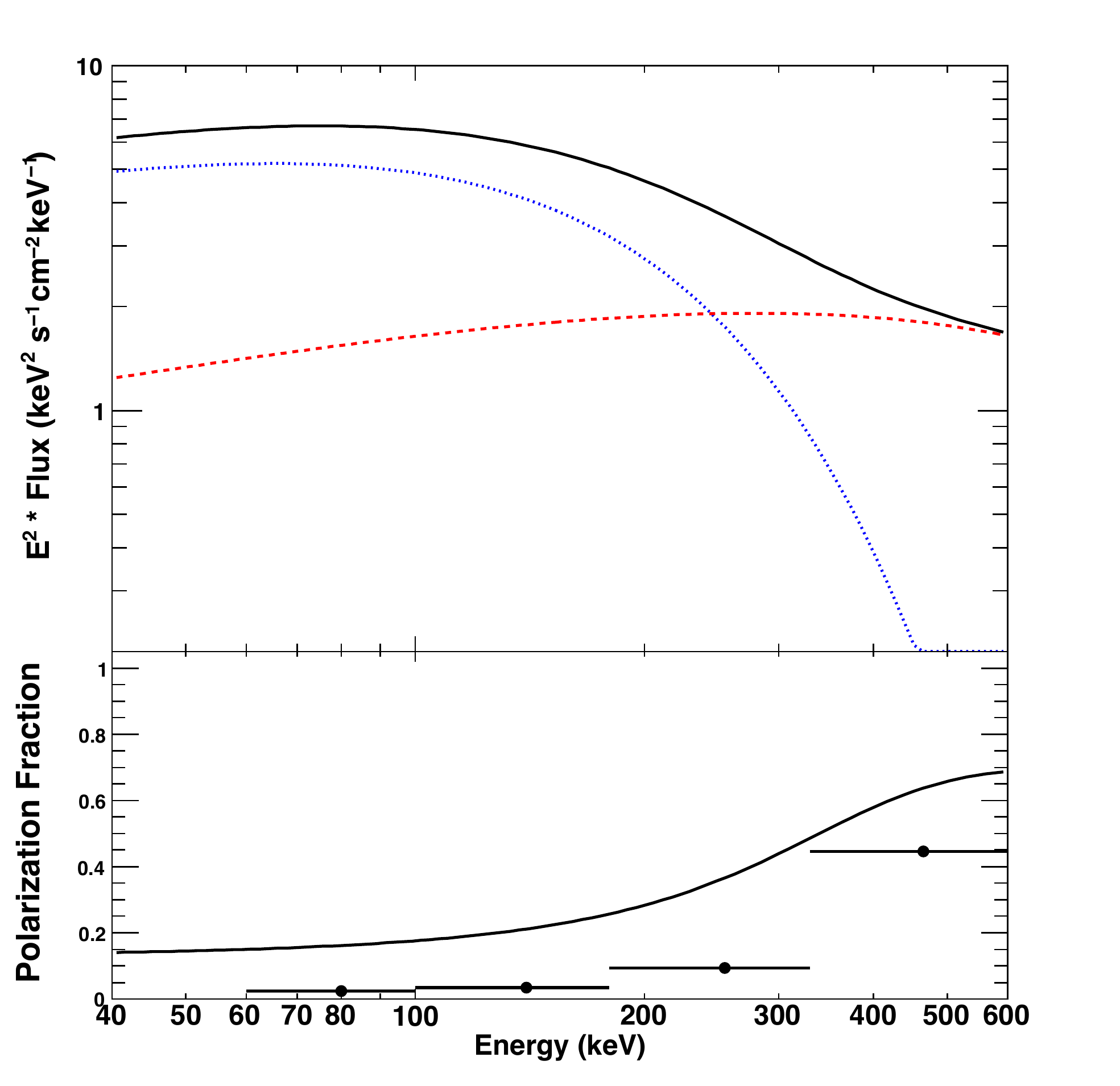}
\includegraphics[width=0.49\textwidth]{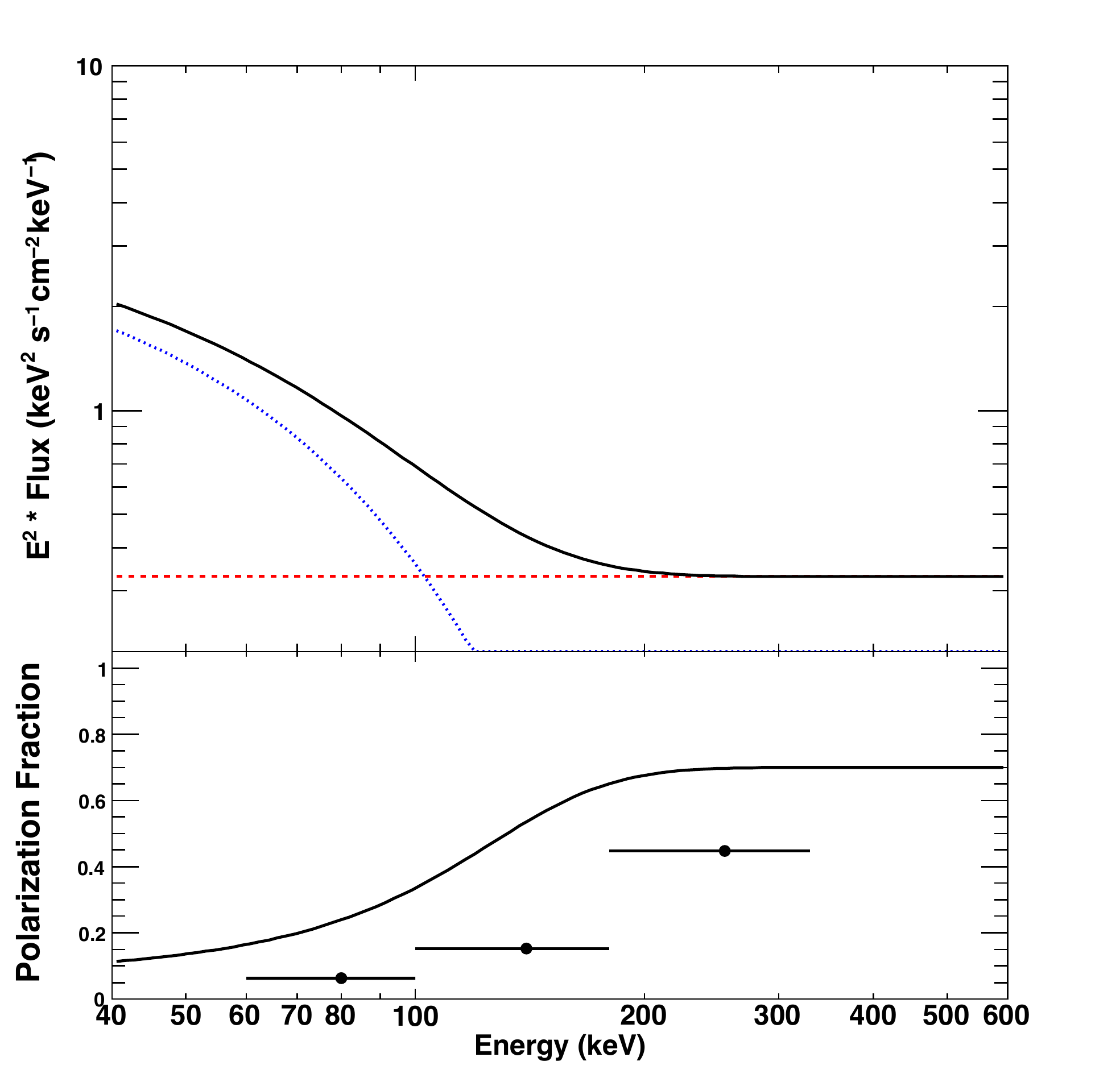}
\caption{Left: Cygnus\,X-1 model spectrum (top panel) and the expected polarization fraction (bottom panel). In the top panel the low energy and high energy components are shown by blue and red curves, respectively. In the bottom panel the ${\rm MDP_{99}}$ is also plotted. Right: The same as the left panel but for GRS\,1915+105.}
\label{fig:cygx1} 
\end{center}
\end{figure} 

A marginal detection of soft X-ray polarization from Cygnus\,X-1 was
reported by \citet{Long80} with the position angle of $162^{\circ} \pm
13^{\circ}$ at 2.6\,keV. Assuming the disk origin of the soft X-ray
polarization signal, and adopting a simplified model for such (with no
relativistic effect taken into account), the position angle of the
polarization is expected to be parallel or perpendicular to the disk
major axis depending on the disk state \citep{Lightman76}; the value
given by \citet{Long80} is actually consistent with the jet position
angle measured in the radio band \citep[about $-21^{\circ}$ to
$-24^{\circ}$;][]{Stirling01}. More recently, \citet{Laurent11}
observed Cygnus\,X-1 with {\it INTEGRAL} IBIS and resolved two
emission components in the 20\,keV--2\,MeV range. They reported that a
second (high-energy) component exhibits a polarization above 400\,keV
with a polarization fraction of $67\% \pm 33\%$. \citet{Jourdain12}
utilized the {\it INTEGRAL} SPI and reinforced the finding by
\citet{Laurent11}. They reported a polarization signal above 230\,keV
with a mean polarization fraction of $76\% \pm 15\%$ and at a position
angle of $42^{\circ}\pm3^{\circ}$. Note that, as described in
\citet{Jourdain12}, the correct position angle measured by
\citet{Laurent11} is not $140^{\circ}$ but in fact
$180^{\circ}-140^{\circ}=40^{\circ}$. All in all, the {\it INTEGRAL}
spectropolarimetry suggests that the hard component, which dominates
the flux above 100\,keV, is due to the jet synchrotron
emission. However, the 
surprisingly high polarization fraction and the apparent
deviation of the position angle from the jet orientation (by about
$65^{\circ}$), make the interpretation quite problematic
\citep[see the discussion in][]{Zdziarski14}. An independent and
precise spectropolarimetry is necessary to deepen our understanding of
high-energy emission from this important source, and the {\it ASTRO-H}
SGD is an ideal instrument for this purpose.

In order to evaluate the SGD sensitivity for polarization
measurements, we first modeled the Cygnus\,X-1 $\gamma$-ray spectrum
based on \citet{Jourdain12}. We represented their low-energy component
(presumably Comptonization, expected to show little or no polarization)
and a high-energy component (presumably highly polarized jet synchrotron
emission) with $F(E)_{\rm soft} = 1.3 \times (E/{\rm keV})^{-1.6}
\times \exp[{-(E/{\rm
160\,keV})^{1.4}}]$\,ph\,s$^{-1}$\,cm$^{-2}$\,keV$^{-1}$ and
$F(E)_{\rm hard} = 0.3 \times (E/{\rm keV})^{-1.6} \times
\exp[{-(E/{\rm 700\,keV})}]$\,ph\,s$^{-1}$\,cm$^{-2}$\,keV$^{-1}$,
respectively. We then evaluated the polarization sensitivity 
across four energy bands in the 60-600\,keV range. The
minimum detectable polarization at the 99\% confidence level was
calculated as
\begin{equation}
{\rm MDP_{99}} = \frac{4.29}{M \times R_{\rm S}} \, \sqrt{\frac{R_{\rm S}+R_{\rm B}}{T}} \, , 
\end{equation}
where $M$ is the modulation factor of the instrument, $T$ is the
observation time, and $R_{\rm S}$ and $R_{\rm B}$ are the source and
the background count rates, respectively \citep{Weisskopf09}. Based on
the high polarization fraction reported by {\it INTEGRAL} IBIS/SPI, we
assumed that the polarization fraction is 70\% for the high energy
(jet) component. The results of the calculation for 100\,ks exposure
are summarized in Figure\,\ref{fig:cygx1} (left panel), indicating
that such an observation with the SGD is able to derive interesting
constraints on the energy-dependent polarization measurements of
Cygnus\,X-1 in the 60--600\,keV range.

GRS\,1915+105 is another promising target for the hard X-ray/soft
$\gamma$-ray polarimetry with the SGD. This source is known for
superluminal jet ejections with bulk velocities of the outflowing
matter exceeding 90\% of the light speed
\citep{Mirabel94}. \citet{Rodriguez08} monitored the source with the
{\it INTEGRAL} IBIS, {\it RXTE}, and the Ryle Telescope for two years,
and observed a possible correlation of the radio flux and that of the
hard power-law component, indicating the emission from the jet in hard
X-rays. \citet{Droulans09} observed the source with the {\it INTEGRAL}
SPI and performed a spectral analysis in soft $\gamma$-rays. They
classified the data into two, the soft sample and the hard sample, and
found that in both samples the spectrum is represented by a thermal
Comptonization component and a high-energy power-law component. We
approximate their hard sample spectrum with $F(E)_{\rm soft} = 4.4
\times (E/{\rm keV})^{-2.1} \times \exp[{-(E/{\rm
60\,keV})^{1.4}}]$\,ph\,s$^{-1}$\,cm$^{-2}$\,keV$^{-1}$ plus
$F(E)_{\rm hard} = 0.33 \times (E/{\rm
keV})^{-2.0}$\,ph\,s$^{-1}$\,cm$^{-2}$\,keV$^{-1}$, and evaluated the
polarization measurement feasibility as we did for Cygnus\,X-1. The
results are summarized in Figure\,\ref{fig:cygx1} (right panel), in
which we assumed the polarization fraction of the hard (jet) component
at the maximum level of 70\%. As shown, the SGD is able to detect
polarization up to 300\,keV. Even if the polarization fraction is
smaller (30\% in the hard component), we can still detect
polarization up to 180\,keV. We therefore are able to robustly
establish (or strongly constrain) the jet emission in hard X-ray/soft
$\gamma$-rays in GRS\,1915+105 for the first time.

\subsection{Blazars}
\label{S:blazars}

Jetted AGN observed at small angle to the line of sight are classified
as blazars. Blazars can be divided further into Flat Spectrum Radio
Quasars (FSRQs) and BL Lacertae Objects (BL Lacs) based on the
emission line diagnostics and other characteristics of their
broad-band spectra \citep{Urry95}. X-ray continua of some BL Lacs (and
in particular of `high frequency-peaked BL Lacs', hereafter HBLs) are
dominated by the synchrotron emission of the highest-energy jet
electrons, and hence a strong X-ray polarization may be expected in
such sources up to the maximum level of 70\%
\citep[e.g.,][]{Pacholczyk67}.

Polarization of blazar synchrotron emission is now routinely detected
at radio and optical frequencies, with the polarization degree and
position angle depending on frequency and the activity level of a
source; a variety of particularly interesting phenomena have recently
been discovered at optical frequencies, consisting of prominent
position angle swings (by $\geq 180$\,deg) and polarization degree
changes (from 0\% up to $>40\%$) apparently related to the overall
source state \citep[as evidenced by coinciding $\gamma$-ray flares, or
ejections of superluminal radio components; see][and references
therein]{Marscher08,LAT3C279}. These new observations allowed for a
novel insight into the internal structure and magnetic field topology
of AGN jets, which are among the most widely debated open questions in
the high-energy astrophysics.

\begin{figure}[!t]
\begin{center}
\includegraphics[width=0.415\textwidth]{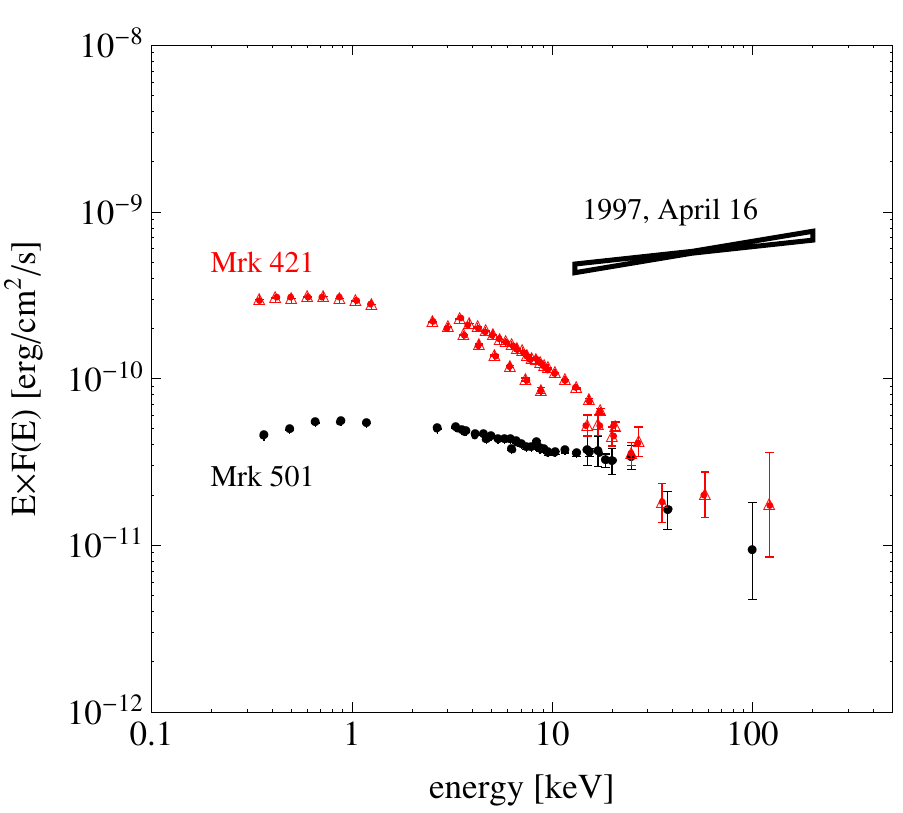}
\includegraphics[width=0.575\textwidth]{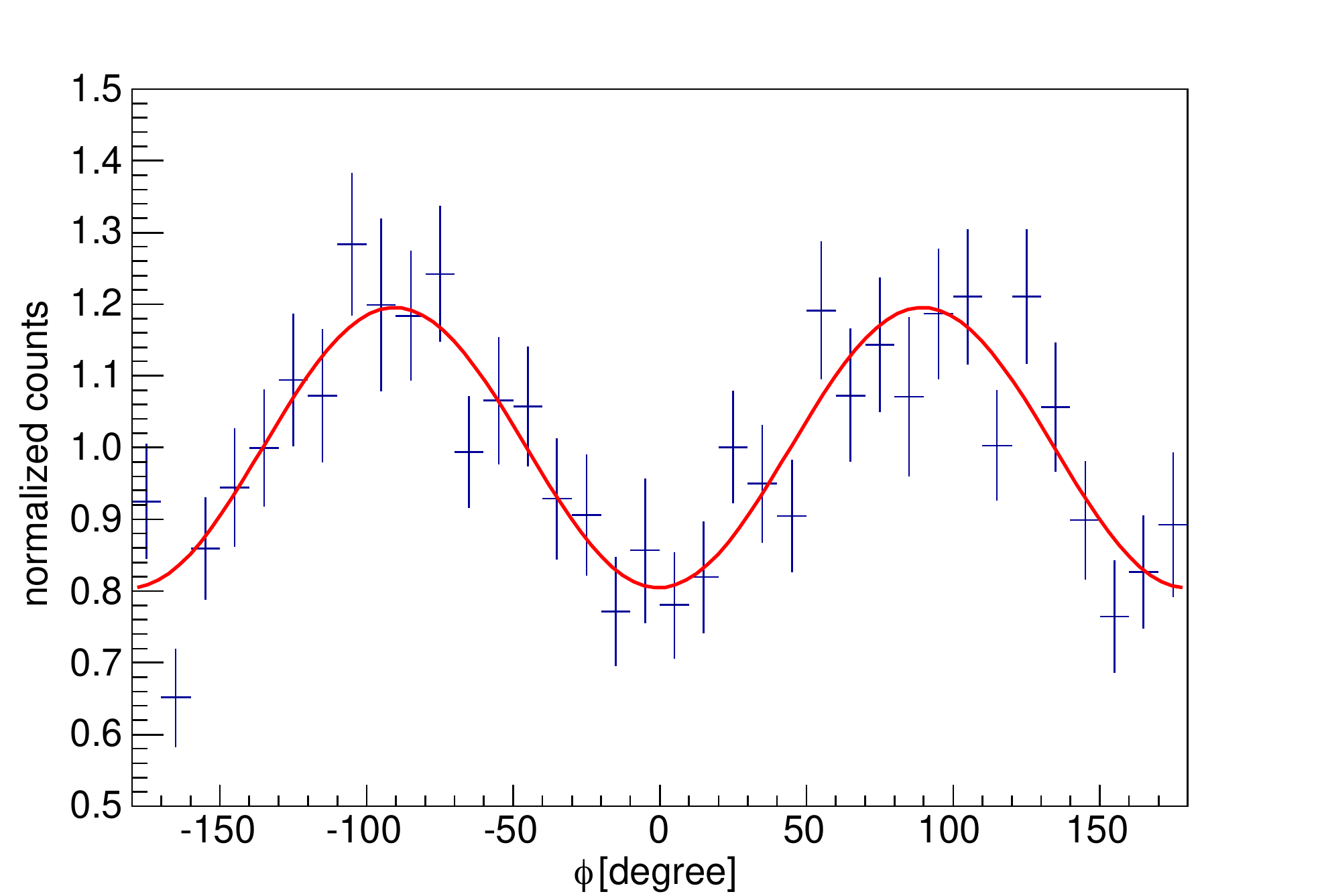}
\caption{Left: Spectral energy distribution of the X-ray brightest HBLs, Mrk\,421 and Mrk\,501, during their quiescence \citep[red triangles and black circles, from][respectively]{LAT421,LAT501}, compared with the April 1997 flare of Mrk\,501 \citep[black bow-tie; from][]{Pian98}. Right: Simulated polarimetry results of 100\,ks SGD observation of Mrk\,501 during the '97 outburst (energy band: 60--100\,keV), assuming the energy flux density of $F_{13-200\,{\rm keV}}=15.9 \times 10^{-10}$\,erg~cm$^{-2}$~s$^{-1}$, the photon index $\Gamma = 1.84$, and the 30\% polarization degree.}
\label{fig:broadband/polarization/observation_mrk501}
\end{center}
\end{figure}

The additional spectropolarimetry of bright blazar sources at X-ray frequencies will allow to constrain further the physics of relativistic jets in active galaxies. The problem is, however, that HBLs are relatively low-power and in addition distant objects, which are therefore much dimmer in X-rays (in their typical quiescence states) than Galactic XRBs or nearby radio-quiet AGN of the Seyfert class. In addition, the quiescence synchrotron continua of the brightest objects of this type, such as Mrk\,421 and Mrk\,501, display a prominent curvature at X-ray energies, with the differential energy flux decreasing rapidly with the increasing photon energy. Still, the SGD can detected at least some HBLs during their flaring states, such as the April 1997 outburst of Mrk\,501 during which the flat-spectrum synchrotron component extended up to $\geq 100$\,keV photon energies, with the persistent (over several days) flux of about $\sim 100$\,mCrab \citep[see Figure\,\ref{fig:broadband/polarization/observation_mrk501}, left panel]{Pian98}. Such flares, although rare, are not unusual, and much can be learned from those. For example, the 2006 flare of HBL PKS\,2155--304 witnessed at very-high energy $\gamma$-rays by the H.E.S.S. telescope \citep{HESS2155} has ignited a vigorous discussion in the community on the location of the dominant $\gamma$-ray emission site in blazar sources, and on the role of magnetic reconnection in accelerating jet particles (electrons) up to $>$\,TeV energies.

We have simulated the performance of SGD during the Mrk\,501 flare
resembling the historical 97' outburst (in a similar manner as
described in \S\,\ref{S:crab}). We have assumed a power-law spectrum
with the energy flux density of $F_{13-200\,{\rm keV}}=15.9 \times
10^{-10}$\,erg~cm$^{-2}$~s$^{-1}$ and the photon index $\Gamma =
1.84$; the degree of polarization was taken conservatively as
30\%. The azimuth angle distribution obtained in the simulation is
shown in Figure\,\ref{fig:broadband/polarization/observation_mrk501}
(right panel). In spite of the relatively conservative value for the
polarization fraction, the simulation result shows clear modulation
curve, yielding the polarization fraction of $35.1\pm 3.7\%$, and the
polarization angle of $-0.4^\circ\pm2.9^\circ$. Thus hard X-ray
polarimetry of blazars of the BL Lac types with SGD is therefore
feasible, although restricted to the brightest outbursts. ToO
observations of a flaring blazar should be therefore triggered by
all-sky monitoring X-ray and $\gamma$-ray instruments ({\it Swift}-XRT
and BAT, MAXI, {\it Fermi}-LAT and GBM), and joint with simultaneous
radio and optical spectropolarimetry, in order to maximize the
detection significance and the scientific gain.

Unlike in the case of low-power HBLs, the X-ray continua of luminous
FSRQ are dominated by the inverse-Compton emission of the
lowest-energy jet electrons \citep[see][]{Sikora09}. Importantly, this
inverse-Compton emission can also be
polarized at some level, with the polarization degree depending on
polarization properties of soft target photons (jet synchrotron
radiation, emission of the accretion disk and circumnuclear gas/dust
external to the jet), on the scattering regime (Thomson
vs. Klein-Nishina), and finally on the geometry of the emitting region
and the jet bulk velocity
\citep[see][]{Bonometto73,Begelman87,Celotti94,Poutanen94,McNamara09,
Krawczynski12,Zhang13}. Model predictions are here still uncertain to
some extent, but the emerging agreement is that the synchrotron
self-Compton emission of blazar sources may be polarized up to even
40\%. Interestingly, the X-ray continua of several bright FSRQs are
expected to be dominated by the SSC emission. Hence, the SGD
observations of such objects may in principle provide interesting
constraints on the blazar emission models as well. The same is true
also in the case of radio galaxies, which are believed to constitute
the parent population of blazars, although the origin of the X-ray
emission of even the brightest objects from this class is much more
controversial, due to the expected significant, possibly even
dominant, contribution from the accreting matter. The issue of
disentangling the jet and disk emission components at X-ray photon
energies is the subject of the following section.

It is important to comment here on the implications of a detected
polarization from any extragalactic sources. Quantum
gravity theories predict an energy-dependent photon velocity, and
therefore a decrease of the polarization signal emitted by
cosmologically distant objects. Currently, the positive detection of
polarization from the Crab nebula provides the strongest limit in
this respect, but any polarization detection of an
\emph{extragalactic} source will enable to revise this limit
significantly \citep[see, e.g.,][]{Maccione08}.

\section{Jet-Disk Coupling in Active Galaxies}
\label{S:jets}

In non-blazar (`misaligned') radio-loud AGN, the observed X-ray
spectra may contain a significant contribution, or even be dominated
by the emission from the accretion disk and (more likely) the
accretion disc coronae. The clearest evidence for this is the
detection of a neutral, narrow Fe-K fluorescence lines in some systems
\citep[see][and references therein]{Fukazawa11a,Fukazawa14}. This line is
produced by the accretion disc corona illuminating distant material
(torus), and its presence in the spectrum means that the X-rays from
the corona are not completely swamped by additional X-rays from the
jet. Thus in these systems we can simultaneously seen the accretion
disk corona and the radio/$\gamma$-ray jet emission. Hence they
constitute an exciting opportunity to investigate the jet-disk
coupling, and therefore the jet launching processes, by means of a
broad-band X-ray spectroscopy, hard X-ray/soft $\gamma$-ray 
polarimetry, and timing analysis joint with
multiwavelength ($\gamma$-ray and high-resolution radio) observations
of the selected bright targets. Yet such an analysis, analogous to
what is routinely done for Galactic microquasars \citep[see,
e.g.,][and references therein]{Fender09}, was hampered in the past by
the fact that the corona and the jet X-ray emission components are not
disentangled well in the majority of cases, with the exception of a
few brightest objects.

Two broad line radio galaxies (BLRGs), namely 3C\,111 and 3C\,120,
have been recently the subject of an extensive monitoring campaign to
study the accretion disc-jet coupling
\citep{Marscher02,Chatterjee09,Chatterjee11}. It was found that the
major X-ray dips in the light curves are followed by ejections of
bright superluminal knots in the radio jets of both targets. Assuming
the dominant disk corona contribution at X-ray frequencies, these
observations imply therefore a direct relation between the disk state
transitions and the jet formation process in AGN. On the other hand,
\citet{Tombesi10b,Tombesi12,Tombesi13} presented a comparison of the
parameters of the disk wind and the jet in radio galaxy 3C\,111 on
sub-pc scales. They found that the superluminal jet coexists with
mildly relativistic outflows and that both of them are powerful enough
to exert a concurrent feedback impact on their surrounding
environment. Interestingly, there are several indications suggesting
that the ultra-fast disk outflows in the studied systems might be
placed within the `X-ray dips/radio outbursts' cycle as well,
providing the necessary pressure support for the efficient collimation
and bulk acceleration of the electromagnetic jet freshly launched from
the rotating SMBH/accretion disk system.

Clearly, {\it ASTRO-H} observations may enable a qualitative progress
in understanding the jet-disk connection in AGN, as its performance
(sensitivity, energy resolution, and energy range) will exceed that of
the previously and currently available X-ray instruments. The
particular targets discussed below in this context ---
the quasar 3C\,273, low-power radio galaxy Centaurus\,A, and luminous
broad-line radio galaxy 3C\,120 --- are all easily detectable with
low- and high-energy instruments onboard the {\it ASTRO-H} in
relatively short exposures, are often monitored with milli-arcsec
radio interferometers, and are all the established
$\gamma$-ray emitters. Hence, while the SXS will constrain
precisely the emission and absorption lines in the soft band in those
sources, thus providing a unique insight into the accretion processes, 
the HXI and SGD will
characterize with unprecedented detail the spectral variability of the
targets in the medium and hard bands, which may be next correlated
with the extensive {\it Fermi}-LAT monitoring at $\gamma$-rays, along
with the high-resolution radio data.

\subsection{Quasar 3C\,273}
\label{S:3C273}

3C\,273 (redshift $ z = 0.158$) is one of the first quasars ever
discovered, is particularly radio-loud, and particularly bright in hard
X-rays \citep{Courvoisier98}. The broad-band spectrum of 3C\,273 is
dominated by the unresolved nucleus, including the accretion disk,
disk corona, and small-scale jet. The source can be considered as a
slightly misaligned FSRQ, observed at intermediate angles of about
$\sim 10$\,deg. The extensive X-ray monitoring of 3C\,273 over about
30 years using different instruments established a variability of the
source on the timescales from years (flux changes by a factor of a
few/several) down to days \citep[a few/few tens
percents;][]{Johnson95,Kataoka02,Courvoisier03,Chernyakova07,Soldi08}. The
compilation of broad-band multi-epoch spectra reveals in addition that
in the hard X-ray regime (above 10\,keV, roughly) the characteristic variability 
timescales are longer and the variability amplitudes are larger than those in the
soft/medium X-ray regime. The X-ray variability properties suggest
therefore a presence of two separate components: one most likely a
Seyfert-like, related to the accreting matter, and the other a
blazar-like jet, produced via inverse-Comptonization of soft photons by
low-energy electrons. This idea is in accord with the fact that
the X-ray spectrum of 3C\,273 in the soft band is steep ($\Gamma >
2$), while it is relatively flat in the medium and hard bands 
($\Gamma \simeq 1.7 \pm0.2$, up to the MeV range). A marginal detection of weak
fluorescent Fe lines in 3C\,273 has been reported only in some epochs.

\begin{figure}[!t]
\begin{center}
\includegraphics[width=0.39\textwidth,angle=90]{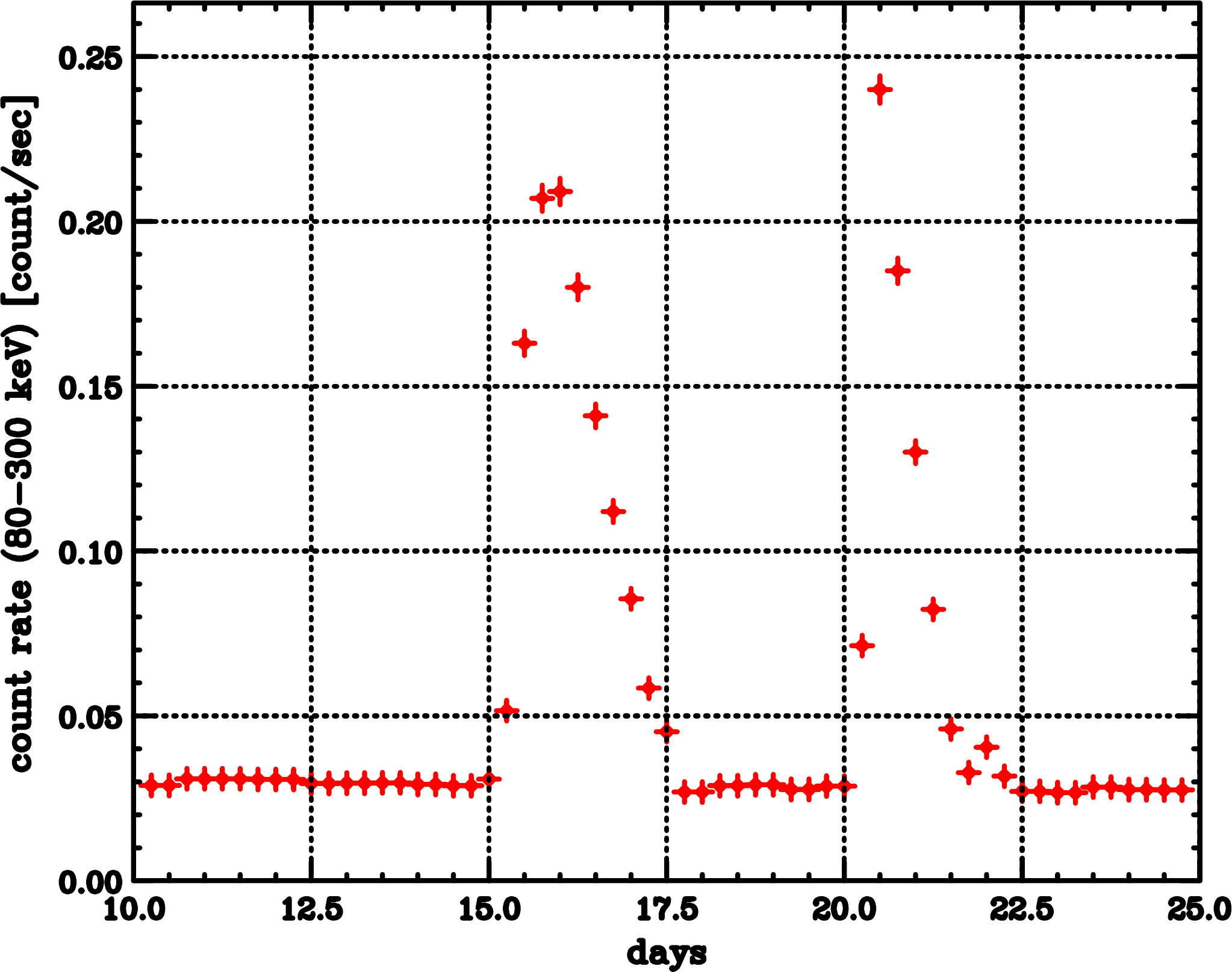}
\includegraphics[width=0.39\textwidth,angle=90]{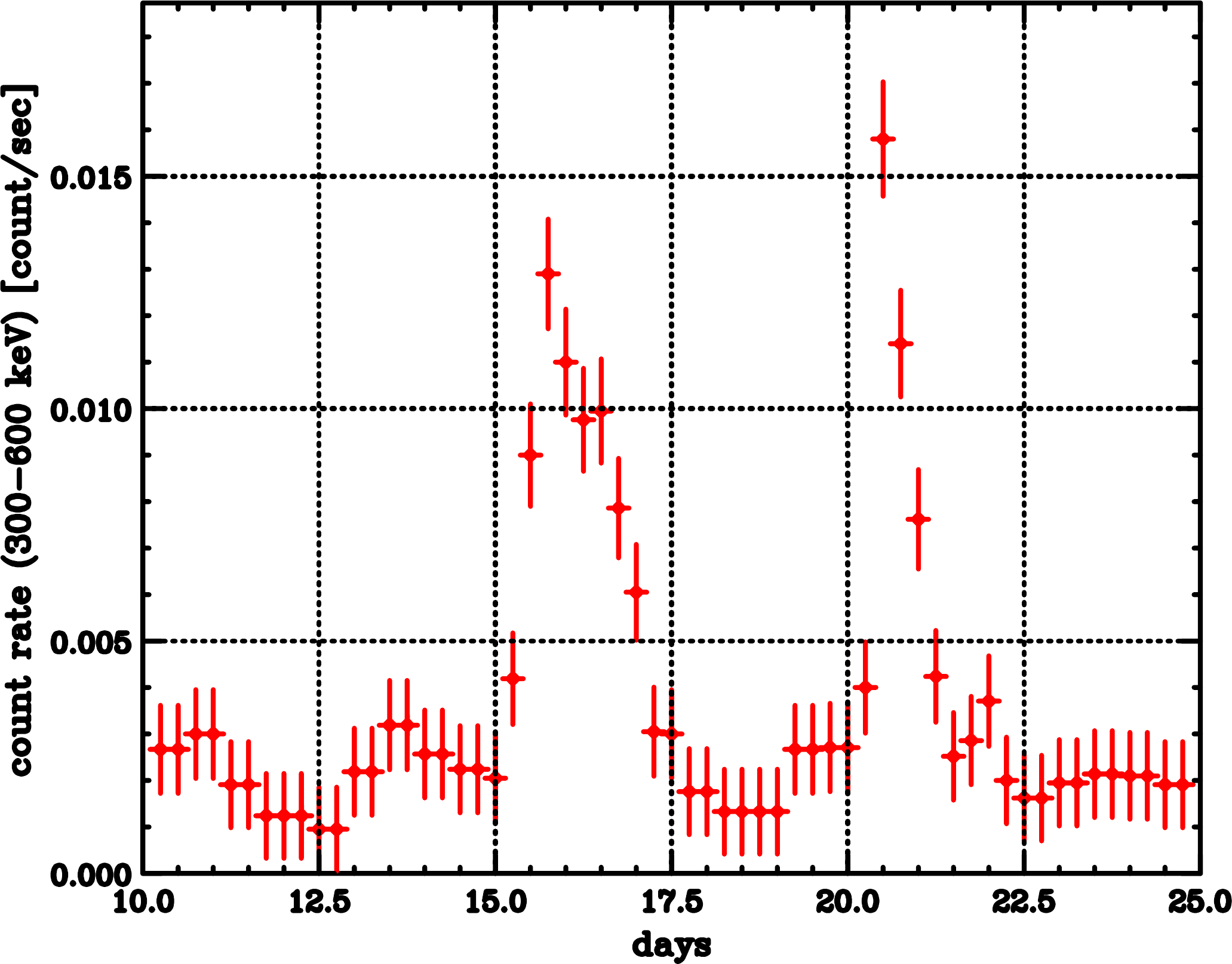}
\caption{Simulated SGD lightcurve 3C\,273 in two energy channels 80--300\,keV (left panel) and 300-600\,keV (right panel), with 6\,h binning, assuming a linear correlation of the X-ray flux and the GeV fluxe as detected with {\it Fermi}-LAT during the 2009 outburst \citep{LAT273}.}
\label{fig:3C273}
\end{center}
\end{figure}

We have simulated the hard X-ray spectrum of 3C\,273 observed with
{\it ASTRO-H} for different exposure times, assuming the average flux
and photon index from the 58 month {\it Swift}-BAT catalog, namely
$\Gamma = 1.73$ and $F_{14-195\,{\rm keV}} = 4.3 \times
10^{-10}$\,erg\,cm$^{-2}$\,s$^{-1}$. The spectrum is characterized well
in a broad energy range with exposures as short as 30\,ks, for
which the joint HXI and SGD fit returned $\Gamma = 1.732-1.743$,
$F_{5-80\,{\rm keV}} = (3.43-3.47) \times
10^{-10}$\,erg\,cm$^{-2}$\,s$^{-1}$, and $F_{40-600\,{\rm keV}} =
(5.69-5.85) \times 10^{-10}$\,erg\,cm$^{-2}$\,s$^{-1}$. In shorter
exposures 3C\,273 is still clearly detected, although its spectral
properties above 200\,keV cannot be accessed robustly. The Fe
line is also easily detectable at the average flux level with the SXS
assuming that this is narrow (from the illuminated torus) rather than
from the illuminated inner disc. 

3C\,273 was established as a high-energy $\gamma$-ray source by the
EGRET instrument onboard the CGRO \citep{Hartman99}. More recently,
continuous monitoring of the source with {\it Fermi}-LAT resulted
in the detection of prominent flares at GeV photon energies
\citep{LAT273}. This $\gamma$-ray emission is widely believed to
originate in the jet, and to constitute the high-energy tail of the
inverse-Compton component which first pops-up in hard X-rays. Hence,
\emph{simultaneous} {\it ASTRO-H} and {\it Fermi}-LAT observations of
3C\,273 may uniquely constrain the spectral evolution of the radiating jet
electrons over a particularly broad range of energies, and on short
timescales which have been never probed before at hard X-rays in this
source or, in fact, in any blazar. We note in this context that
3C\,273 is much more prominent in X-rays than even the brightest FSRQs
(in their quiescence), since even though the jet emission in FSRQs is
more significantly beamed due to their small jet inclinations, they
are more distant objects.

In order to illustrate this point, we have simulated the SGD light
curve of 3C\,273 in two energy channels, assuming a linear correlation
of the X-ray flux with the GeV flux as detected with {\it Fermi}-LAT
during the large outburst in 2009 \citep{LAT273}. The results are
presented in Figure\,\ref{fig:3C273}. As shown, the SGD observations
will indeed allow for a unique characterization of the hard X-ray
variability in the source down to the timescales of 6\,h, with
one-to-one comparison between the lightcurves in different X-ray and
$\gamma$-ray bands. The jet variability robustly constrained in this
way at high-energies may be next juxtaposed with the variability of
the accretion-related soft X-ray continuum probed with the SXS.

\subsection{Radio Galaxy Centaurus\,A}
\label{S:CenA}

\begin{figure}[!t]
\begin{center}
\includegraphics[width=0.55\textwidth]{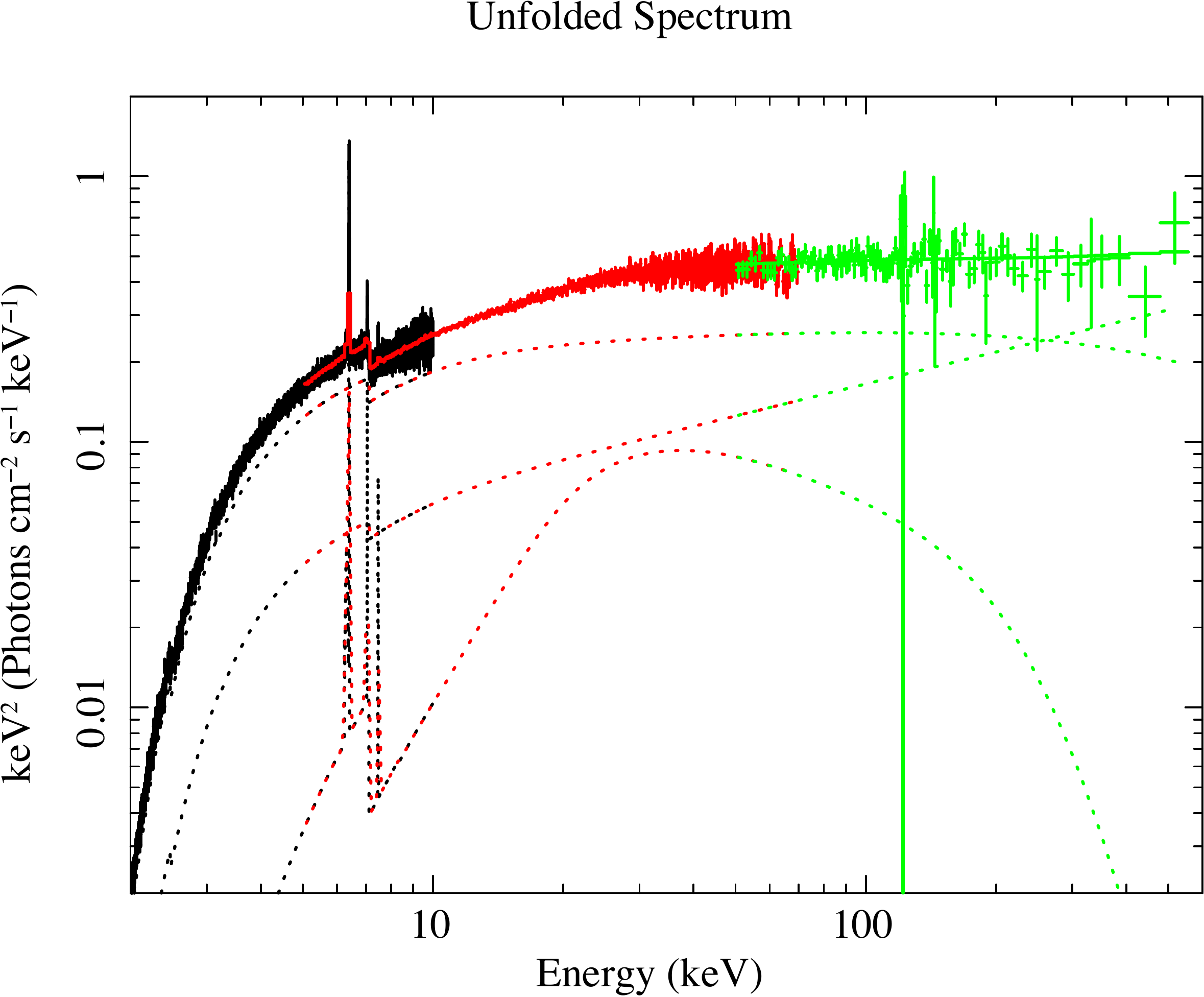}
\includegraphics[width=0.55\textwidth]{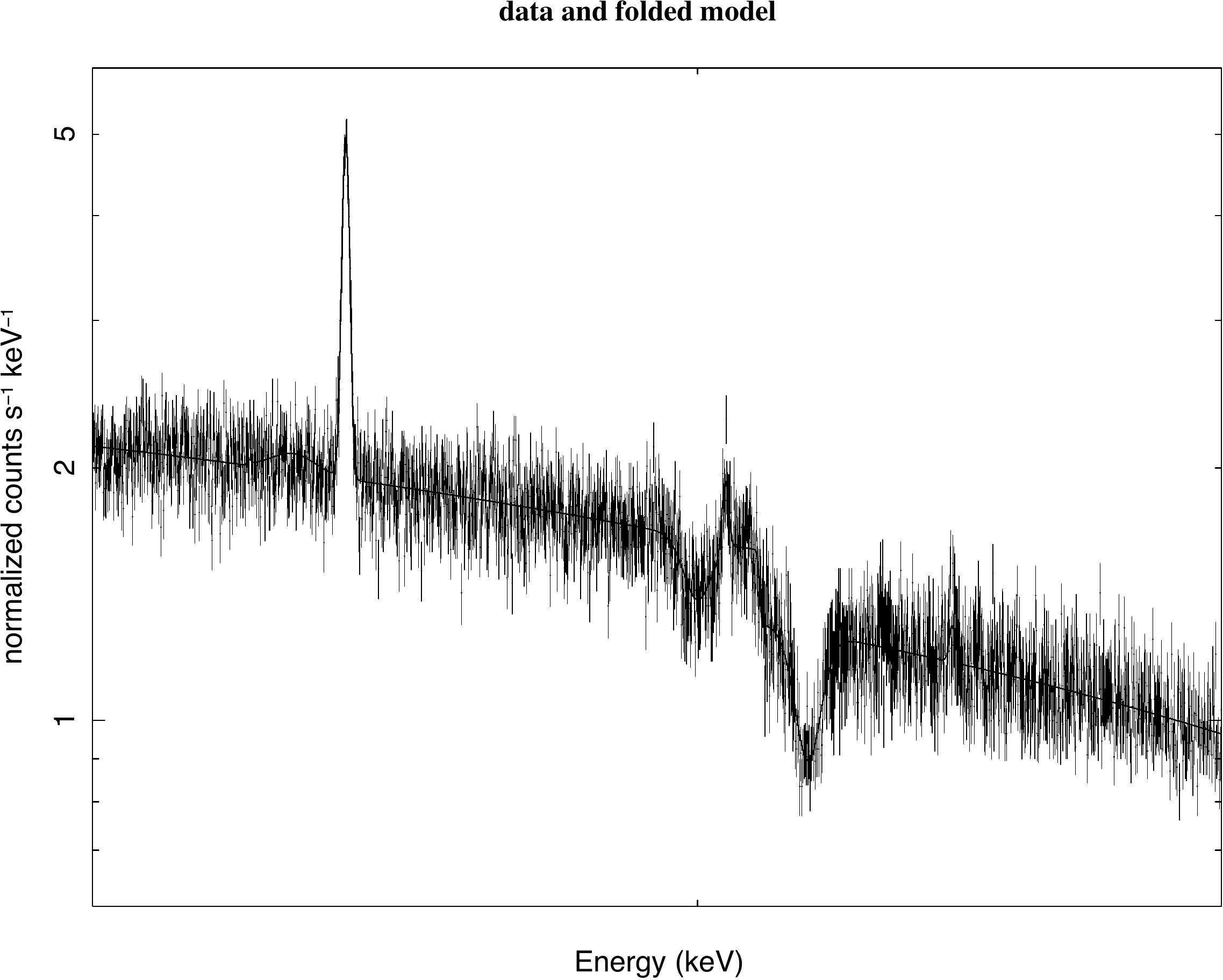}
\caption{Upper: Simulated broad-band spectrum of Centaurus\,A with 100\,ks exposure, based on the spectral parameters taken from \citet{Fukazawa11b}, with SXI shown in black, HXI in red, and SGD in green. The fitted model shown with dotted curves includes the absorbed power-law representing the disk/corona component, the additional power-law related to the jet, and the torus reflection component. Lower: 6--8\,keV SXS simulation of blue-shifted absorption lines in the Centaurus\,A spectrum for 100~ks exposure. Here we assumed two absorption lines at 7.0\,keV and 7.2\,keV with the line width of 25\,eV and the normalizations of $-5.6\times10^{-5}$\,ph\,cm$^{-2}$\,s$^{-1}$.}
\label{fig:cena}
\end{center}
\end{figure}

The low-power radio galaxy Centaurus\,A is the nearest AGN (distance
of 3.7\,Mpc), known for its complex radio structure and post-merger
host galaxy \citep[see][]{Israel98}. It is one of the brightest hard
X-ray/soft $\gamma$-ray emitter in the sky, detected by all the
instruments onboard the CGRO during the period 1991-1995
\citep[see][and references therein]{Steinle98}. More recently, the
central regions of Centaurus\,A have been also resolved with {\it
Fermi}-LAT \citep{LATCenA} and H.E.S.S. \citep{HESSCenA} at high and
very-high energy $\gamma$-rays, respectively. The origin of the
observed $\gamma$-rays is puzzling: on the one hand, the steep
$\gamma$-ray continuum seems to resemble a misaligned blazar
component, while on the other hand this continuum joins smoothly with
the X-ray continuum, which is in fact most likely related to the
accreting matter in the source; in addition, the TeV emission seems in
excess of the extrapolated GeV continuum, suggesting either a separate
spectral component, or a non-standard emission spectrum \citep[see the
discussion in][]{LATCenA}. The broad-band characteristics of the X-ray
emission of Centaurus\,A are therefore of a particular importance for
understanding this archetypal AGN, and several observational studies
have been dedicated in the past to this problem
\citep{Evans04,Markowitz07,Fukazawa11b,Beckmann11}. For example, based on the
variability analysis from multiple deep {\it Suzaku} pointings,
\citet{Fukazawa11b} argued for a dominant jet contribution only
above100\,keV; however, the limit of $\sim 250$\,keV on the 
energy range of the {\it Suzaku} XIS and HXD 
precluded any definitive conclusions.

The broad-band coverage and high sensitivity of the {\it ASTRO-H}
HXI and SGD instruments will allow us to search for a spectral cutoff
in the hard X-ray/soft $\gamma$-ray spectrum of Centaurus\,A, expected
in the case of a dominant disk corona component, to look for polarization
signatures around 100\,keV photon energies, which may be expected in 
the case of a pronounced jet component, and in general to constrain 
precisely the slope of the X-ray continuum in the hard X-ray/soft 
$\gamma$-ray range. Figure~\ref{fig:cena} (left) presents the simulated
broad-band {\it ASTRO-H} spectrum of the source in 100\,ks exposure
\citep[based on the parameters from][]{Fukazawa11b}. 

However, detailed soft X-ray spectroscopy of the target with the SXS
are expected to be equally revealing.  First, the Fe-K line in
Centaurus\,A is the second brightest after that in the Circinus
galaxy, with the equivalent of about 80\,eV. The SXS can measure
precisely the line width and Compton shoulder, which in turn, when
assessed jointly with constraints from the broad-band continuum study
of the reflection bump around 50\,keV, will constrain the location and
geometry of the neutral reflector in the source.

Second, we note that the presence of the two absorption lines from Fe
XXV He\,$\alpha$ at $E= 6.7$\,keV (EW\,$\simeq 10$\,eV) and Fe XXVI
Ly\,$\alpha$ at $E= 6.96$\,keV (EW\,$= 15$\,eV) was reported in the
three {\it Suzaku} observations of Centaurus\,A performed in 2009
\citep{Tombesi14}. The lines are detected with a high significance of
$> 99.99\%$. However, due to the limited energy resolution of the {\it
Suzaku} XIS, these lines were not resolved and also the outflow
velocity of the absorber could not be constrained. Hence the line
origin (outflowing wind versus the extended ionized absorber with no
net velocity) remains an open question. Accordingly to our simulations
based on the line modeling with an XSTAR table, the unprecedented
energy resolution of the SXS will allow us to resolve these lines with
high accuracy in the exposures as short as 100\,ks (see
Figure~\ref{fig:cena}, right panel), and to measure the velocity of
the absorber, thus providing robust constraints on the line origin.

\subsection{UFOs and Radio-Loud AGN}
\label{S:UFOs}

A systematic analysis of the {\it Suzaku} spectra of a small sample
of five bright BLRGs performed by \citet{Tombesi10a} revealed
blue-shifted Fe XXV/XXVI K-shell absorption lines at $>7$\,keV
energies at least in three of them, namely 3C\,111, 3C\,120 and
3C\,390.3. These lines originate from highly ionized and high column
density gas outflowing with very high velocities, from $\sim
1,000$\,km\,s$^{-1}$ up to mildly-relativistic values of the order of
40\% of the speed of light, which are referred to as ultra-fast
outflows (UFOs). Successive studies found similar outflows also in
3C\,445 and 4C$+$74.26 \citep{Reeves10,Gofford13}. Later, the search
for such absorption lines was extended to a large sample of 26
radio-loud AGN observed with {\it XMM-Newton} and {\it Suzaku}
\citep{Tombesi14}. Combining the results of this analysis with those
in the literature and correcting for the number of spectra with
insufficient signal-to-noise, the incidence of UFOs in radio-loud AGN
was estimated to be in the range 30-70\%. This indicates the presence
of complex accretion disk winds in a significant fraction of sources.

Ultra-fast outflows found in radio-loud AGN are similar to the UFOs
reported in $>40\%$ of radio-quiet AGN \citep[e.g.,][see the ``AGN
Winds'' {\it ASTRO-H} White Paper]{Tombesi10b,Gofford13}, suggesting
that any jet-related radio-quiet/radio-loud dichotomy might 
not apply to AGN winds. The typical location of the UFOs is estimated
to be in the range $\sim 10^2-10^4 \, r_s$ from the central SMBH,
where $r_s=G \, M_{BH}/c^2$ is the black hole gravitational
radius. The study of UFOs in radio-loud AGN can therefore provide
important insights into the characteristics of the innermost jet
launching regions, where the collimation and the bulk acceleration of
initially broad and only mildly-relativistic electromagnetic jets is
expected to take place
\citep[e.g.,][]{McKinney06,Tchekhovskoy11,Sadowski14}.

One of the most relevant and promising targets for exploring the
presence and the energetics of the disk winds in the particular
context of the jet/disk coupling in AGN is the nearby radio galaxy
3C\,120 ($z=0.033$). Previous X-ray monitoring of the source, augmented
by high-resolution radio data, hint at a close link between the
disk state transitions and the jet formation processes, manifesting as
major X-ray dips in the accretion disk corona lightcurve followed by
ejections of bright superluminal knots in the radio jet
\citep[see][]{Marscher02,Chatterjee09}. More recently,
\citet{Lohfink13} presented the in-depth study of the central engine
in the source using a multi-epoch analysis of the deep {\it
XMM-Newton} observation and the two deep {\it Suzaku} pointings. The
authors discussed a composite source model consisting of a
truncated/refilling disk during the {\it Suzaku} observations, and a
complete disk extending down to the innermost stable circular orbit
(ISCO) during the {\it XMM-Newton} observation. The ejection of a new
jet knot was observed approximately one month after the analyzed {\it
Suzaku} pointings, and \citet{Lohfink13} argued that this might
indicate the timescale for 
propagation of a disturbance from the disk into the jet.

The $2-10$\,keV flux of 3C\,120 varied over the
range $2-6 \times10^{-11}$\,erg~cm$^{-2}$~s$^{-1}$ in 
the long-term {\it RXTE} monitoring analyzed in
\citet{Chatterjee09}. In our
simulations below we therefore assume conservatively the lowest flux
of the target. We consider the combined disk-jet model of
\citet{Lohfink13}, in which the jet X-ray emission is parameterized by
a steep power-law with the photon index of $2.5-4$, while the disk
corona continuum is represented by a power-law component with
the photon index of $1.7-2.4$ with a high energy cut-off at
$150$\,keV.  In the first considered case, the disk extends down to
the ISCO. This is modeled in XSPEC with a relativistically blurred
(kdblur) ionized reflection component (reflionx) for the inner disk
radius $r_{\rm in} \simeq r_g \simeq 0.86 \times 10^{13}$\,cm,
$q\simeq 7$, inclination $i \simeq 15^{\circ}$, and the ionization
$\xi \simeq 200$\,erg~cm~s$^{-1}$; the neutral distant reflector
(pexmon) is included with the reflection fraction $R \simeq 2$. In the
second case, which is related to the launch of the jet knot, the inner
disk is disrupted and the inner radius recedes to $r_{\rm in} \simeq
38\,r_g$; the relativistic blurred reflection has now a more standard
$q$ value of $\simeq 3.5$ and the neutral reflection factor is $R
\simeq 0.26$. The ultra-fast outflow with the velocity $v_{out} \simeq
0.161\,c$ was detected by \citep{Tombesi14} in this latter state. The
outflow is modeled below with the XSTAR table assuming a turbulent
velocity of 3,000\,km\,s$^{-1}$, ionization $\log \xi \simeq
4.9$\,erg~cm~s$^{-1}$, and column density $N_{\rm H} \simeq 5
\times 10^{22}$\,cm$^{-2}$. When modeled with an inverted Gaussian
absorption line, this is equivalent to energy  $E \simeq
8.23$\,keV, intrinsic line width of $\sigma \simeq 110$\,eV, and 
EW\,$\simeq -20$\,eV.

\begin{figure}[!t]
\begin{center}
\includegraphics[width=0.45\textwidth]{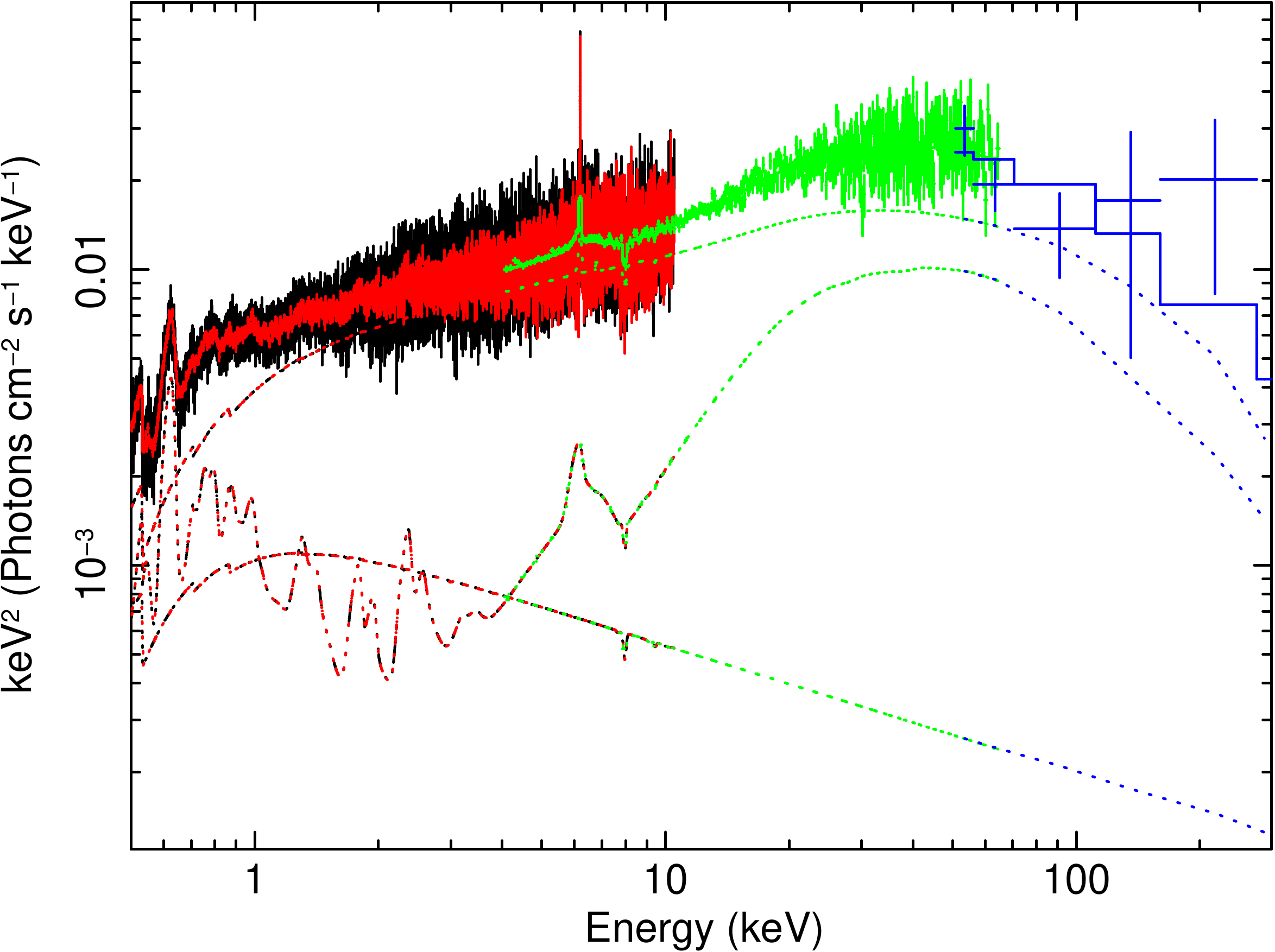}
\includegraphics[width=0.45\textwidth]{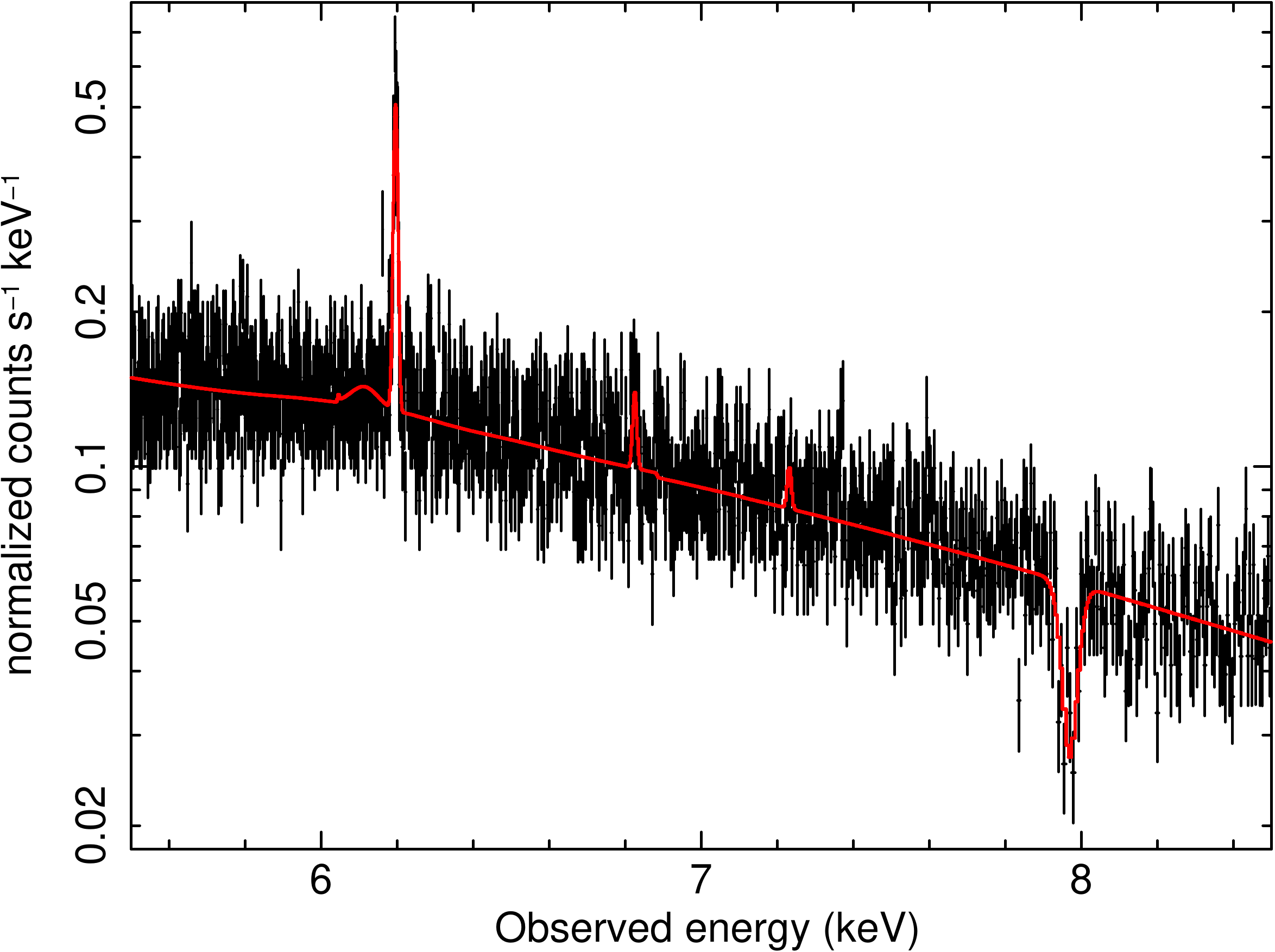}
\caption{Left: Simulated 100\,ks {\it ASTRO-H} broad-band spectrum of 3C\,120 in the range $0.5-300$\,keV, corresponding to the state in which the jet knot is ejected and the inner disk is disrupted (the inner disk radius receds from $r_{in} \simeq r_g$ up to $r_{in} \simeq 38\,r_g$). The SXS spectrum is shown in black, the SXI in red, the HXI in green, and the SGD in blue. Right: Zoomed SXS spectrum in the Fe K band (5--8.5\,keV).}
\label{fig:3C120}
\end{center}
\end{figure}

The simulated {\it ASTRO-H} broad-band spectrum of 3C\,120 in 100\,ks
exposure, corresponding to the latter state described above, is shown
in Figure\,\ref{fig:3C120}. The source is 
detected at high significance with all the instruments onboard, from
0.5\,keV up to 300\,keV. This will simultaneously constrain
the corona power-law continuum, the jet emission component, and the
neutral/ionized reflection. In particular, the accuracy in the
determination of the power-law slopes, cut-off energy, 
reflection fraction, and the ionization parameter, will be of 2\%,
25\%, 10\% and 10\%, respectively. In addition, the inner disk radius
will be measured with the 15\% accuracy. Moreover, the presence of the
disk wind will be determined by means of the Fe K UFO
at the $>$5$\sigma$ level. The energy, width, and the equivalent width
of the line will be determined at the 0.5\%, 20\% and 20\% levels,
respectively. Finally, the velocity and ionization of the UFO will be
constrained at the 2\% and 10\% levels, respectively. These will allow
for a direct and truly simultaneous comparison between the activity
level of the intermittent jet, the state of the accretion disk, and
the energetics of the disk wind. We note that 3C\,120 is
systematically monitored on a daily basis in $\gamma$-rays with {\it
Fermi}-LAT \citep[see in this context][]{Kataoka11}, approximately
once per week at optical frequencies with KANATA, and roughly in the
monthly cadence at radio frequencies with high and low-resolution
interferometers through several projects, such as the MOJAVE or the
F-Gamma projects.

\section{Starburst-AGN Connection in Nearby Galaxies}
\label{S:agn}

Several nearby galaxies host both a prominent AGN and a nuclear
starburst. Disentangling the two emission components in such sources
is essential for constraining their overall energetics, and for
understanding galaxy evolution processes in general. However, this task
requires simultaneous, truly broad-band, and high-quality
X-ray spectra, which are rarely available even for
the brightest systems. As argued below, this may change significantly
in the near future with {\it ASTRO-H}.

Some of the starburst/AGN systems are now also established
$\gamma$-ray emitters, including the two cases discussed below, namely
NGC\,4945 and Circinus galaxy
\citep[respectively]{LATstarburst,Hayashida13}. The observed
$\gamma$-ray emission of such sources is believed to originate
predominantly in the starburst regions via cosmic rays
produced by supernova explosions interacting with the interstellar
medium. Yet in some cases the origin of the $\gamma$-ray photons
detected with {\it Fermi}-LAT is an open issue, as for example the
Circinus galaxy appears over-luminous at GeV energies with respect to
the expected starburst emission given its far-infrared and radio
luminosities \citep[see][]{Hayashida13}. The difficulty is, however,
that the starformation rate in this
system, estimated from
the radio-to-IR data, is rather uncertain, translating to large
uncertainties in the expected cosmic ray luminosity. {\it ASTRO-H}
observations may help to resolve this problem, which is very relevant
as $\gamma$-ray observations are now widely
considered as an important, newly emerging window into the action of
starformation processes in the evolving galaxies.

\subsection{Circinus Galaxy}
\label{S:Circinus}

The Circinus galaxy, at a distance of $\sim 4$\,Mpc, hosts a
Compton-thick AGN and shows evidence of starburst activity. It is thus
an excellent laboratory for studying the X-ray reprocessor in a
heavily obscured system, as well as studying the interplay between AGN activity
and star-formation. The broad-wavelength {\it ASTRO-H} coverage, from
the SXI to HXI and SGD, together with the high resolution capabilities
of the SXS will provide unprecedented insight into the
obscuring medium, and constrain the relative contribution of
photo-ionization and collisional ionization to the soft X-ray
emission.

\begin{figure}[!b]
\centering
\includegraphics[scale=0.55]{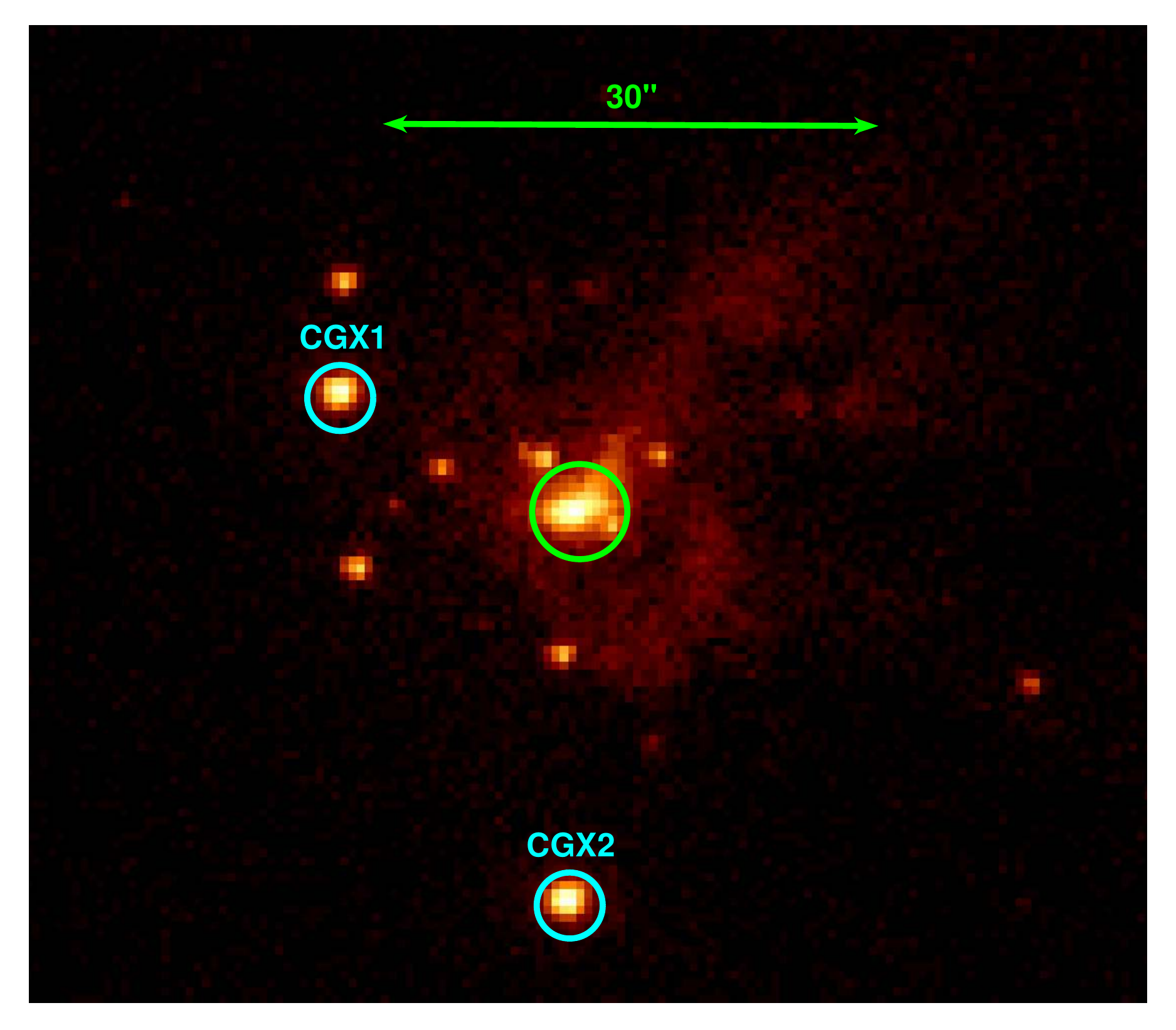}
\caption{{\it Chandra} ACIS image of Circinus, where the green circle indicates the nucleus and the cyan circles show strong X-ray point sources that will contribute to the emission observed by {\it ASTRO-H}. Diffuse emission from the circumnuclear starburst ring and the north-west ionization cone is also visible.}
\label{ch_image} 
\end{figure}

With the advent of models, such as MYTorus \citep{Murphy09} that
self-consistently treat the effects of the transmitted, Compton
scattered and fluorescent line emission in obscured AGN, we are able
to analyze the X-ray spectrum of Compton-thick AGN in a physically
meaningful way. Currently, the most comprehensive X-ray analysis of
Circinus involves simultaneous {\it NuSTAR} and {\it XMM-Newton}
observations \citep{Arevalo14}, which included archival {\it Chandra}
data to model the nuclear only region (3$^{\prime\prime}$) and
contamination within the {\it NuSTAR} FOV from an X-ray binary (CGX1)
and supernova remnant (CGX2, see Figure\,\ref{ch_image}).  The X-ray
binary is highly variable in X-rays: for the best constraints on
modeling the AGN emission and details of the X-ray reprocessor, a
simultaneous {\it Chandra} ACIS-S observation is required. In
combination with coverage from 0.5--500\,keV with {\it ASTRO-H} this will
provide the best measurements on physically meaningful modeling. This
will enable us to 
derive constraints on the line-of-sight and global column
density, from which we can calculate the intrinsic X-ray luminosity,
extending the bandpass visible by {\it NuSTAR} to lower energies
($<$3\,keV) and higher energies ($>$80\,keV).

\begin{figure}[!t]
\centering
\includegraphics[width=0.6\hsize]{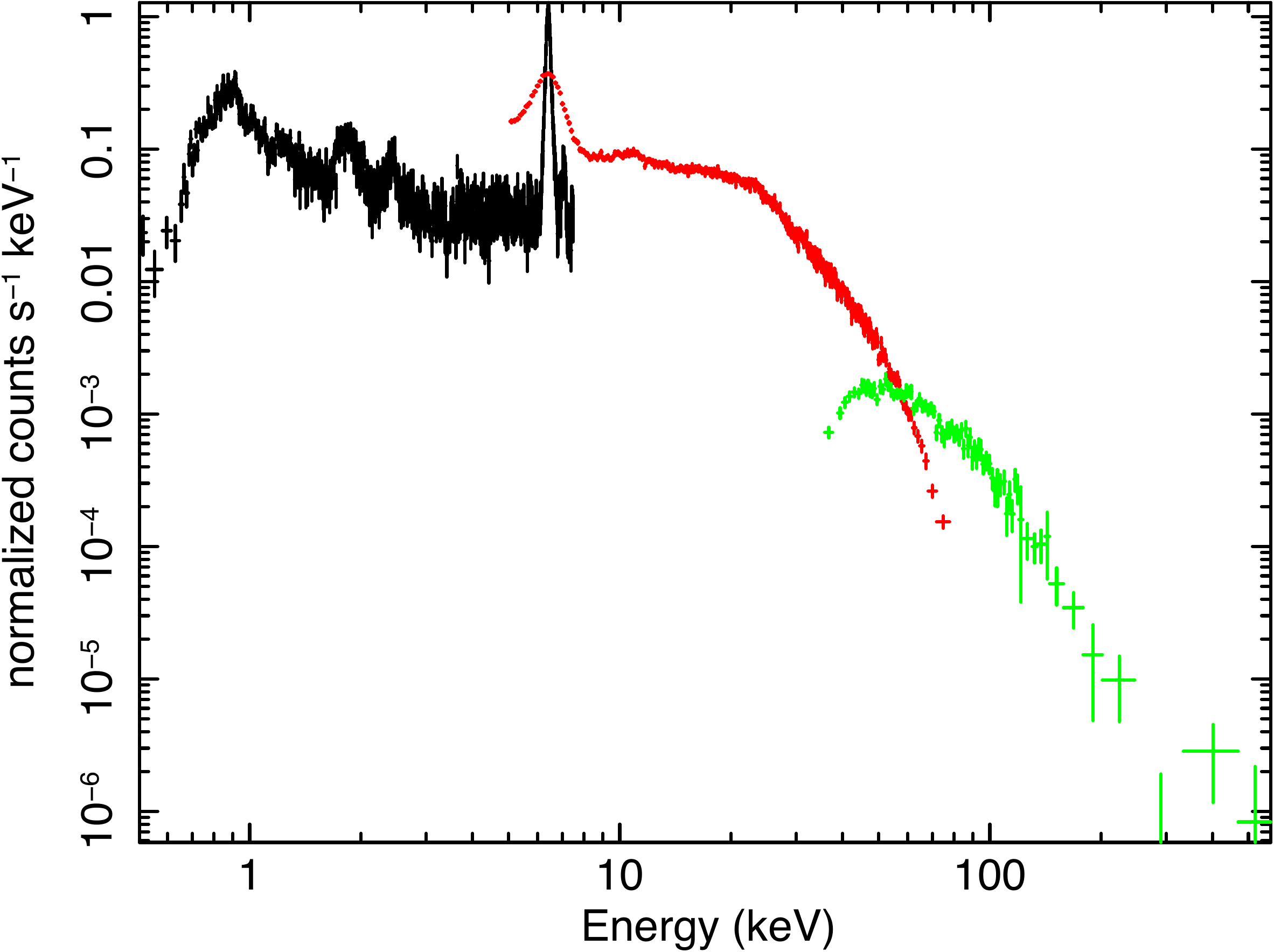}
\caption{Simulated low-resolution 100\,ks {\it ASTRO-H} spectrum of Circinus (SXI shown in black, HXI in red, and SGD in green) based on the modeling by \citet{Massaro06}, \citet{Arevalo14}, and \citet{Yang09}.}
\label{circ_lowres}
\end{figure}

To estimate the low-resolution {\it ASTRO-H} spectra of Circinus, we
set up models using published fits to archival data, where the
low-resolution SXI spectrum is based on the XMM-{\it Newon} fit
reported in \citet{Massaro06} and the HXI spectrum uses the joint {\it
NuSTAR}, {\it Suzaku} PIN, {\it BeppoSAX} and {\it Swift}-BAT fit from
\citet{Arevalo14} as a template. As shown in
Figure\,\ref{circ_lowres}, a 100\,ks {\it ASTRO-H} observation would
provide extremely high signal-to-noise in the low-resolution spectra. 
The high-resolution SXS spectrum will enable the Compton-shoulder
around the Fe K$\alpha$ 6.4\,keV line to be modeled, imparting
information about the opening angle that may not be apparent with
lower resolution spectra (see the ``AGN Reflection'' {\it ASTRO-H}
White Paper). Neither CGX1 or CGX2 have strong neutral Fe K$\alpha$
emission \citep{Bauer01} and would therefore not contaminate the
nuclear Fe K$\alpha$ and Compton-shoulder emission.

As depicted in\,Figure\,\ref{ch_image}, {\it Chandra} observations of
Circinus reveal that diffuse emission is associated with the active
nucleus as well as an ionization cone extending north-west above the
AGN which contributes $\sim$15\% to the extended emission
\citep{Sambruna01a}. Modeling the circumnuclear and extended diffuse
emission in the low-resolution {\it Chandra} ACIS spectra reveals both
thermal and non-thermal contributions, where the observed Fe K$\alpha$
emission likely results from Compton-scattering of the AGN continuum
\citep{Arevalo14}. The simultaneous 0.5--500\,keV {\it ASTRO-H}
coverage will enable the starburst contribution to be disentangled
from the AGN component: the 2--500\,keV spectrum will allow the
tightest constraints on modeling the AGN continuum, isolating the
starburst component at lower energies.

\begin{figure}[!t]
\centering
\includegraphics[scale=0.43]{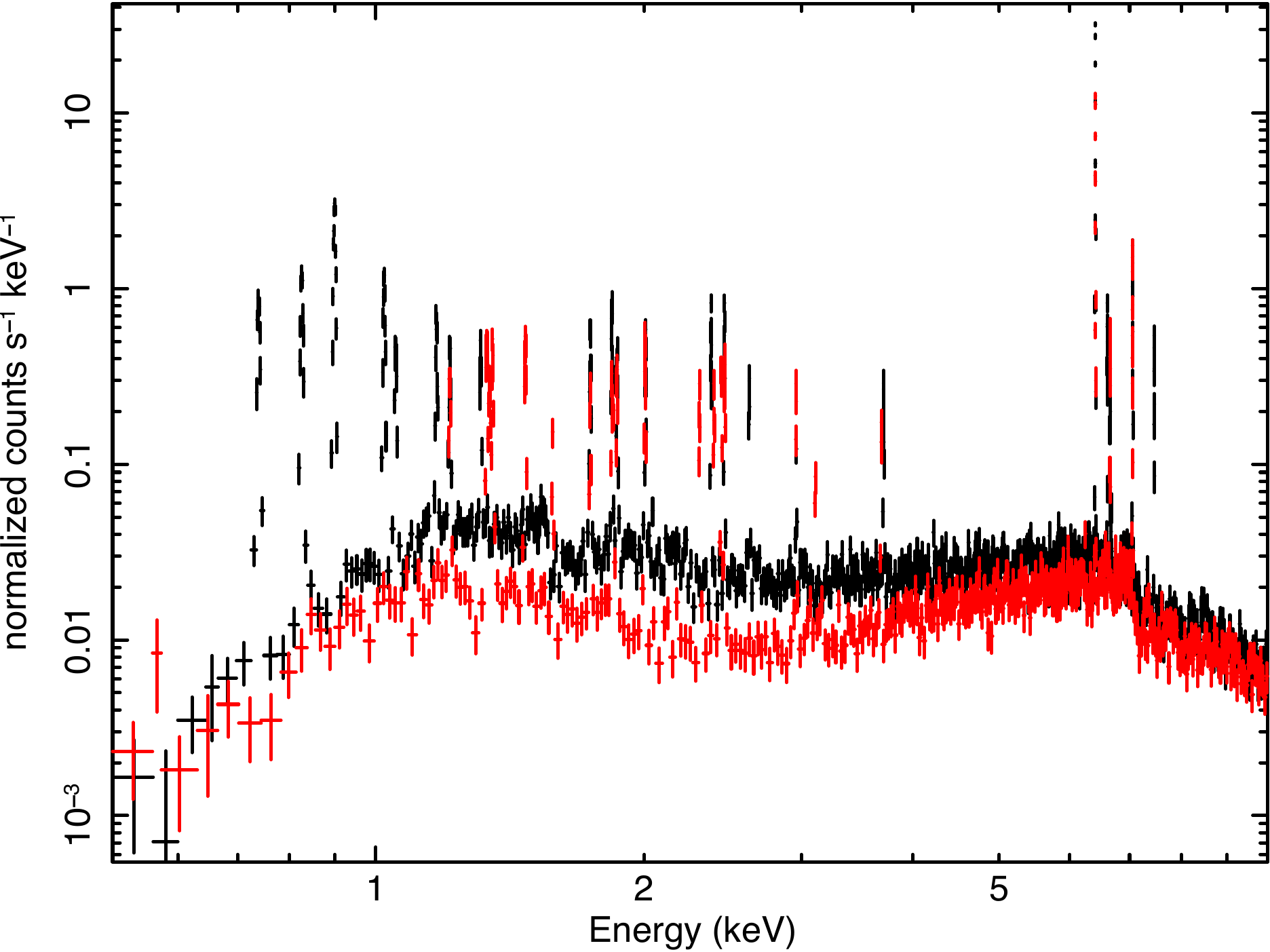}
{\includegraphics[scale=0.45]{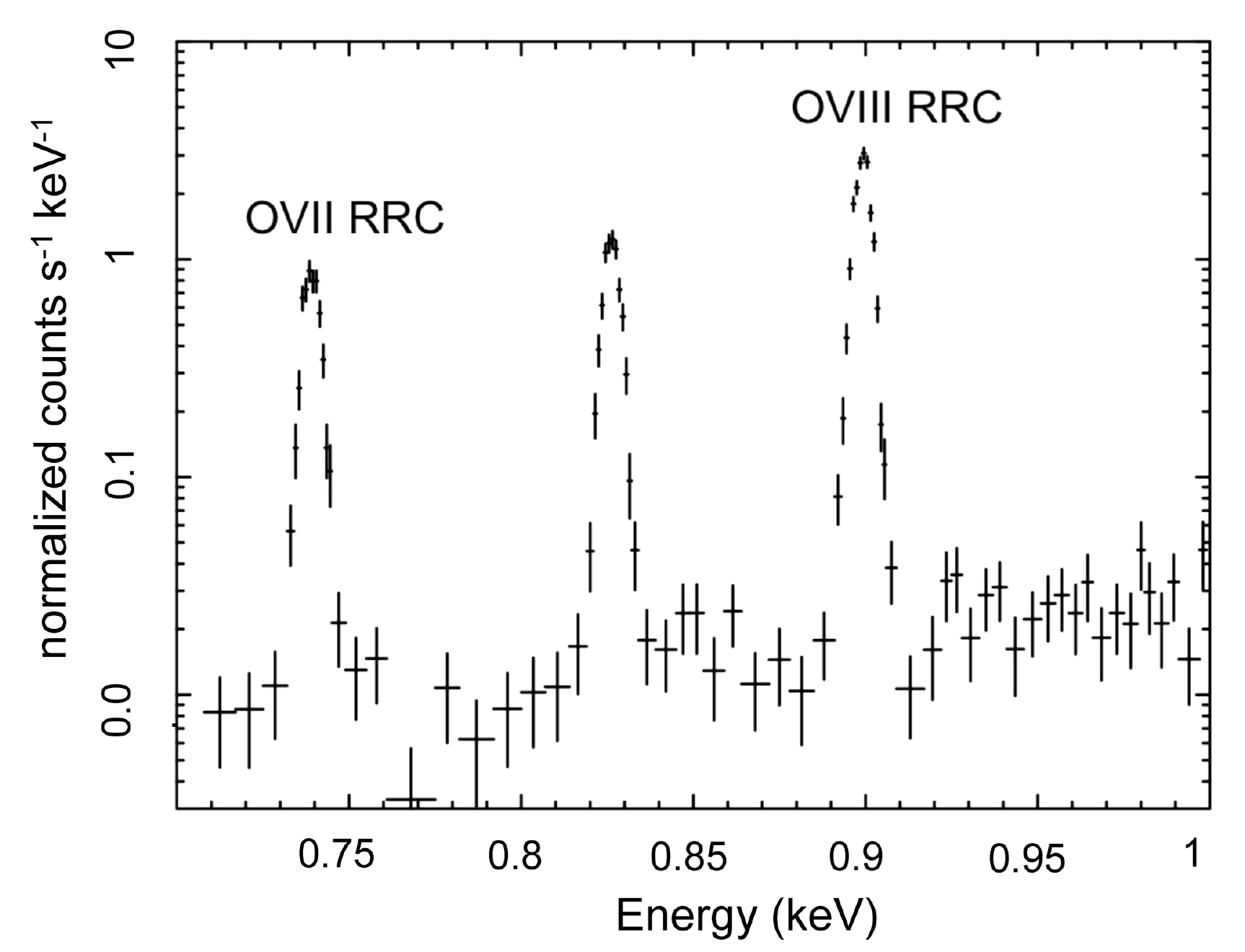}}
\caption{Left: Simulated 10\,ks SXS spectra of Circinus, where the black spectrum is based on the {\it XMM-Newton} RGS fit and the red spectrum is based in the {\it Chandra} HETG fit presented in \citet{Massaro06}. Right: Close-up of the OVII RRC and OVIII RRC features which will be prominent in a 100\,ks observation. The widths of these lines will diagnose the relative starburst to AGN contribution to the observed X-ray emission.}
\label{circ_hires}
\end{figure}

Further details of the relative amount of starburst to AGN activity
can be provided by high-resolution spectroscopy from SXS. {\it Chandra} grating analysis reveals that the circumnuclear emission
results from both photoionization and photoexcitation, i.e., reprocessed
AGN emission rather than star-formation \citep{Sambruna01b}. As shown
in Figure\,\ref{circ_hires}, a 100\,ks {\it ASTRO-H} observation,
based on the {\it Chandra} and {\it XMM-Newton} modeling of the
circumnuclear region \citep{Massaro06}, will provide a spectrum with
high signal-to-noise in the emission lines. Using diagnostics such as
the radiative recombination continua (RRC) and the ratios of forbidden
($f$), inter combination ($i$), and resonance ($r$), lines in the OVII
triplet, gives the relative amount of photoionization and collisional
ionization.  For instance, a narrow RRC feature
indicates that the plasma is photoionized \citep{Liedahl96} while the
ratio $G=(f+i)/r$ exceeds 4 in photo ionized plasma and is $\sim 1$
when collisionally ionized \citep{Porquet00}.  Based on the {\it
XMM-Newton} RGS spectra of Circinus, the OVII RRC (0.739\,keV) and
OVIII RRC (0.899\,keV) features will be prominent in a 100\,ks SXS
observation, as shown in Figure\,\ref{circ_hires} (right). The plasma
diagnostics that can be studied with SXS will represent an improvement
upon previous HEG and RGS analysis.

\subsection{NGC\,4945}
\label{S:4945}

NGC\,4945 is the brightest radio-quiet Seyfert~2 in hard X-rays
\citep{Done96}, as well as the closest bona fide Compton-thick AGN
\citep{Iwasawa93}. It lies in a nearby (distance of 3.7\,Mpc) edge-on
spiral galaxy with a highly obscured nucleus containing a composite
starburst/AGN system \citep{Moorwood94}. The nucleus also appears to
be obscured at mid-infrared wavelengths \citep{Krabbe01,Asmus14,Gandhi14}, with
the starburst being energetically dominant over the full optical and
infrared regime \citep{Moorwood96,Goulding09}. A starburst ring $\sim
200$\,pc in diameter has been imaged in the system
\citep{Marconi00}. At subarcsecond scales, a compact radio source is
detected, along with a 5\,pc--long jet \citep{Elmouttie97,Lenc09}. A
megamaser is known to exist in the nucleus, and VLBI mapping has
determined a black hole mass of $1.4 \times 10^6\,M_{\odot}$
\citep{Greenhill97}. NGC\,4945 was also detected with {\it Fermi}-LAT
\citep{LATstarburst}. Although the origin of the observed $\gamma$-ray
emission is not fully resolved, the detected GeV flux is consistent
with expectations from starburst emission.

\begin{figure}[!t]
\begin{center}
\includegraphics[scale=1.1]{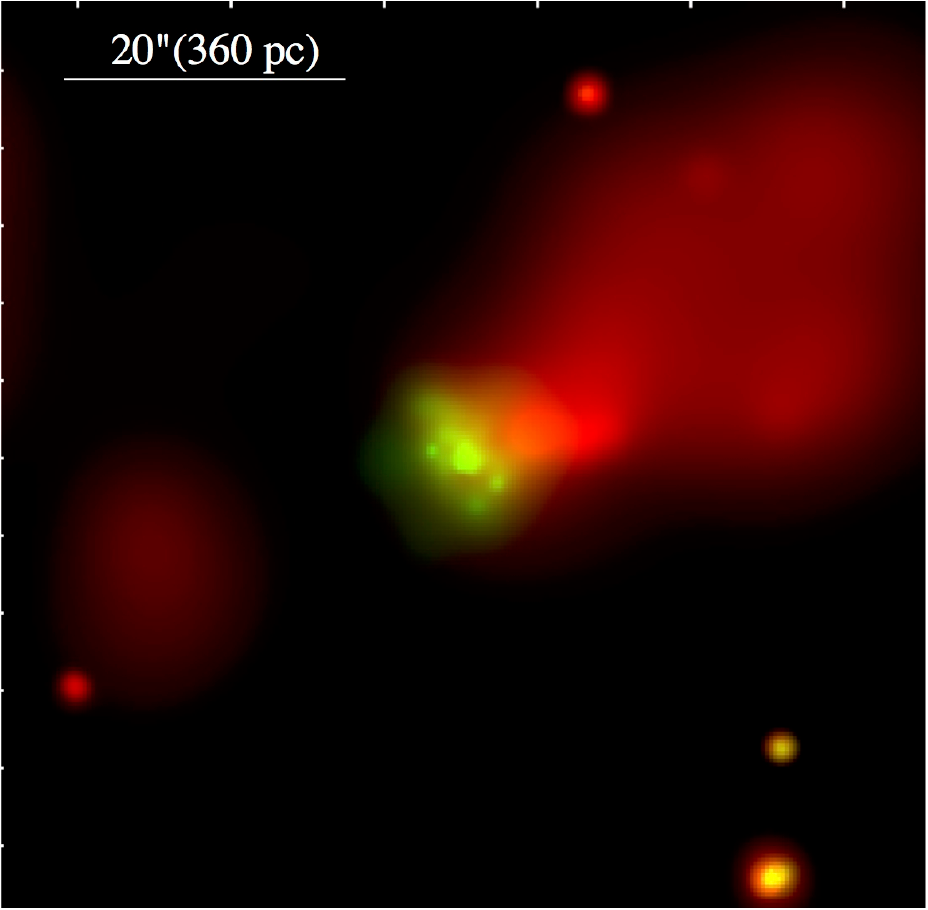}
\caption{Direct imaging of the core of NGC\,4945 with {\it Chandra} (red: 0.3--2\,keV is dominated by extended emission, presumably associated with the starburst; green: 2--10\,keV emission is associated with the reflector). Note that the reflector is resolved. The field of view is 1\,arcmin on a side (Credit: \citealt{Marinucci12}).}
\label{fig:4945chandra}
\end{center}
\end{figure} 

Despite its proximity, NGC\,4945 remains one of the least understood
of Compton-thick AGN. It is the best example of only a handful of
Compton-thick AGN known to vary strongly above 10\,keV. This variation
implies that scattering from circumnuclear clouds does not
completely suppress the direct component at hard X-ray energies
which, in turn, argues in favor of a small covering factor of the
circumnuclear `torus' \citep{Madejski00}. This has been confirmed from
broadband spectral analyses in a number of studies
\citep[e.g. ][]{Itoh08,Yaqoob12,Puccetti14}. Direct imaging with {\it Chandra} has resolved the reflector around the nucleus and found it to
be located at large nuclear distances ($\sim 30$\,pc) rather than
being associated with a classical sub-pc scale torus
\citep{Marinucci12}. This is shown in
Figure\,\ref{fig:4945chandra}. Clumpy reflecting clouds appear to be
mixed in with the extended starburst ring. A {\it Chandra}/HEG
observation has found a broad component to the Fe K$\alpha$ line at a
full-width-half-maximum (FWHM) of $2780_{-740}^{+1110}$\,km\,s$^{-1}$
(1-$\sigma$ uncertainty; \citealt{Shu11}). As discussed by
\citet{Yaqoob12}, this broadening may be related to the spatial extent
of the line, and the presence of a classical sub-pc scale torus has
not been definitely established in NGC\,4945; if it exists, it appears
to be geometrically-thin, a unique property amongst the nearby
Compton-thick AGN.

The source shows accretion episodes with luminosities of up to at
least 30\% of
the Eddington luminosity \citep{Puccetti14}. This is a
factor of about 10 larger than the mean Eddington rates inferred for
the local {\it Swift}-BAT AGN population \citep{Vasudevan10}. It is
unknown whether the Compton-thick obscurer on large scales is
providing the fuel necessary for this growth, or whether strong
radiative feedback is responsible for driving away matter and
rendering the torus geometrically-thin and clumpy. The high energy
cut-off ($E_{\rm cut}$) to the direct power-law which can constrain
the shape of the intrinsic continuum has not yet been securely detected in
NGC\,4945. Using {\it NuSTAR}, \citet{Puccetti14} determined a loose
constraint on $E_{\rm cut} = 190_{-40}^{+200}$\,keV in a high flux
state, but could only place lower limits of 200--300\,keV on $E_{\rm
cut}$ in median and low flux states. It remains to be determined
whether $E_{\rm cut}$ is dependent upon source flux. Note that these
lower limits lie beyond the energies that can be easily constrained
with {\it Swift}-BAT.

The unknown coronal properties at hard X-rays and poorly understood
reflector properties at soft energies make NGC\,4945 an ideal target
for broadband observations with {\it ASTRO-H}. The high resolution
spectroscopic capability of SXS will allow resolved measurements of
the iron line fluxes and widths, and can provide the first constraints on the
profile shapes. From the broadest component of the Fe K$\alpha$ line,
the SXS will locate the distance ($r$) of the innermost
reflector, which can be compared with torus models. This will give a
direct answer to the question of whether a classical sub-pc
Compton-thick torus exists in NGC\,4945 or not. Grating spectrometer
observations from the {\it Chandra}/HEG cannot distinguish spatial
broadening of lines from spectral broadening. The 5\,eV spectral
resolution of SXS will allow searches for sub-structure to within
1/10$^{\rm th}$ the line width of about 2800\,km\,s$^{-1}$ inferred
from HEG observations by \citet{Shu11}, and so will quantify the
strength of any single broad profile due to the inner torus
vs. broadening due to the extended reflector.

\begin{figure}[!t]
\begin{center}
\includegraphics[scale=0.325]{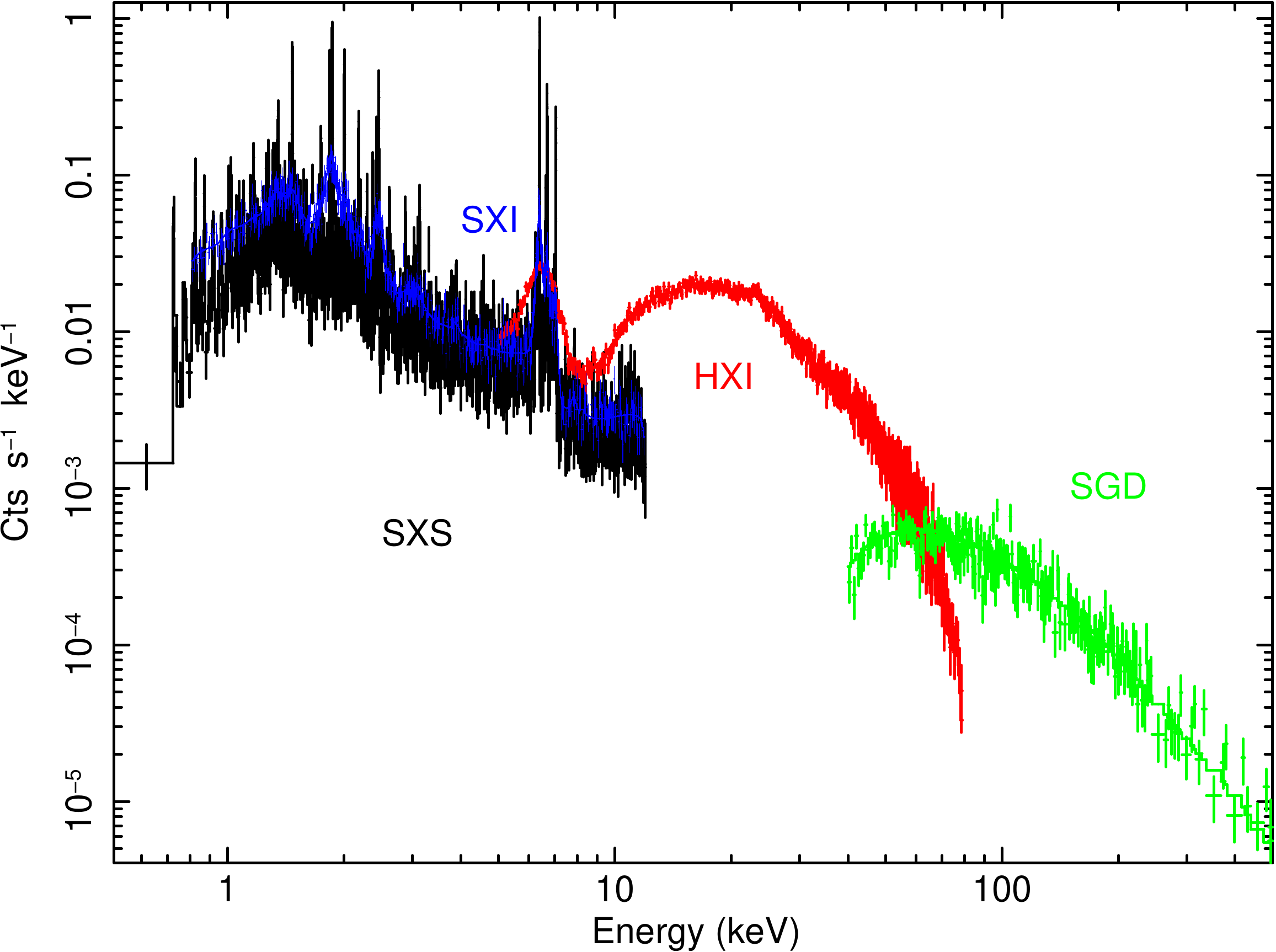}
\includegraphics[scale=0.325]{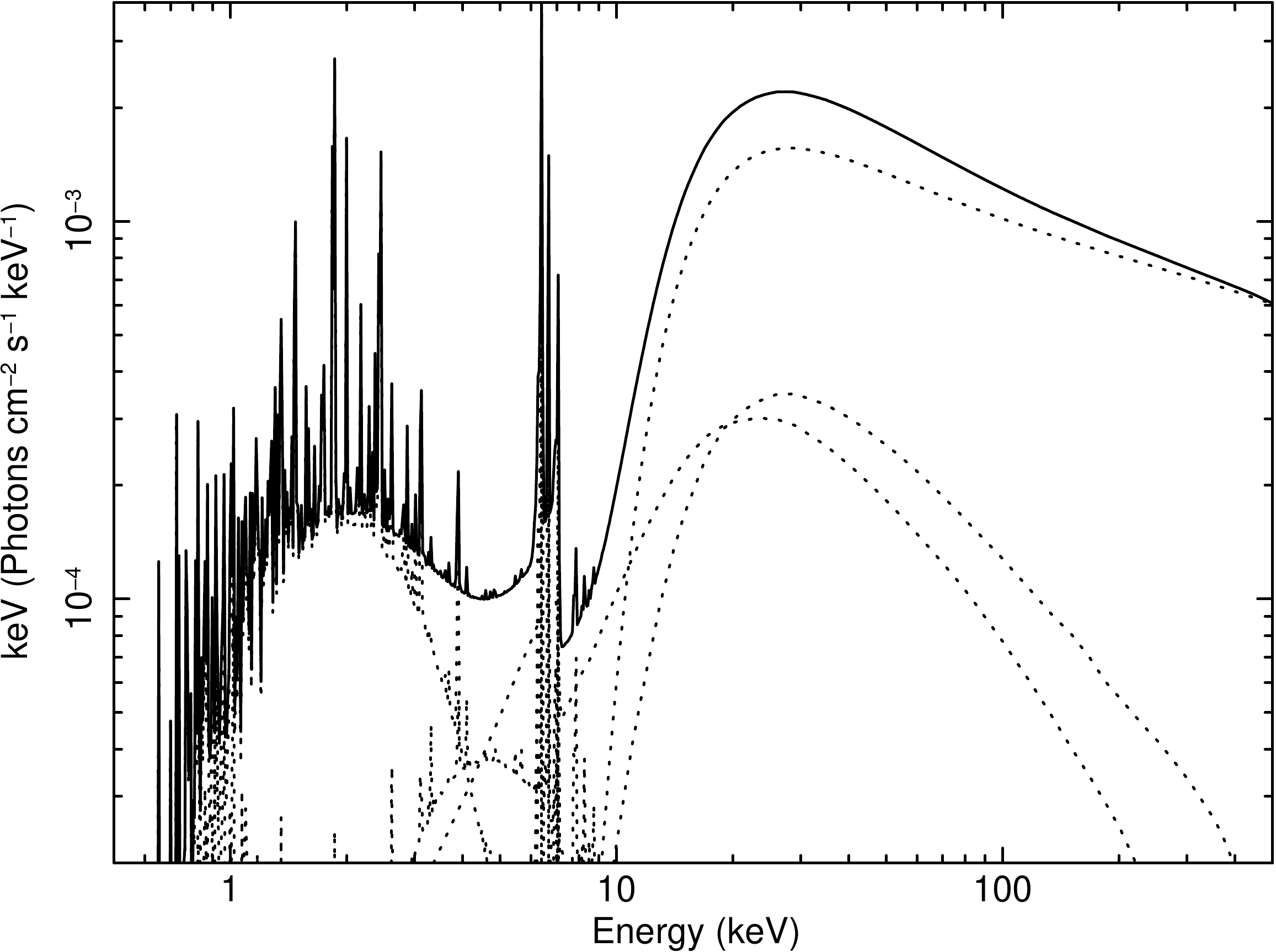}
\caption{{\it ASTRO-H} simulation of NGC\,4945 with 150\,ks of exposure, assuming the median flux model M from the {\it NuSTAR} observation reported in \citet{Puccetti14}. The counts spectra for SXS, SXI, HXI and SGD are shown on the left, and the model is shown on the right, in $F_{\rm E}$ units. The model includes three thermal components at soft energies, a power-law intrinsic continuum and two decoupled MYTorus components (edge-on and pole-on) for the Compton-thick absorber/reflector. A simple absorbed power-law ($\Gamma=2.2$; not shown) can describe the contribution of contaminating sources within the host galaxy at the level of 50\%\ of the 4--6 keV flux. These contaminants display no Fe line emission.}
\label{fig:4945sim}
\end{center}
\end{figure} 

In addition, the SXS will shed light on the power source of the ionized
lines as due to either a collisionally-ionized or photo-ionized
plasma, by measurement of ionized line ratios. If the reflector
itself is strongly photo-ionized, this may partly explain the low
reflection fraction. With regard to the low covering factor of the
reflector, this could be related either to torus clumpiness or to a
geometrically-thin configuration, or a combination of the two. The
iron line profiles will constrain the geometrical distribution of the
reflector. The Fe K$\alpha$ and $\beta$ lines will determine the
ionization state. Combined with the reflector location, the ionization
parameter ($\xi=L/nr^2$), and hence density ($n$) can be
determined. Clumpiness can then be constrained by combining the
observed line luminosity, $n$ and column density ($N_{\rm H}=nr$). 


Measurement of the iron line central energies and line profiles will
constrain torus kinematics. If there is significant radiative feedback
as a result of the high Eddington fraction, this may affect the matter
distribution on large scales, e.g. resulting in an outflow. Recently,
interferometric observations have revealed dusty outflows in
mid-infrared emission \citep{Hoenig13}. Dust velocities cannot be
measured in the mid-infrared though these are expected to be similar
to warm absorber velocities of several~hundred km\,s$^{-1}$. If the
dust is entrained with the X-ray--fluorescing gas, iron line
kinematics will directly measure the outflow velocities. Although some
constraints on the line kinematics are available from the {\it Chandra}/HEG analysis by \citet{Shu11}, these are based on single
Gaussian fits to the entire K$\alpha$ complex and include the narrow
doublet, Compton shoulder and any underlying broad component. The SXS will
enable deconvolution of these various components and allow much better
constraints on the kinematics. Finally, and importantly, {\it ASTRO-H}
will provide a sensitive measurement of the intrinsic continuum shape,
especially the high energy cutoff.

\begin{figure}[!t]
\begin{center}
\includegraphics[width=0.45\hsize]{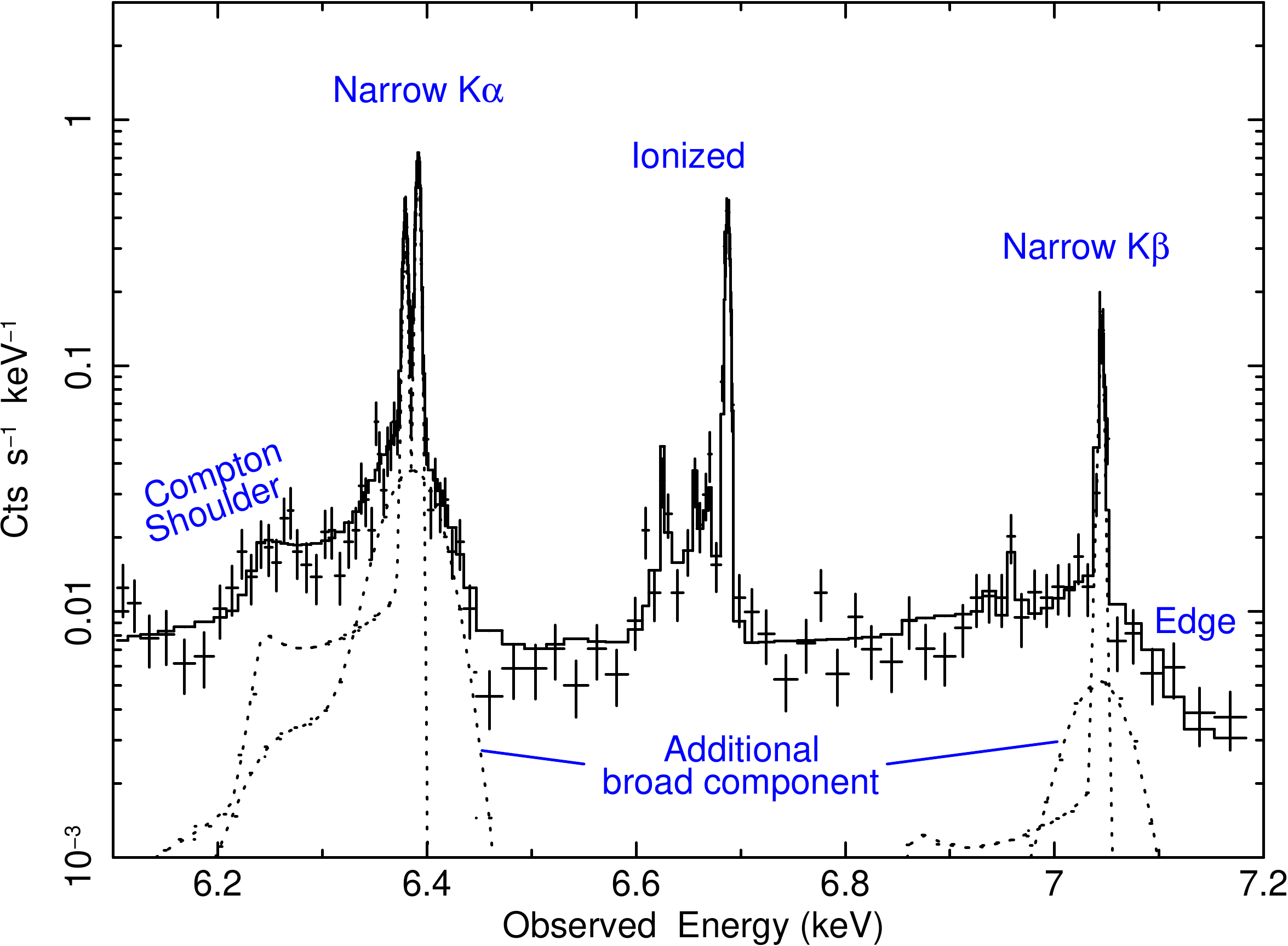}
{\includegraphics[width=0.45\hsize]{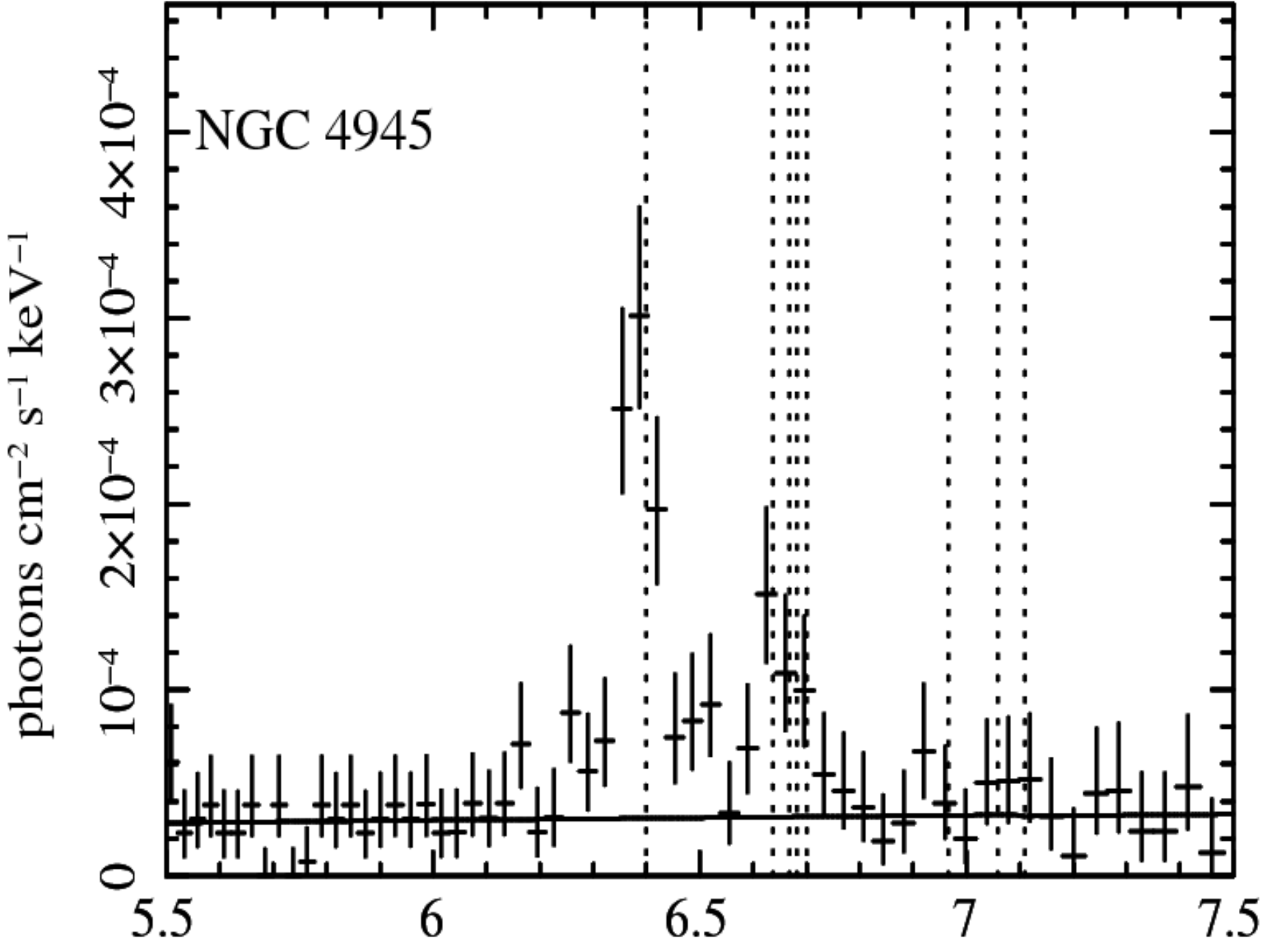}}
\caption{Left: Zoom-in on the iron line complex including neutral K$\alpha$, K$\beta$ and the ionized lines for the 150\,ks broadband simulation shown in Figure\,\ref{fig:4945sim}. The narrow line components are assumed to have intrinsic widths (full-widths-at-half-maximum; [FWHM]) of 100\,km\,s$^{-1}$, as may be expected on extended ($\sim 30$\,pc) scales where the bulk of the reflector is seen to be located with {\it Chandra} \citep{Marinucci12}. In addition, a potential underlying broad component simulating fluorescence from the walls of the inner torus (whose existence is uncertain thus far), has been included as a MYTorusL (fluorescence) component smoothed with a Gaussian of FWHM = 2800\,km\,s$^{-1}$ and normalized to contain half of the total K$\alpha$ line flux. This inner torus emission can clearly be distinguished from the narrower extended components in K$\alpha$. The most prominent K$\alpha$ and $\beta$ model components are shown as the dotted lines. Right: The best spectral resolution data on the iron complex available until now ({\it Chandra} High Energy Grating observation of NGC\,4945, as reported by \citealt{Shu11}). 
\label{fig:4945fe}}
\end{center}
\end{figure} 

We simulated a 150\,ks exposure with {\it ASTRO-H}, assuming the model
parameters of the median flux state {\it NuSTAR} observation reported
by \citet{Puccetti14}. The latest instrument responses available on
the {\tt astroh-cal} wiki page as of May 2014 were employed. This
model includes three thermal components to account for emission from
the superwind known in this system \citep{Schurch02}, and a direct
power-law of photon-index $\Gamma=1.88$ obscured by a column density
$N_{\rm H}=3.6\times 10^{24}$~cm$^{-2}$. The torus reflector is
modeled with the MYTorus code \citep{Murphy09} with decoupled
transmission and scattering normalizations to account for the source
complex emission properties (e.g. clumpiness, non-solar abundances
etc.), resulting in an effective covering fraction of 0.16. Additional
emission from Fe XXV is also included, as reported by
\citet{Puccetti14}. The result is shown in
Figure\,\ref{fig:4945sim}, and a zoom-in on the Fe complex is shown in
Figure\,\ref{fig:4945fe}.

The SXS will clearly measure the Fe line central energies and
widths. Towards the red side of the Fe K$\alpha$ line, the Compton
shoulder will be easily detectable and will shed light on the origin of
the iron fluorescence emission as either reflection or
transmission. Previous high spectral resolution observations with {\it Chandra}'s High Energy Grating are also shown in the figure. The
simulated SXS spectrum has far superior sensitivity to the {\it Chandra} data in terms of both line detection as well as
characterization.

The origin of the broad component is unclear from the {\it Chandra}
data, and one promising origin may be in an inner torus reflector
characterized by line velocities similar to the Broad Line Region. In
order to test whether such a component can be separated from the
narrow components, we added in a broad component with FWHM\,$=
2800$\,km\,s$^{-1}$ to the baseline model inferred from {\it NuSTAR}. This was simulated using a smoothed MYTorusL fluorescence
spectrum. The normalization of this inner torus broad component is
taken to be half of the total Fe K$\alpha$ line flux
(\citealt{Marinucci12} estimated that about half of the Fe line seen
with {\it Chandra} may be spatially unresolved). As shown in
Figure\,\ref{fig:4945fe}, the inner (broad) and outer (narrow) torus
reflection components can be clearly distinguished. The fractional
error on the FWHM of this broad component when fitting the simulated
spectrum is $\approx 0.13$, i.e. 13\%\ of the assumed intrinsic FWHM,
which is 350\,km\,s$^{-1}$ (90\%\ confidence). The observation will be
sensitive enough to detect the presence of sub-structure in addition
to any single broad component.

The ionized line around 6.7\,keV is also apparent in the spectrum,
with the resonance, forbidden and intercombination lines all
potentially resolvable if the velocity spread is low. Flux ratios
between these lines can distinguish between collisionally ionized
and photo-ionized plasma, and thus shed light on the connection
between the AGN and starburst emission. Finally, the Fe edge and Ni
K$\alpha$ (latter not shown) are also detectable and will place
important constraints on elemental abundances.

\begin{figure}[!t]
\begin{center}
\includegraphics[width=0.55\hsize]{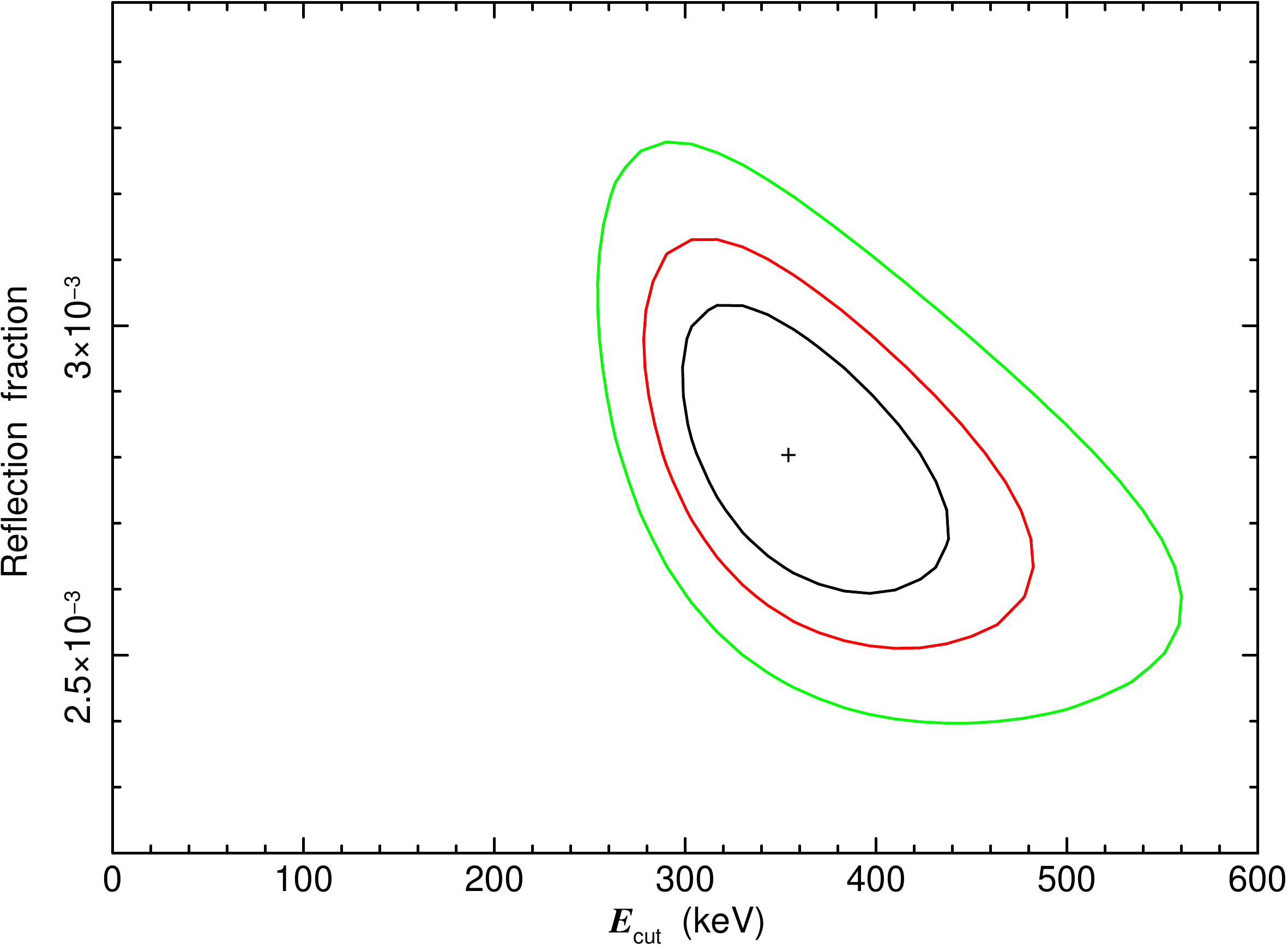}
\caption{Confidence contours for a fit to the refection fraction and $E_{cut}$ for the simulated 150 ks exposure of NGC\,4945, assuming $E_{\rm cut} = 400$\,keV. The contours are for $\Delta \chi^2$ changes corresponding to 68\%, 90\% and 99\% confidence, for two parameters of interest. $E_{cut}$ can be constrained well beyond energies constrained thus far. 
\label{fig:4945contours}}
\end{center}
\end{figure} 

The combination of HXI and SGD will constrain $E_{\rm cut}$ to higher
energies than possible before. The MYTorus model above does not
include a high energy cutoff. So, in order to simulate the cutoff, we
approximated the continuum above 10 keV as a combination of the
absorbed power-law from \citet{Puccetti14} and a pure reflection
PEXRAV component. Such a model was found to be fully acceptable in the
{\it Suzaku} analysis reported in \citet{Itoh08}, with a reflection
fraction (defined as the ratio between the normalizations of the
reflector and the intrinsic power-law) of $3\times 10^{-3}$. We
simulated various values of $E_{\rm cut}$. Note that $E_{\rm cut}$ can
be sensitive to calibration uncertainties, especially the
cross-normalization between HXI and SGD, so we included a 10\%\
systematic uncertainty in the SGD simulation, as a conservative
approximation of our ignorance of such extra uncertainty.
Figure\,\ref{fig:4945contours} shows the result of the constraints on
the cutoff when $E_{\rm cut} = 400$\,keV is assumed for the input
spectrum. The figures shows joint confidence intervals for 
the reflection fraction and $E_{\rm cut}$. It is clear that $E_{\rm
cut}$ will be significantly constrained despite the degeneracy with
the reflection component. If the source happens to be caught in a high
flux state with a much harder power-law slope, as in the {\it Suzaku}
observation of \citet{Itoh08} where $\Gamma=1.5-1.6$ was measured, the
confidence interval on $E_{\rm cut}$ will shrink substantially. These
improvements over previous data are a consequence of the fact that
NGC\,4945 is the brightest hard X-ray Sy 2 in the sky, and is thus
ideal for SGD observation.

The nucleus of NGC\,4945 is the brightest point source in the field
above 2 keV. The brightest nearby ULX lies at angular distance of
approximately 1~armin from the nucleus \citep{Walton11} and has an
observed 0.2--12 keV luminosity about 10 times lower than the other
components (thermal emission and AGN reflection) modeled by
\citet{Puccetti14} in {\it Suzaku} and {\it NuSTAR} data. Some other
binaries within the central few tens of arcsec contribute at most
50\%\ (combined) of the {\it Suzaku} flux \citep{Puccetti14}. But
importantly, their spectrum can be represented as simple absorbed
power-law ($\Gamma=2.2$) and they show no evidence of Fe line
emission. Therefore, they will not contaminate the detailed analysis
of the iron line properties to be carried out with SXS, and
simultaneous {\it Chandra}/{\it XMM-Newton} observations are not
critical with {\it ASTRO-H}. A short ($\sim 10-20$\,ks), strictly
simultaneous {\it NuSTAR} observation can be considered in order to
mitigate the effect of unknown absolute calibration uncertainties in
the early part of the {\it ASTRO-H} mission. This may prove worthwhile
given the strong hard X-ray variability of the source. The {\it NuSTAR} flux can be used to anchor the HXI flux, thus removing one
degree of freedom from the instrumental cross-normalization
uncertainties.

\section{A Space for New Discoveries with {\it ASTRO-H}}
\label{S:other}

The hard X-ray/soft $\gamma$-ray sky is still a poorly explored territory
due to limited sensitivities of previous and current instruments
operating in this range. On the other hand, non-thermal processes in
various types of both transient and steady sources manifest (most)
clearly in exactly this domain. In this section some selected topics dealing
with non-thermal phenomena in hard X-rays are discussed briefly, to
emphasize the space for new discoveries with hard X-ray instruments
onboard the {\it ASTRO-H}.

\subsection{Tidal Disruption Events}
\label{S:TDEs}

As an example of a source class that may provide surprises as well
as crucial information when observed with {\it ASTRO-H}, we now consider the
so-called ``tidal disruption events'' \citep[TDEs; see][for recent 
reviews]{Gezari09,Saxton12}. Two of the observational ``clues'' often referred 
to in current scenarios for how and when supermassive black holes grow
and possibly interact with their host galaxies are
(i) the amounts of AGN (black hole accretion) activity and star formation activity
in the Universe appear to evolve in the same manner as a function of time,
and (ii) today there appears to be a correlation, the $M-\sigma$ relation,
between the central velocity dispersion of a massive galaxy and the mass of
supermassive black hole that always seems to be found at its center.
While we can come up with plausible models and even computer simulations
for supermassive black hole evolution that are consistent with these
clues, the bottom line remains that we still cannot deal confidently with
the relevant physics. Moreover, the clues themselves may not be quite right:
we may still be missing significant star formation and black hole activity due to 
obscuration, black hole growth can be intrinsically dark
(due to low radiative efficiency or the importance of black hole-black
hole mergers), and the $M-\sigma$ relation shows significant scatter and is
based on a relatively small sample of systems. For these reasons,
an accurate, theorist-independent, census of supermassive black holes and
their activity (``black hole demographics'') is still a holy grail,
especially now that we realize that supermassive black holes can release enough energy
to radically alter their surroundings (``feedback''). Cutting through
obscuration using hard X-ray observations or measuring black hole
mass and spin by constraining the size of the innermost stable orbit via
lower energy X-ray observations, e.g., by the SXS, are discussed
extensively in the ``AGN Reflection'' {\it ASTRO-H} White Paper. 
None of these approaches help,
however, if the supermassive black hole is simply shut off, i.e., not
accreting or radiating strongly -- which is actually the case most of the
time given the rarity of AGN activity, especially at redshifts $z<1$.

Tidal disruption events, once properly understood, offer a potentially
powerful way to make progress on the black hole demographics problem.
Even if a supermassive black hole, like the one at the center of our galaxy,
is presently starved of gas, there is a non-zero chance that a star orbiting
in the host galaxy nucleus will get scattered onto an orbit that takes it very
close to the black hole. If the star gets close enough (crossing
the tidal disruption radius), the tidal force on the star due to the
black hole's gravity overwhelms the self-gravity of the star, shredding it and 
sending a significant fraction of its mass into the black hole on a
timescale of months according to current estimates. Since a mass accretion rate
$\sim 1\,M_{\odot}$\,yr$^{-1}$ corresponds to that of a powerful quasar,
a tidal disruption event should be a fairly obvious and unambiguous signature
of the presence of a supermassive black hole -- if we happen to be
looking at the right time.  Unfortunately, while tidal disruption effects may
be dramatic, they are also rare.  The rate at which stars scatter into
the black hole depends sensitively on the dynamical state of the
host galaxy nucleus, but it is estimated to 1 per $10^4$ to $10^6$ years 
\citep{Merritt05,Vasiliev13}. In other words, if we hope to see one disruption 
event in a year, we need to monitor at least one million galaxies
on at least $\sim$\,month timescales. This is still a tall order, indeed an
impossible one for a narrow field-of-view instrument like {\it ASTRO-H}, but it is
one that is becoming increasingly feasible given the advent of 
increasingly sensitive large-area surveys. Already the reports of TDE candidates 
are a few per year, and the arrival of LSST at the tail end of the {\it ASTRO-H} 
mission could boost that rate to $\gtrsim 100$ per year if our current 
understanding of TDEs is correct.

\begin{figure}[!t]
\begin{center}
\includegraphics[scale=0.175]{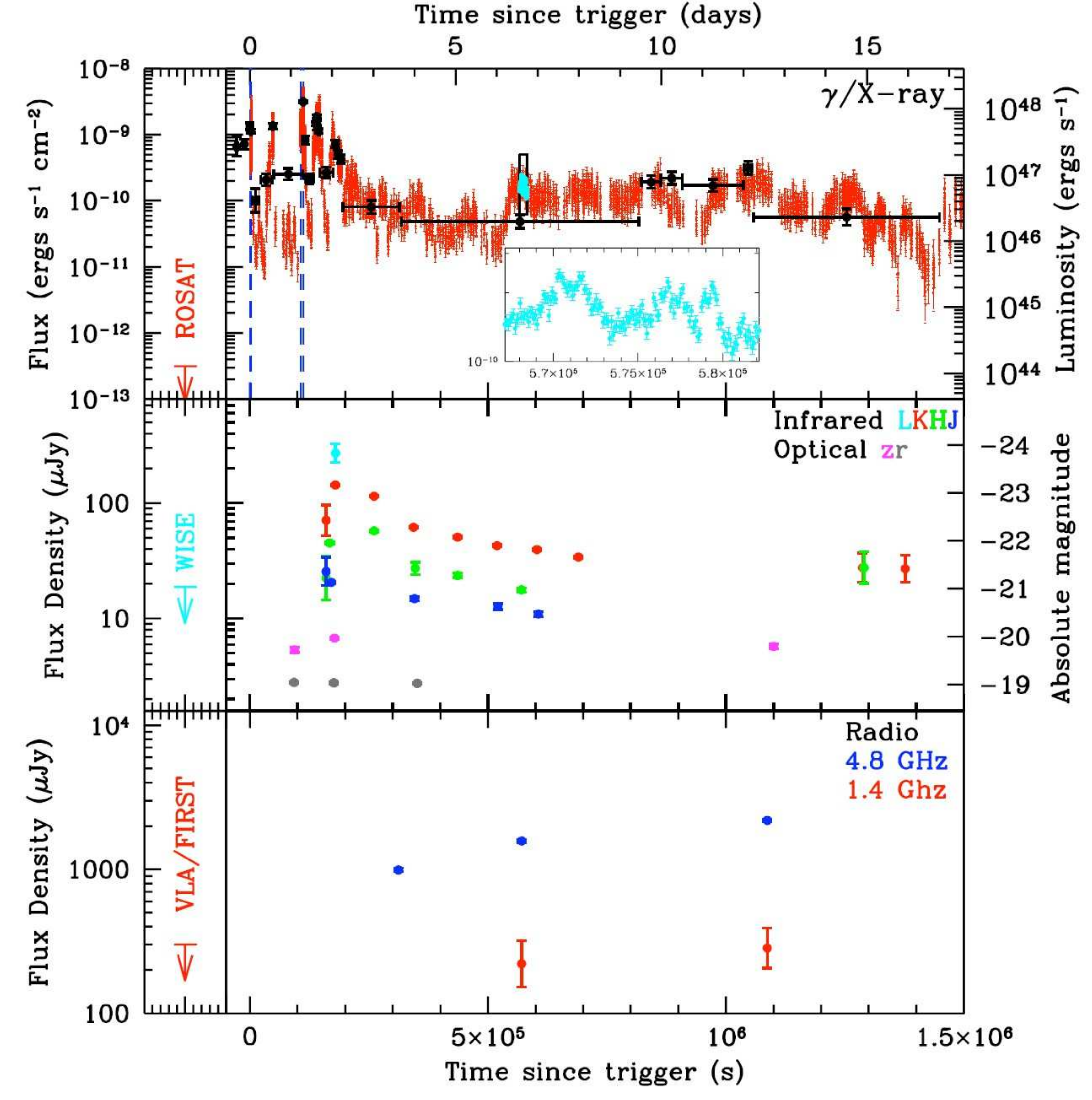}
\includegraphics[scale=0.2]{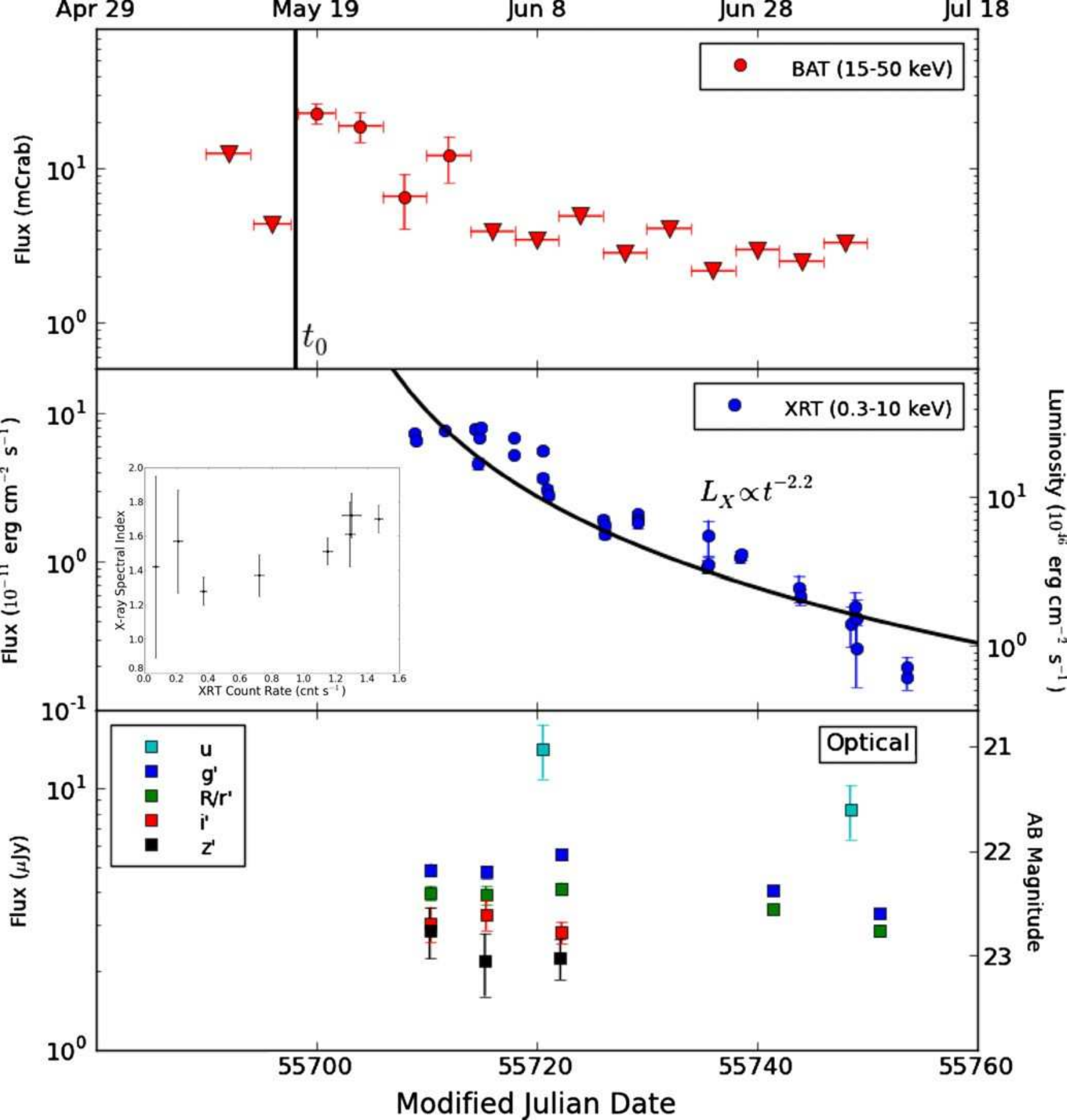}
\includegraphics[scale=0.4]{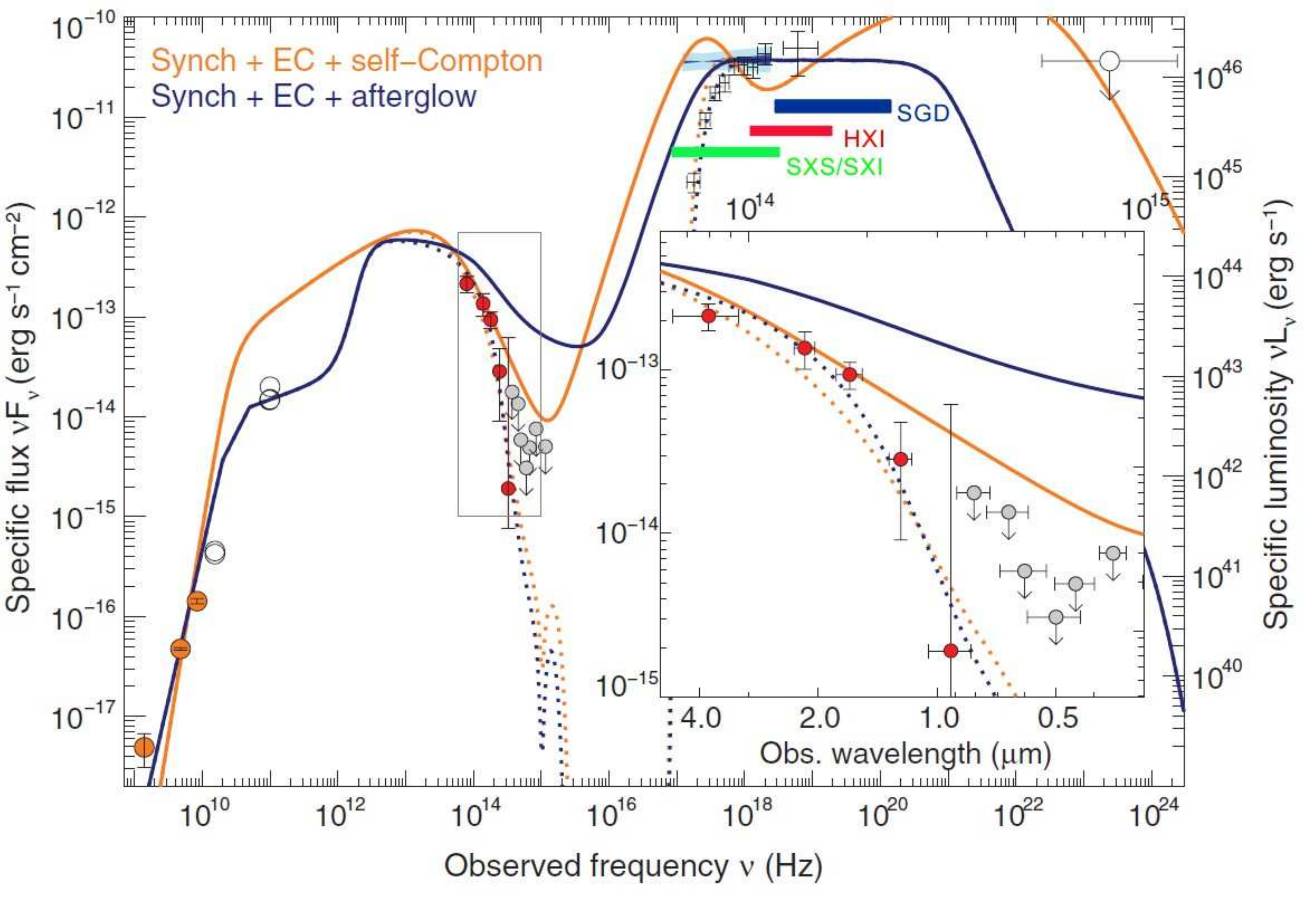}
\caption{Upper left panel: Light curves at multiple energies
for the unusual TDE candidate, SWIFT\,1644+57 \citep{Levan11}. Note the
high flux at hard X-ray energies, particularly at the beginning of the
burst. Lower panel: A snapshot of the spectral energy distribution
of SWIFT\,1644+57 and possible emission models to explain it,
adapted from \cite{Bloom11}. The broad energy coverage provided by the
{\it ASTRO-H} instruments (shown by the red, green, and blue horizontal bars)
would play an important role in constraining
possible emission mechanisms, particularly by measuring or placing lower
limits on the cutoff energy of the high-energy peak. Upper right panel:
Light curves, adapted from \cite{Cenko12},
for another ``high-energy'' TDE candidate, SWIFT\,2058+05,
that appears similar to SWIFT\,1644+57. This event was found not long after
SWIFT\,1644+57, indicating that events of this type may actually not be 
that rare.}
\label{fig:TDE}
\end{center}
\end{figure}

While {\it ASTRO-H} may not be able to find TDEs, it can play a key
role in confirming that a candidate event is a TDE and then helping
to unravel the astrophysics of the event using its broadband suite of instruments.
As evidenced by the long debate over the starburst vs. black hole origin
of AGN activity, the fact that supernova searches studiously avoid the centers of
galaxies, and that luminous blue variable stars like Eta Carinae
instead tend to be found near centers of galaxies and can flare by
many magnitudes on the timescale of years, it is actually quite
easy to confuse processes related to stars with those arising from
black holes. The arguments usually end, however, with the detection of
intense hard X-ray ($>10$\,keV) flux that also varies rapidly on $\sim$\,hour 
to day timescales. As the most sensitive hard X-ray telescopes operating,
{\it NuSTAR} and {\it ASTRO-H} therefore have an important role to play.
Indeed Figure\,\ref{fig:TDE} shows two recent (and unexpected!) examples of 
TDE candidates where hard X-ray
measurements played a key role in arguing for a supermassive black hole
origin of the emission. The event SWIFT\,1644+57, in fact, was originally
discovered as a gamma-ray burst \citep[which then refused to fade, see][for more 
discussion of this source]{Levan11,Bloom11}.
For that object, not only would {\it ASTRO-H} have seen a strong signal
in all instruments, allowing it to measure and track the shape of the
high-energy spectrum much better than {\it Swift} 
(particularly at energies above $100$\,keV, where a cutoff might start becoming
visible), but for the first few days, the source was so bright
($\sim 10^{-9}$\,erg~cm$^{-2}$\,s$^{-1}$) that SGD would
either measure or strongly constrain the polarization of the high-energy flux,
in turn placing important constraints on the physical emission mechanism.
Given that supermassive black holes are known to produce powerful
relativistic jets, in retrospect
the detection of a TDE at $\sim$\,MeV energies might not be so surprising,
despite the standard prediction for TDE events that their emission peaks
at the energies typical of an optically thick AGN accretion disk,
at UV to soft X-ray energies. Even if most TDEs follow the standard prediction,
essentially every black hole system we know of has coronal X-ray
emission that also excites atomic features such as iron fluorescence
lines. Given that within 100 Mpc of us, there are over 50,000 known
galaxies and probably more unknown that have not had their redshifts
measured yet, the odds of having a bright, nearby TDE occur during the
{\it ASTRO-H} are not small.
In this case, prompt and repeated observations by {\it ASTRO-H}'s instrument suite,
including the SXS, can uniquely help us track and probe the response and
evolution of the inner accretion flow through a large change in mass accretion rate.
Not only may TDEs turn out to be key probes of black hole demographics,
but well-observed TDEs may
also represent some of the best laboratories to probe AGN accretion physics.
The analogs of state transitions in stellar mass black holes, if they occur at
all for AGN, would take ten of thousands to millions of years if we simply
scaled by mass from the $\sim$ month-year timescales for stellar mass
black holes. In the case of a TDE, however, a star on a plunging orbit that then
disrupts close to the black hole can deliver gas
much faster to the black hole, circumventing the need for gas to move (slowly)
through the outer parts of the accretion disk.

\subsection{High-redshift Blazars}
\label{S:high-z}

\begin{figure}[!t]
\begin{center}
\hspace{-1cm}
\includegraphics[scale=0.43]{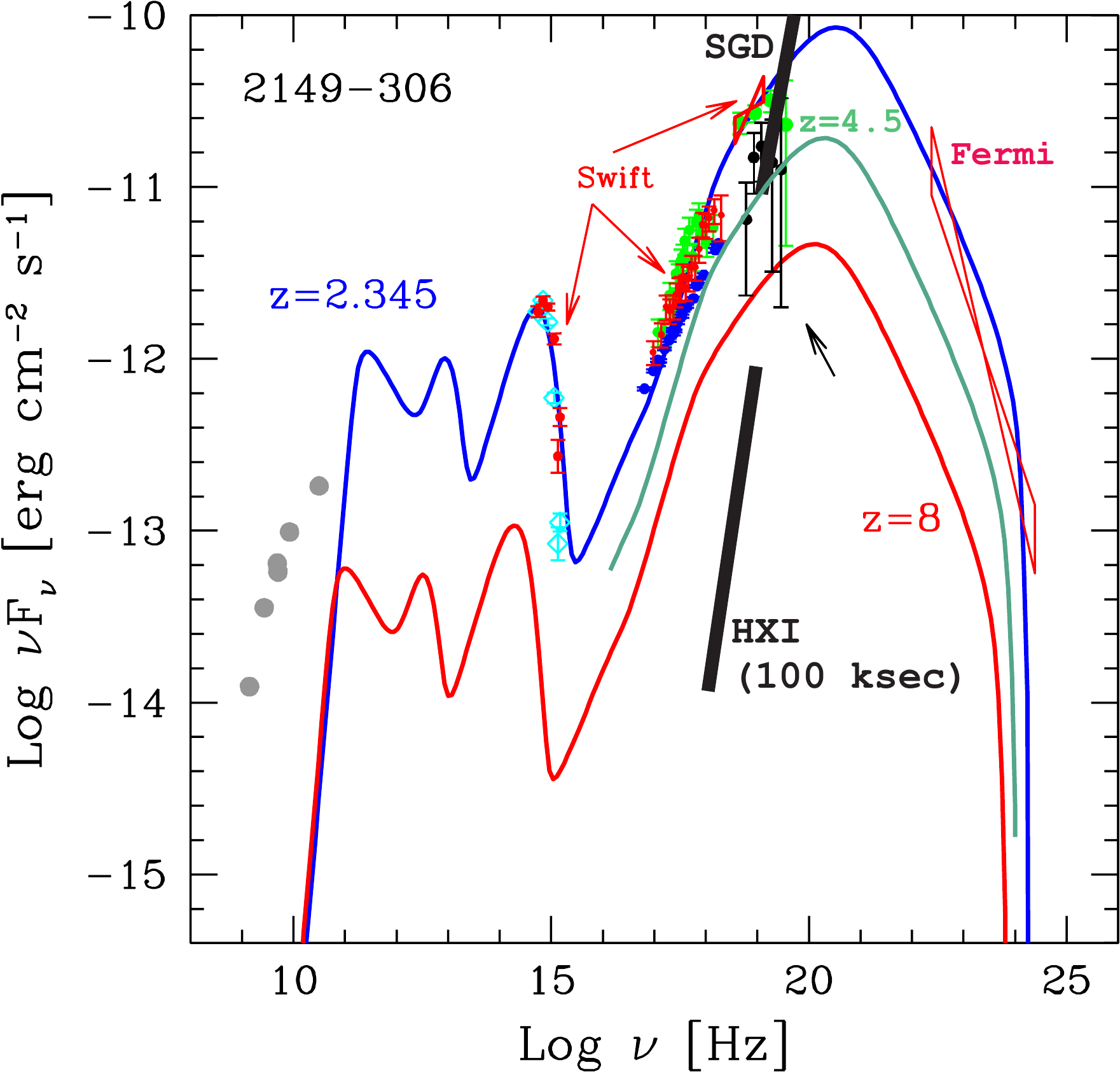}
\caption{Broad-band spectral energy distribution of the $z=2.345$ blazar 2149--306, along with the spectral model fit (blue curve) adopted from \cite{Ghisellini10}. In addition, the green and red curves illustrate the source spectrum shifted to the redshifts $z=4.5$ and 8, respectively. The HXI and SGD sensitivity curves corresponding to the 100\,ks exposure are shown as thick black lines.}
\label{fig:high-z-blaz}
\end{center}
\end{figure}

The discovery of bright (mostly radio-quiet) quasars at the redshifts
$z \sim 6$ and beyond \citep{Fan06,Willott10,Mortlock11} implies that
SMBHs with masses $\geq 10^9 \, M_{\odot}$ were already assembled and
fully developed when the Universe was less than $1$\,Gyr-old. The
mechanism by which such systems formed during the first Gyr after the
Big Bang is currently debated. In addition to this, the all-sky
{\it Swift}-BAT survey identified about 
10 luminous blazars at redshifts greater
than 2. Although accessing the energetics of jets
(as well as black hole masses) in blazar sources is rather
model-dependent, the high-$z$ blazars detected by BAT seem to be
particularly powerful and produced by particularly massive black holes
accreting at Eddington rates
\citep{Ghisellini11,Ghisellini13}. Importantly, the
number of such objects can be used to constrain the parent population
of jetted AGN in the early Universe in general. As discussed in
\citet{Volonteri11}, a comparison between the high-$z$ blazar
sample and quasars detected in the Sloan Digital Sky Survey (SDSS),
indicates a serious deficit of `misaligned' radio-loud AGN at $z >
3$. Possible explanations for this disagreement elaborated in
\citeauthor{Volonteri11} include lower bulk Lorentz factors of
high-redshift quasar jets when compared with their low-redshift
analogs, or the incompleteness of the utilized SDSS sample. {\it
ASTRO-H} can contribute to the study of high-$z$ blazars by means of a
detailed spectral characterization of their broad-band X-ray continua,
allowing for a precise insight into the source energetics 
\cite[see Figure\,\ref{fig:high-z-blaz}; see also in this context][for the 
recent {\it NuSTAR} results]{Sbarrato13}. 
It is important to note here that the majority of the discussed objects
release the bulk of their radiative power at MeV frequencies, and as
such are particularly luminous in hard X-rays; moreover, luminous
blazars are often characterized by flat or very flat hard
X-ray spectra \citep[photon indices $\Gamma < 2$, see][]{Sikora09}.

\subsection{Non-thermal Emission of Novae}
\label{S:novae}

A nova eruption is a thermonuclear runaway on the surface of a white
dwarf that ejects much of the hydrogen-rich envelope that the star has
accreted. Many novae have been detected in the 0.5--10\,keV band as
sources of optically-thin, collisionally-excited thermal plasma
emission \citep{Mukai08}. This indicate that shocks are ubiquitous in
novae.  The recent discovery of GeV emission first from V407\,Cyg
\citep{LATnova} and subsequently several others with {\it Fermi}-LAT
indicates that nova shocks are also capable of particle
acceleration. In addition, the detection of apparent synchrotron
emission in the radio \citep[see, e.g.,][]{Rupen14} in some novae is
further evidence for a population of accelerated electrons.  We will
gain a better understanding of the accelerated particles in nova
shocks if we can detect and characterize the associated hard
X-rays. This has so far only been claimed in one nova, V2491\,Cyg
\citep{Takei09}, using non-imaging {\it Suzaku} HXD/PIN data, but this 
erupted shortly before the launch of {\it
Fermi}, so there is no constraint on its  GeV emission. 
If a sufficiently 
bright nova is discovered during the {\it ASTRO-H} era, particularly
one that is detected with {\it Fermi}-LAT, it will be an excellent
target for {\it ASTRO-H} HXI observation for this possible non-thermal
hard X-ray component, while the SXS will be able to provide an
in-depth characterization of the thermal X-ray emission.

\subsection{Unidentified Hard X-ray/soft $\gamma$-ray Sources}
\label{S:unid}

COMPTEL instrument onboard the CGRO provided the first complete all-sky survey in the energy range 0.75 to 30\,MeV. The First COMPTEL Catalogue includes 32 steady sources and about 50 transients; among the continuum sources are spin-down pulsars, XRBs, supernova remnants, interstellar clouds, AGN, gamma-ray bursts, and solar flares \citep{Schoenfelder00}. The third IBIS/ISGRI catalog, based on $>40$\,Ms observations performed during 3.5 yrs of the {\it INTEGRAL} operation, includes $>400$ sources detected in the 17--100\,keV range \citep{Bird07}. Finally, the {\it Swift}-BAT 70-month survey has detected 1171 hard X-ray sources in the 14--195\,keV band down to a significance level of $4.8\sigma$, associated with 1210 counterparts including various types of AGN, clusters, cataclysmic variables, pulsars, stars, supernova remnants, and XRBs \citep{Baumgartner13}. Thus, the hard X-ray/soft $\gamma$-ray sky is crowded indeed, being populated by a variety of high-energy emitters. Among these, some sources still remain to be identified, and many to be characterized precisely (regarding their spectral and timing properties) at $>10$\,keV photon energies. This constitutes a space for potential new exciting discoveries with the high-energy instruments HXI and SGD onboard the {\it ASTRO-H} in the near future.

\begin{multicols}{2}
\end{multicols}


{\footnotesize

\begin{thebibliography}{99}

\bibitem[Abdo et al.(2010a)]{LATnova} Abdo, A.~A., Ackermann, M., Ajello, M., et al.\ 2010a, Science, 329, 817 
\bibitem[Abdo et al.(2010b)]{LAT3C279} Abdo, A.~A., Ackermann, M., Ajello, M., et al.\ 2010b, Nature, 463, 919 
\bibitem[Abdo et al.(2010c)]{LAT273} Abdo, A.~A., Ackermann, M., Ajello, M., et al.\ 2010c, ApJL, 714, L73 
\bibitem[Abdo et al.(2010d)]{LATCenA} Abdo, A.~A., Ackermann, M., Ajello, M., et al.\ 2010d, ApJ, 719, 1433 
\bibitem[Abdo et al.(2011a)]{LAT501} Abdo, A.~A., Ackermann, M., Ajello, M., et al.\ 2011a, ApJ, 727, 129 
\bibitem[Abdo et al.(2011b)]{LAT421} Abdo, A.~A., Ackermann, M., Ajello, M., et al.\ 2011b, ApJ, 736, 131 
\bibitem[Ackermann et al.(2012)]{LATstarburst} Ackermann, M., Ajello, M., Allafort, A., et al.\ 2012, ApJ, 755, 164 
\bibitem[Aharonian et al.(2007)]{HESS2155} Aharonian, F., Akhperjanian, A.~G., Bazer-Bachi, A.~R., et al.\ 2007, ApJL, 664, L71 
\bibitem[Aharonian et al.(2009)]{HESSCenA} Aharonian, F., Akhperjanian, A.~G., Anton, G., et al.\ 2009, ApJL, 695, L40 
\bibitem[Arevalo et al.(2014)]{Arevalo14} Ar{\'e}valo, P., Bauer, F.~E., Puccetti, S., et al.\ 2014, ApJ, 791, 81
\bibitem[Asmus et al.(2014)]{Asmus14} Asmus, D., H{\"o}nig, S.~F., Gandhi, P., Smette, A., \& Duschl, W.~J.\ 2014, MNRAS, 439, 1648 
\bibitem[Bauer et al.(2001)]{Bauer01} Bauer, F.~E., Brandt, W.~N., Sambruna, R.~M., et al.\ 2001, AJ, 122, 182 
\bibitem[Baumgartner et al.(2013)]{Baumgartner13} Baumgartner, W.~H., Tueller, J., Markwardt, C.~B., et al.\ 2013, ApJS, 207, 19 
\bibitem[Beckmann et al.(2011)]{Beckmann11} Beckmann, V., Jean, P., Lubi\'nski, P., et al.\ 2011, A\&A, 531, 70 
\bibitem[Begelman \& Sikora(1987)]{Begelman87} Begelman, M.~C., \& Sikora, M.\ 1987, ApJ, 322, 650 
\bibitem[Begelman et al.(1984)]{Begelman84} Begelman, M.~C., Blandford, R.~D., \& Rees, M.~J.\ 1984, Reviews of Modern Physics, 56, 255 
\bibitem[Beilicke et al.(2012)]{Beilicke12} Beilicke, M., Baring, M.~G., Barthelmy, S., et al.\ 2012, American Institute of Physics Conference Series, 1505, 805 
\bibitem[Bellazzini et al.(2010)]{Bellazzini10} Bellazzini, R., Costa, E., Matt, G., \& Tagliaferri, G.\ 2010, X-ray Polarimetry: A New Window in Astrophysics by Ronaldo Bellazzini, Enrico Costa, Giorgio Matt and Gianpiero Tagliaferri.~Cambridge University Press, 2010
\bibitem[Bird et al.(2007)]{Bird07} Bird, A.~J., Malizia, A., Bazzano, A., et al.\ 2007, ApJS, 170, 175 
\bibitem[Black et al.(2010)]{Black10} Black, J.~K., Deines-Jones, P., Hill, J.~E., et al.\ 2010, Space Telescopes and Instrumentation, 7732,  
\bibitem[Blandford et al.(2002)]{Blandford02} Blandford, R., Agol, E., Broderick, A., et al.\ 2002, Astrophysical Spectropolarimetry, 177 
\bibitem[Bloom et al.(2011)]{Bloom11} Bloom, J.~S., Giannios, D., Metzger, B.~D., et al.\ 2011, Science, 333, 203 
\bibitem[Bloser et al.(2009)]{Bloser09} Bloser, P.~F., Legere, J.~S., McConnell, M.~L., et al.\ 2009, Nuclear Instruments and Methods in Physics Research A, 600, 424 
\bibitem[Bonometto \& Saggion(1973)]{Bonometto73} Bonometto, S., \& Saggion, A.\ 1973, A\&A, 23, 9 
\bibitem[Celotti \& Matt(1994)]{Celotti94} Celotti, A., \& Matt, G.\ 1994, MNRAS, 268, 451 
\bibitem[Cenko et al.(2012)]{Cenko12} Cenko, S.~B., Krimm, H.~A., Horesh, A., et al.\ 2012, ApJ, 753, 77 
\bibitem[Chatterjee et al.(2009)]{Chatterjee09} Chatterjee, R., Marscher, A.~P., Jorstad, S.~G., et al.\ 2009, ApJ, 704, 1689
\bibitem[Chatterjee et al.(2011)]{Chatterjee11} Chatterjee, R., Marscher, A.~P., Jorstad, S.~G., et al.\ 2011, ApJ, 734, 43
\bibitem[Chernyakova et al.(2007)]{Chernyakova07} Chernyakova, M., Neronov, A., Courvoisier, T.~J.-L., et al.\ 2007, A\&A, 465, 147 
\bibitem[Collins et al.(2013)]{Collins13} Collins, P., Kyne, G., Lara, D., et al.\ 2013, Experimental Astronomy, 36, 479 
\bibitem[Costa et al.(2010)]{Costa10} Costa, E., Bellazzini, R., Tagliaferri, G., et al.\ 2010, Experimental Astronomy, 28, 137 
\bibitem[Courvoisier(1998)]{Courvoisier98} Courvoisier, T.~J.-L.\ 1998, A\&ARv, 9, 1 
\bibitem[Courvoisier et al.(2003)]{Courvoisier03} Courvoisier, T.~J.-L., Beckmann, V., Bourban, G., et al.\ 2003, A\&A, 411, L343 
\bibitem[Dean et al.(2008)]{Dean08} Dean, A.~J., Clark, D.~J., Stephen, J.~B., et al.\ 2008, Science, 321, 1183 
\bibitem[Done et al.(1996)]{Done96} Done, C., Madejski, G.~M., \& Smith, D.~A.\ 1996, ApJL, 463, L63 
\bibitem[Droulans \& Jourdain(2009)]{Droulans09} Droulans, R., \& Jourdain, E.\ 2009, A\&A, 494, 229 
\bibitem[Elmouttie et al.(1997)]{Elmouttie97} Elmouttie, M., Haynes, R.~F., Jones, K.~L., et al.\ 1997, MNRAS, 284, 830 
\bibitem[Evans et al.(2004)]{Evans04} Evans, D.~A., Kraft, R.~P., Worrall, D.~M., et al.\ 2004, ApJ, 612, 786 
\bibitem[Fan et al.(2006)]{Fan06} Fan, X., Strauss, M.~A., Richards, G.~T., et al.\ 2006, AJ, 131, 1203 
\bibitem[Fender et al.(2009)]{Fender09} Fender, R.~P., Homan, J., \& Belloni, T.~M.\ 2009, MNRAS, 396, 1370 
\bibitem[Forot et al.(2008)]{Forot08} Forot, M., Laurent, P., Grenier, I.~A., Gouiff{\`e}s, C., \& Lebrun, F.\ 2008, ApJL, 688, L29 
\bibitem[Fukazawa et al.(2011a)]{Fukazawa11a} Fukazawa, Y., Hiragi, K., Mizuno, M., et al.\ 2011a, ApJ, 727, 19 
\bibitem[Fukazawa et al.(2011b)]{Fukazawa11b} Fukazawa, Y., Hiragi, K., Yamazaki, S., et al.\ 2011b, ApJ, 743, 124
\bibitem[Fukazawa et al.(2014)]{Fukazawa14} Fukazawa, Y., Finke, J., Stawarz, \L ., et al.\ 2014, ApJ, in press (arXiv:1410.2733) 
\bibitem[Gandhi et al.(2014)]{Gandhi14} Gandhi, P., Yamada, S., Ricci, C., et al.\ 2014, MNRAS, submitted (arXiv:1408.4453)
\bibitem[Gezari et al.(2009)]{Gezari09} Gezari, S., Strubbe, L., Bloom, J.~S., et al.\ 2009, astro2010: The Astronomy and Astrophysics Decadal Survey, 2010, 88 
\bibitem[Ghisellini et al.(2010)]{Ghisellini10} Ghisellini, G., Della Ceca, R., Volonteri, M., et al.\ 2010, MNRAS, 405, 387 
\bibitem[Ghisellini et al.(2011)]{Ghisellini11} Ghisellini, G., Tagliaferri, G., Foschini, L., et al.\ 2011, MNRAS, 411, 901 
\bibitem[Ghisellini et al.(2013)]{Ghisellini13} Ghisellini, G., Nardini, M., Tagliaferri, G., J., et al.\ 2013, MNRAS, 428, 1449 
\bibitem[Gofford et al.(2013)]{Gofford13} Gofford, J., Reeves, J.~N., Tombesi, F., et al.\ 2013, MNRAS, 430, 60
\bibitem[Goulding \& Alexander(2009)]{Goulding09} Goulding, A.~D., \& Alexander, D.~M.\ 2009, MNRAS, 398, 1165 
\bibitem[G{\"o}tz et al.(2009)]{Goetz09} G{\"o}tz, D., Laurent, P., Lebrun, F., Daigne, F., \& Bo{\v s}njak, {\v Z}.\ 2009, ApJL, 695, L208 
\bibitem[Greenhill et al.(1997)]{Greenhill97} Greenhill, L.~J., Moran, J.~M., \& Herrnstein, J.~R.\ 1997, ApJL, 481, L23
\bibitem[Hartman et al.(1999)]{Hartman99} Hartman, R.~C., Bertsch, D.~L., Bloom, S.~D., et al.\ 1999, ApJS, 123, 79 
\bibitem[Hayashida et al.(2013)]{Hayashida13} Hayashida, M., Stawarz, {\L}., Cheung, C.~C., et al.\ 2013, ApJ, 779, 131
\bibitem[Hester et al.(1995)]{Hester95} Hester, J.~J., Scowen, P.~A., Sankrit, R., et al.\ 1995, ApJ, 448, 240  
\bibitem[Hill et al.(2008)]{Hill08} Hill, J.~E., McConnell, M.~L., Bloser, P., et al.\ 2008, American Institute of Physics Conference Series, 1065, 331 
\bibitem[H{\"o}nig et al.(2013)]{Hoenig13} H{\"o}nig, S.~F., Kishimoto, M., Tristram, K.~R.~W., et al.\ 2013, ApJ, 771, 87 
\bibitem[Israel(1998)]{Israel98} Israel, F.~P.\ 1998, A\&ARv, 8, 237 
\bibitem[Itoh et al.(2008)]{Itoh08} Itoh, T., Done, C., Makishima, K., et al.\ 2008, PASJ, 60, 251 
\bibitem[Iwasawa et al.(1993)]{Iwasawa93} Iwasawa, K., Koyama, K., Awaki, H., et al.\ 1993, ApJ, 409, 155
\bibitem[Johnson et al.(1995)]{Johnson95} Johnson, W.~N., Dermer, C.~D., Kinzer, R.~L., et al.\ 1995, ApJ, 445, 182 
\bibitem[Jourdain et al.(2012)]{Jourdain12} Jourdain, E., Roques, J.~P., Chauvin, M., \& Clark, D.~J.\ 2012, ApJ, 761, 27 
\bibitem[Kataoka et al.(2002)]{Kataoka02} Kataoka, J., Tanihata, C., Kawai, N., et al.\ 2002, MNRAS, 336, 932 
\bibitem[Kataoka et al.(2011)]{Kataoka11} Kataoka, J., Stawarz, {\L}., Takahashi, Y., et al.\ 2011, ApJ, 740, 29 
\bibitem[Kormendy \& Ho(2013)]{Kormendy13} Kormendy, J., \& Ho, L.~C.\ 2013, ARA\&A, 51, 511 
\bibitem[Krabbe et al.(2001)]{Krabbe01} Krabbe, A., B{\"o}ker, T., \& Maiolino, R.\ 2001, ApJ, 557, 626 
\bibitem[Krawczynski(2012)]{Krawczynski12} Krawczynski, H.\ 2012, ApJ, 744, 30 
\bibitem[Krawczynski et al.(2011)]{Krawczynski11} Krawczynski, H., Garson, A., Guo, Q., et al.\ 2011, Astroparticle Physics, 34, 550 
\bibitem[Laurent et al.(2011)]{Laurent11} Laurent, P., Rodriguez, J., Wilms, J., et al.\ 2011, Science, 332, 438 
\bibitem[Lei et al.(1997)]{Lei97} Lei, F., Dean, A.~J., \& Hills, G.~L.\ 1997, Space Science Reviews, 82, 309 
\bibitem[Lenc \& Tingay(2009)]{Lenc09} Lenc, E., \& Tingay, S.~J.\ 2009, AJ, 137, 537 
\bibitem[Levan et al.(2011)]{Levan11} Levan, A.~J., Tanvir, N.~R., Cenko, S.~B., et al.\ 2011, Science, 333, 199 
\bibitem[Liedahl \& Paerels(1996)]{Liedahl96} Liedahl, D.~A., \& Paerels, F.\ 1996, ApJL, 468, L33 
\bibitem[Lightman \& Shapiro(1976)]{Lightman76} Lightman, A.~P., \& Shapiro, S.~L.\ 1976, ApJ, 203, 701 
\bibitem[Lohfink et al.(2013)]{Lohfink13} Lohfink, A.~M., Reynolds, C.~S., Jorstad, S.~G., et al.\ 2013, ApJ, 772, 83
\bibitem[Long et al.(1980)]{Long80} Long, K.~S., Chanan, G.~A., \& Novick, R.\ 1980, ApJ, 238, 710 
\bibitem[Maccione et al.(2008)]{Maccione08} Maccione, L., Liberati, S., Celotti, A., Kirk, J.~G., \& Ubertini, P.\ 2008, Phys.Rev.D, 78, 103003 
\bibitem[Madejski et al.(2000)]{Madejski00} Madejski, G., {\.Z}ycki, P., Done, C., et al.\ 2000, ApJL, 535, L87
\bibitem[Marconi et al.(2000)]{Marconi00} Marconi, A., Oliva, E., van der Werf, P.~P., et al.\ 2000, A\&A, 357, 24 
\bibitem[Markowitz et al.(2007)]{Markowitz07} Markowitz, A., Takahashi, T., Watanabe, S., et al.\ 2007, ApJ, 665, 209 
\bibitem[Marinucci et al.(2012)]{Marinucci12} Marinucci, A., Risaliti, G., Wang, J., et al.\ 2012, MNRAS, 423, L6 
\bibitem[Marscher et al.(2002)]{Marscher02} Marscher, A.~P., Jorstad, S.~G., G{\'o}mez, J.-L., et al.\ 2002, Nature, 417, 625 
\bibitem[Marscher et al.(2008)]{Marscher08} Marscher, A.~P., Jorstad, S.~G., D'Arcangelo, F.~D., et al.\ 2008, Nature, 452, 966 
\bibitem[Massaro et al.(2006)]{Massaro06} Massaro, F., Bianchi, S., Matt, G., D'Onofrio, E., \& Nicastro, F.\ 2006, A\&A, 455, 153
\bibitem[Mayer et al.(2013)]{Mayer13} Mayer, M., Buehler, R., Hays, E., et al.\ 2013, ApJL, 775, L37
\bibitem[McGlynn et al.(2007)]{McGlynn07} McGlynn, S., Clark, D.~J., Dean, A.~J., et al.\ 2007, A\&A, 466, 895 
\bibitem[McKinney(2006)]{McKinney06} McKinney, J.~C.\ 2006, MNRAS, 368, 1561
\bibitem[McNamara et al.(2009)]{McNamara09} McNamara, A.~L., Kuncic, Z., \& Wu, K.\ 2009, MNRAS, 395, 1507 
\bibitem[Merritt \& Wang(2005)]{Merritt05} Merritt, D., \& Wang, J.\ 2005, ApJL, 621, 101
\bibitem[Mirabel \& Rodr{\'{\i}}guez(1994)]{Mirabel94} Mirabel, I.~F., \& Rodr{\'{\i}}guez, L.~F.\ 1994, Nature, 371, 46 
\bibitem[Moorwood \& Oliva(1994)]{Moorwood94} Moorwood, A.~F.~M., \& Oliva, E.\ 1994, ApJ, 429, 602 
\bibitem[Moorwood et al.(1996)]{Moorwood96} Moorwood, A.~F.~M., van der Werf, P.~P., Kotilainen, J.~K., Marconi, A., \& Oliva, E.\ 1996, A\&A, 308, L1 
\bibitem[Moran et al.(2013)]{Moran:2013} Moran, P., et al., 2013, arXiv:1302.3622v1 
\bibitem[Mortlock et al.(2011)]{Mortlock11} Mortlock, D.~J., Warren, S.~J., Venemans, B.~P., et al.\ 2011, Nature, 474, 616
\bibitem[Mukai et al.(2008)]{Mukai08} Mukai, K., Orio, M., \& Della Valle, M.\ 2008, ApJ, 677, 1248 
\bibitem[Murphy \& Yaqoob(2009)]{Murphy09} Murphy, K.~D., \& Yaqoob, T.\ 2009, MNRAS, 397, 1549 
\bibitem[Odaka et al.(2010)]{Odaka:2010} Odaka, H., et al., 2010, Nucl.\ Instr.\ Meth.\ A, 624, 303
\bibitem[Orsi et al.(2011)]{Orsi11} Orsi, S., \& Polar Collaboration 2011, Astrophysics and Space Sciences Transactions, 7, 43 
\bibitem[Ozaki et al.(2012)]{Ozaki:2012} Ozaki, M., et al., 2012, Proc. SPIE, 8443, 844356
\bibitem[Pacholczyk \& Swihart(1967)]{Pacholczyk67} Pacholczyk A.~G., Swihart T.~L., 1967, ApJ, 150, 647 
\bibitem[Pearce et al.(2012)]{Pearce12} Pearce, M., Flor{\'e}n, H.-G., Jackson, M., et al.\ 2012, arXiv:1211.5094 
\bibitem[Pian et al.(1998)]{Pian98} Pian, E., Vacanti, G., Tagliaferri, G., et al.\ 1998, ApJL, 492, L17 
\bibitem[Porquet \& Dubau(2000)]{Porquet00} Porquet, D., \& Dubau, J.\ 2000, A\&AS, 143, 495 
\bibitem[Poutanen(1994)]{Poutanen94} Poutanen, J.\ 1994, ApJS, 92, 607 
\bibitem[Puccetti et al.(2014)]{Puccetti14} Puccetti, S., Comastri, A., Fiore, F., et al.\ 2014, ApJ, 793, 26 
\bibitem[Rodriguez et al.(2008)]{Rodriguez08} Rodriguez, J., Shaw, S.~E., Hannikainen, D.~C., et al.\ 2008, ApJ, 675, 1449 
\bibitem[Rupen et al.(2014)]{Rupen14} Rupen, M.~P., Mioduszewski, A.~J., Chomiuk, L., et al.\ 2014, The Astronomer's Telegram, 5884, 1 
\bibitem[Sambruna et al.(2001a)]{Sambruna01a} Sambruna, R.~M., Brandt, W.~N., Chartas, G., et al.\ 2001a, ApJL, 546, L9 
\bibitem[Sambruna et al.(2001b)]{Sambruna01b} Sambruna, R.~M., Netzer, H., Kaspi, S., et al.\ 2001b, ApJL, 546, L13 
\bibitem[Saxton \& Komossa(2012)]{Saxton12} Saxton, R., \& Komossa, S.\ 2012, European Physical Journal Web of Conferences, 39, 00001 
\bibitem[S{\c a}dowski et al.(2014)]{Sadowski14} S{\c a}dowski, A., Narayan, R., McKinney, J.~C., \& Tchekhovskoy, A.\ 2014, MNRAS, 439, 503
\bibitem[Sbarrato et al.(2013)]{Sbarrato13} Sbarrato, T., Tagliaferri, G., Ghisellini, G., et al.\ 2013, ApJ, 777, 147 
\bibitem[Sch{\"o}nfelder et al.(2000)]{Schoenfelder00} Sch{\"o}nfelder, V., Bennett, K., Blom, J.~J., et al.\ 2000, A\&AS, 143, 145 
\bibitem[Schurch et al.(2002)]{Schurch02} Schurch, N.~J., Roberts, T.~P., \& Warwick, R.~S.\ 2002, MNRAS, 335, 241 
\bibitem[Shu et al.(2011)]{Shu11} Shu, X.~W., Yaqoob, T., \& Wang, J.~X.\ 2011, ApJ, 738, 147 
\bibitem[Sikora et al.(2009)]{Sikora09} Sikora, M., Stawarz, {\L}., Moderski, R., Nalewajko, K., \& Madejski, G.~M.\ 2009, ApJ, 704, 38 
\bibitem[S{\l}owikowska et al.(2009)]{Slowikowska09} S{\l}owikowska, A., Kanbach, G., Kramer, M., \& Stefanescu, A.\ 2009, MNRAS, 397, 103
\bibitem[Smith et al.(1988)]{Smith88} Smith, F.~G., Jones, D.~H.~P., Dick, J.~S.~B., \& Pike, C.~D.\ 1988, MNRAS, 233, 305
\bibitem[Soffitta et al.(2013)]{Soffitta13} Soffitta, P., Barcons, X., Bellazzini, R., et al.\ 2013, Experimental Astronomy, 36, 523 
\bibitem[Soldi et al.(2008)]{Soldi08} Soldi, S., T{\"u}rler, M., Paltani, S., et al.\ 2008, A\&A, 486, 411 
\bibitem[Steinle et al.(1998)]{Steinle98} Steinle, H., Bennett, K., Bloemen, H., et al.\ 1998, A\&A, 330, 97 
\bibitem[Stirling et al.(2001)]{Stirling01} Stirling, A.~M., Spencer, R.~E., de la Force, C.~J., et al.\ 2001, MNRAS, 327, 1273 
\bibitem[Striani et al.(2011)]{Striani11} Striani, E., Tavani, M., Piano, G., et al.\ 2011, ApJL, 741, L5 
\bibitem[Tajima et al.(2010a)]{Tajima:2010a} Tajima, H.\ et al., 2010, Proc. SPIE, 7732, 773216
\bibitem[Tajima et al.(2010b)]{Tajima:2010b} Tajima, H.\ et al., 2010, X-ray Polarimetry: A New Window in Astrophysics by Cambridge University Press, p.\ 275
\bibitem[Takahashi et al.(2013)]{Takahashi13} Takahashi, T., Uchiyama, Y., \& Stawarz, {\L}.\ 2013, Astroparticle Physics, 43, 142 
\bibitem[Takeda et al.(2010)]{Takeda:2010} Takeda, S., et al., 2010, Nucl.\ Instr.\ Meth.\ A, 622, 619
\bibitem[Takei et al.(2009)]{Takei09} Takei, D., Tsujimoto, M., Kitamoto, S., et al.\ 2009, ApJL, 697, L54 
\bibitem[Tavani et al.(2011)]{Tavani11} Tavani, M., Bulgarelli, A., Vittorini, V., et al.\ 2011, Science, 331, 736 
\bibitem[Tchekhovskoy et al.(2011)]{Tchekhovskoy11} Tchekhovskoy, A., Narayan, R., \& McKinney, J.~C.\ 2011, MNRAS, 418, L79
\bibitem[Tombesi et al.(2010a)]{Tombesi10a} Tombesi, F., Sambruna, R.~M., Reeves, J.~N., et al.\ 2010a, ApJ, 719, 700
\bibitem[Tombesi et al.(2010b)]{Tombesi10b} Tombesi, F., Cappi, M., Reeves, J.~N., et al.\ 2010b, A\&A, 521, A57
\bibitem[Tombesi et al.(2012)]{Tombesi12} Tombesi, F., Sambruna, R.~M., Marscher, A.~P., et al.\ 2012, MNRAS, 424, 754
\bibitem[Tombesi et al.(2013)]{Tombesi13} Tombesi, F., Reeves, J.~N., Reynolds, C.~S., Garc{\'{\i}}a, J., \& Lohfink, A.\ 2013, MNRAS, 434, 2707
\bibitem[Tombesi et al.(2014)]{Tombesi14} Tombesi, F., Tazaki, Mushotzky, R. F., Ueda, Y., et al.\ 2014, MNRAS, 443, 2154
\bibitem[Reeves et al.(2010)]{Reeves10} Reeves, J.~N., Gofford, J., Braito, V., \& Sambruna, R.\ 2010, ApJ, 725, 803
\bibitem[Urry \& Padovani(1995)]{Urry95} Urry, C.~M., \& Padovani, P.\ 1995, PASP, 107, 803 
\bibitem[Vasiliev \& Merritt(2013)]{Vasiliev13} Vasiliev, E., \& Merritt, D.\ 2013, ApJ, 774, 87
\bibitem[Vasudevan et al.(2010)]{Vasudevan10} Vasudevan, R.~V., Fabian, A.~C., Gandhi, P., Winter, L.~M., \& Mushotzky, R.~F.\ 2010, MNRAS, 402, 1081
\bibitem[Volonteri et al.(2011)]{Volonteri11} Volonteri, M., Haardt, F., Ghisellini, G., \& Della Ceca, R.\ 2011, MNRAS, 416, 216 
\bibitem[Walton et al.(2011)]{Walton11} Walton, D.~J., Roberts, T.~P., Mateos, S., \& Heard, V.\ 2011, MNRAS, 416, 1844
\bibitem[Watanabe et al.(2012)]{Watanabe:2012} Watanabe, S., et al., 2012, Proc.\ SPIE, 8443, 844326
\bibitem[Watanabe et al.(2014)]{Watanabe:2014} Watanabe, S., et al., 2014, Nucl.\ Instr.\ Meth.\ A, in press
\bibitem[Weisskopf et al.(1976)]{Weisskopf76} Weisskopf, M.~C., Cohen, G.~G., Kestenbaum, H.~L., et al.\ 1976, ApJL, 208, L125 
\bibitem[Weisskopf et al.(2009)]{Weisskopf09} Weisskopf, M.~C., Elsner, R.~F., Kaspi, V.~M., et al.\ 2009, Astrophysics and Space Science Library, 357, 589 
\bibitem[Willott et al.(2010)]{Willott10} Willott, C.~J., Delorme, P., Reyl{\'e}, C., et al.\ 2010, AJ, 139, 906
\bibitem[Yang et al.(2009)]{Yang09} Yang, Y., Wilson, A.~S., Matt, G., Terashima, Y., \& Greenhill, L.~J.\ 2009, ApJ, 691, 131 
\bibitem[Yaqoob(2012)]{Yaqoob12} Yaqoob, T.\ 2012, MNRAS, 423, 3360 
\bibitem[Zdziarski et al.(2014)]{Zdziarski14} Zdziarski, A.~A., Pjanka, P., Sikora, M., \& Stawarz, L.\ 2014, 442, 3243 
\bibitem[Zhang \& Boettcher(2013)]{Zhang13} Zhang, H., Boettcher, M.\ 2013, ApJ, 774, 18 

\end{thebibliography}
}
\end{document}